\documentclass[12pt,a4paper]{elsarticle}
\usepackage[top=25mm,bottom=25mm,left=2.4cm,right=2.4cm,asymmetric]{geometry}

\newcommand{\eref}[1]{(\ref{#1})}

\newcommand{\vect}[1]{\mathbf{#1}}
\newcommand{\vecsym}[1]{\boldsymbol{#1}}
\newcommand{\mtrx}[1]{\mathds{#1}}

\newcommand{\sref}[1]{Section~\ref{#1}}

\usepackage{chngcntr}
\usepackage{amssymb}
\usepackage{amsmath}
\usepackage{bbm}
\usepackage{bm}
\usepackage{dsfont}
\usepackage[nodots]{numcompress}
\usepackage{url}
\usepackage{natbib}
\usepackage{units}
\usepackage{graphicx}
\usepackage{subfigure}
\usepackage{caption}
\usepackage{color}
\usepackage[titletoc,toc,title]{appendix}
\usepackage{geometry}

\biboptions{authoryear}

\let\today\relax
\journal{International Journal}
\makeatletter
\def\ps@pprintTitle{%
	\let\@oddhead\@empty
	\let\@evenhead\@empty
	\let\@oddfoot\@empty
	\let\@evenfoot\@oddfoot
}
\makeatother

\begin{document}

\begin{frontmatter}

\title{A strengthening model of particle-matrix interaction based on an axisymmetric strain gradient plasticity analysis}

\author{Mohammadali Asgharzadeh}
\author{Jonas Faleskog\corref{CORR}}

\address{Solid Mechanics, Department of Engineering Mechanics, KTH Royal Institute of Technology, 10044 Stockholm, Sweden}

\cortext[CORR]{Corresponding author. E-mail: faleskog@kth.se}

\begin{abstract}
Precipitation of fine particles into the base material of a metal is a potent strengthening mechanism. This is numerically analyzed within a continuum framework based on a higher order strain gradient plasticity theory and by use of an axi-symmetric unit cell model. The unit cell contains a spherical particle which is resilient to inelastic deformation and embedded in a homogeneous matrix material. An interface with special characteristics, that separates the particle from the matrix, plays a key role for the overall strengthening. Based on a systematic parametric study a closed form relation is deduced and proposed for the increase in the overall yield stress. This relation is limited to materials containing elastic particles with spacing smaller than the material length scale and volume fractions less than 10 $\%$. It these conditions are met, the plastic strain field in the material becomes essentially constant on the scale of particle spacing. The character of the solution suggests that this result is general despite the simplicity of the unit cell employed in the parametric study. Predictions from the closed form relation is compared with published experimental results, and good agreement is observed in some metal matrix composites and alloys. The influence of a mismatch in elastic modulus between particles and matrix is elucidated, and effects on post yield strain hardening are discussed. 
\end{abstract}


\end{frontmatter}

\section{Introduction}

\noindent The mechanical behavior of multiphase metals on the macro scale depends on their complex structures on the micro scale, ranging from nanometers to microns. Upon loading, the state of deformation on the micro scale will be highly heterogeneous, even if a homogeneous load is applied on the macro scale. The presence of heterogeneous deformation introduces size scale effects. The well-known Hall-Petch relation between grain size and strength is an example of this phenomenon. There, the underlying mechanism and source of micro scale heterogeneity is due to grain boundaries, which raises the resistance against plastic deformation carriers, which in turn results in increased strength of the material. Another example that introduces size scale effects is precipitation hardening.

Precipitation is a potent strengthening mechanism in metals and relies on introduction of fine particles into the base material. As dislocations glide, they frequently hit those particles intersecting with the slip plane. So, particles hinder the movement of dislocations in the crystal lattice. Since dislocations are the dominant carriers of plasticity, this serves to harden the material. Depending on the strength of the particle, which is related to its size, misfit, and coherency, particles can be categorized in two main groups: `hard` particles, which are bypassed by dislocations, and `soft` particles, which are penetrated by dislocations when the material starts to deform plastically (\cite{Gladman99}). The bypassing occurs either by Orowan looping or cross-slip (\cite{Orowan48}), and the penetration occurs by particle shearing (\cite{friedel2013dislocations}). In the former, dislocations may be assumed to be more homogeneously distributed, and in the latter, dislocations appear to be more organized into bands that cut through the particles (\cite{Poole05}). The bypassing of dislocations generates geometrically necessary dislocation (GND) loops around particles. The net strengthening effect of this process has been observed to correlate to the inverse of the average distance between the particles (\cite{Orowan48}, \cite{Ashby70}). A comprehensive account of precipitation hardening is given in \cite{Nembach97}.

The study of precipitation hardening got boosted after the development of the concept of dislocations, and dislocation mechanics has been the dominant framework to study this phenomenon. The looping mechanism suggested by \cite{Orowan48} and the analytical solution introduced by \cite{friedel2013dislocations} for the cutting mechanism are both pioneering works within this approach. In recent years, more sophisticated methods have been used to study precipitation hardening, such as crystal plasticity (\cite{Ohashi04}, \cite{wulfinghoff2015gradient}, \cite{mayeur2015micropolar}) or dislocation dynamics (\cite{Chang12}, \cite{monnet2015multiscale}, \cite{hu2021modeling}).

Continuum mechanics-based theories have to some extent been used to study this phenomenon. Based on classical $J_2$ plasticity theory, \cite{Bao91} analyze how rigid particles reinforce metal matrix materials against plastic flow, but do not capture the accentuated effect of particles at small scales. However, the development of strain gradient plasticity (SGP) theories has overcome this shortcoming, see e.g. \cite{Aifantis87}, \cite{Fleck93}, \cite{Gao99}, \cite{Fleck01}, \cite{Gudmundson04}, \cite{Gurtin05a}, and \cite{Fleck09}, see also the review by \cite{Lubarda16}. Particles reinforcing a matrix is investigated in \cite{Fleck97}. The model by \cite{Gao99} is used to study particle size effects with an axisymmetric model in \cite{Xue02}, and to investigate size effects of a distribution of particles in a plane strain model by \cite{Yan07}. \cite{Borg06} use the SGP theory of \cite{Fleck01} to investigate strengthening in metals reinforced by rigid particles, and the theory of \cite{Gudmundson04} is employed to study size effects in metal matrix composites by \cite{Azizi14}. 

In a SGP framework, the plastic strain gradients represent GNDs, and thus contribute to the strengthening and hardening of the material, cf. \cite{Ashby70}. Since several of the aforementioned SGP theories predict that strengthening depends inversely on the characteristic length of the analyzed structure, at least one of primary features associated with precipitation hardening would be in place if modelling is based on these SGP theories.

\begin{figure}[!htb]
\begin{center}
\includegraphics[width=0.35\textwidth]{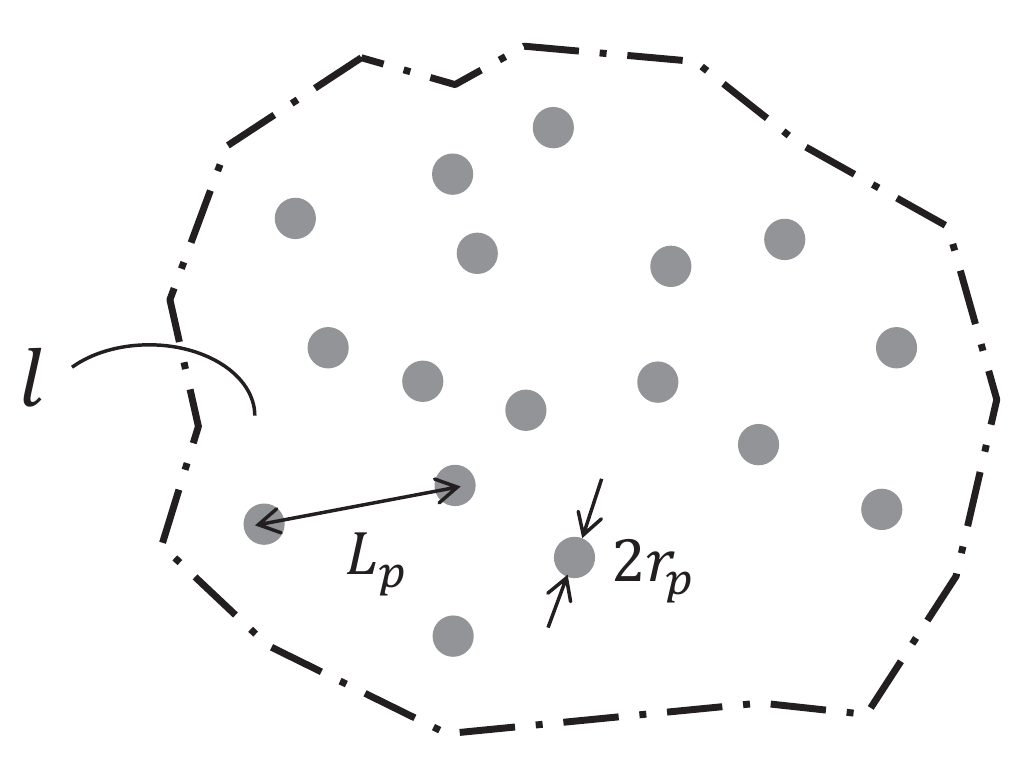}
\caption{Schematic of a micro-structure containing a distribution of particles, with the key parameters indicated; particle average radius ($r_{p}$), average distance between particles ($L_{p}$), and matrix material characteristic length scale ($\ell$).}
\label{fig:Intro}
\end{center}
\end{figure}

Figure \ref{fig:Intro} visualizes a precipitation strengthened material containing a distribution of small particles, where the typical average size $2r_{p}$ is found in the range from a few to several hundreds of nanometers. Several micro-structural parameters are important for the strength that can be achieved. These are: the average distance between particles $L_{p}$, the volume fraction of particles $f$, the mismatch in elastic modulus between the matrix and the particles, and inhomogeneity in the particle spacing distribution (clustering).  The inelastic deformation processes that lead to strengthening occurs on a length scale characteristic of the matrix material $\ell$, as indicated in Figure \ref{fig:Intro}. The impact of the length parameters characterizing the micro-structure, i.e. $L_{p}$ and $r_{p}$, must therefore be related to $\ell$.

In this work, a micro-mechanical model suitable for analyzing different parameters that affect precipitation hardening is proposed. A 2D axisymmetric model is employed that contains one particle embedded in a matrix modeled as a higher order SGP material. It will be shown that such a simple model is able to capture essential features of the parameters discussed above affecting strengthening of materials reinforced by small particles. A key element is an interface model that can be tuned to account for GNDs accumulating at the particle surface. The formulation of the interface model necessitates a higher order SGP theory that includes higher order stress and moment like variables that are captured by additional balance equations and boundary conditions (\cite{Gudmundson04}; \cite{Gurtin05a}; \cite{Gurtin09}; \cite{Fleck14}). Based on a systematic parametric study, a closed form expression is deduced and proposed that describes the influence of volume fraction and size of particles on strengthening. The model is then applied to predict the increase in yield stress in experiments carried out on precipitated materials and metal matrix composites published in the literature. 

The plan of this paper is as follows. In Section 2, the micro-mechanical model is presented and constitutive models for both bulk and interface are described. The results are given in section 3, where the influence of different parameters on the strengthening and hardening of the material is covered. Finally, concluding remarks are given in Section 4.

\section{Problem definition and modeling}
\label{sec:2}

\noindent The problem of interest here is the increase in yield strength and hardening due to the presence of precipitates in an otherwise homogeneous material. Focus will be on precipitates that behave as so called hard particles that act as strong obstacles for dislocation motion, causing dislocations to loop around the particles which leads to the formation of geometrically necessary dislocations (GND), i.e. the Orowan mechanism (\cite{Orowan48}). A key parameter in this strengthening mechanism is the average distance between particles, as the absolute increase in yield strength seems to scale with the inverse of this distance. To bring in a material specific length scale, a micro-mechanical model will be analyzed within the framework of the isotropic, higher order SGP theory proposed by \cite{Gudmundson04}. In this SGP theory, the interface between the hard particles and the surrounding material plays a crucial role for the strengthening, as it qualitatively controls the amount of GNDs that can be harbored at the interface. The analysis was carried out numerically by a modified and extended version of the SGP finite element method proposed by \cite{Dahlberg13a}.

\subsection{Particle distribution and micro-mechanical model}
\begin{figure}[!htb]
\begin{center}
\includegraphics[width=0.8\textwidth]{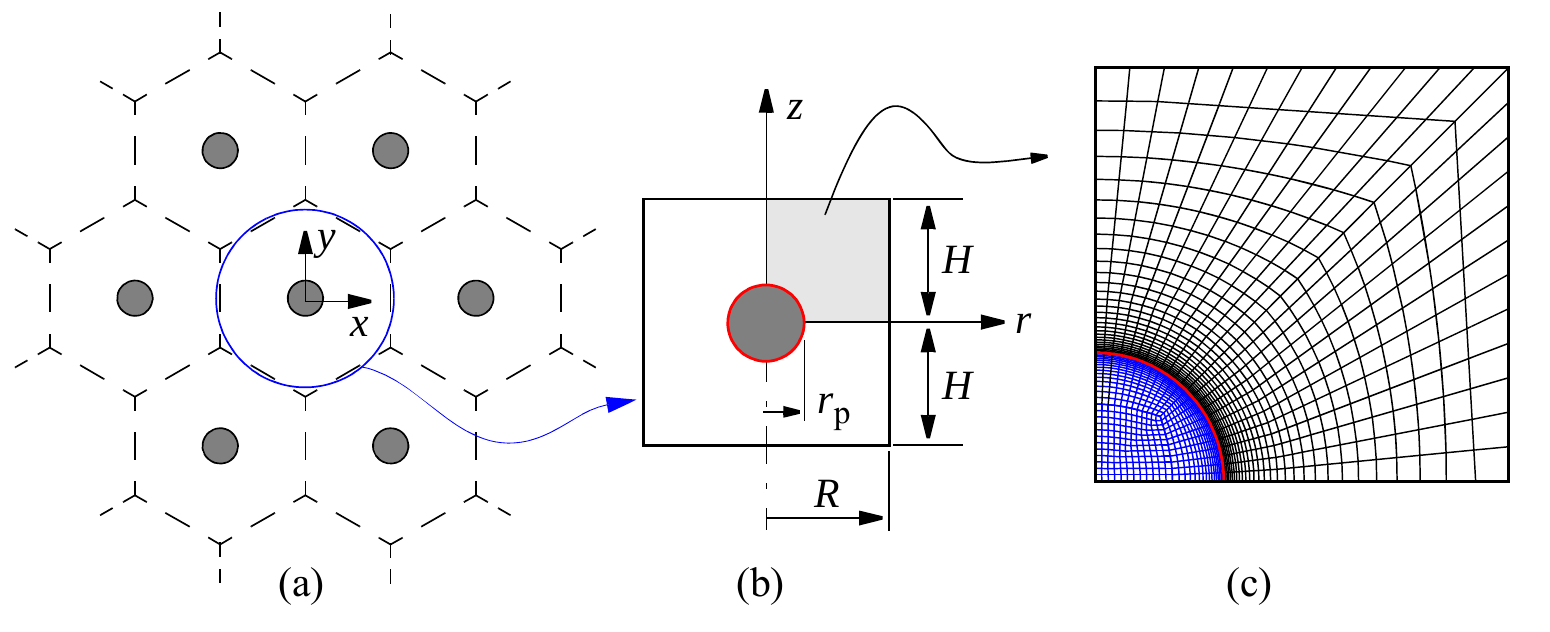}
\caption{(a) Particles are distributed in a hexagonal pattern in planes that are stacked on top of each other. (b) The corresponding axi-symmetric model used in this study illustrated for the special case of $xi = H/R = 1$. (c) A representative finite elements mesh used in the simulations illustrated by $ \xi = 1 $ and a volume fraction of particles equal to 2\%. The interface elements are highlighted in red.}
\label{fig:AxiModel}
\end{center}
\end{figure}

\noindent The particles considered in the study are regularly distributed and embedded in a homogeneous matrix material, and assumed to form a hexagonal pattern in planes that are stacked on top of each other, as illustrated in Figure \ref{fig:AxiModel}a. Hence, perpendicular to the planes, the particles form columns. This type of micro-structure can be analyzed by use of an axi-symmetric approximation, and thus a 2D axi-symmetric unit cell model will be employed in this work as shown in Figure \ref{fig:AxiModel}b.  As can be seen, the unit cell consists of three parts: a solid cylinder with radius $R$ and height $2H$ representing the matrix; a spherical particle with radius $r_{\rm p}$ in the center of the cell representing the regular distribution of the particles; and an interface layer between the former two to mimic the particle/matrix interaction (line in color red). Since the hard particle is impenetrable to dislocations, ruling out inelastic deformation, it is modeled by an isotropic, linear elastic material. Hence, all inelastic deformation is confined to the matrix material, which is modeled by an isotropic, elastic-SGP material with isotropic power law hardening.

To evaluate the strengthening effect, the unit cell may be subjected to either uniaxial tension or biaxial tension. The two types of loading differ only in stress triaxiality. The outcome of the analysis should preferably be interpreted based on the statistical measures of the particle distribution at hand. These measures are evaluated by considering that each particle is the center point of a 3D material unit defined by Voronoi tessellation. In the context of a regular hexagonal pattern, such a 3D unit will have 8 surface facets, which define the number of neighboring particles for each particle. It is then straightforward to evaluate the average distance $L_{\rm p}$, the standard deviation $S_{\rm D}$, and the coefficient of variance $C_{\rm V}$. It has been suggested that the latter is a promising measure for describing the effects of non-regularity in a microstructure on mechanical properties, cf. \cite{Yang01, Deng06}.

Two sets of parameters were considered: set C-C and set S-S. When evaluating the parameters for set C-C, the center-to-center distance between particles was used, whereas parameters for set S-S was evaluated using the surface-to-surface distance between particles. By introducing the volume fraction of particles, given by $f = (2/(3\xi)) \cdot (r_{\rm p}/R)^3$, where $\xi = H/R$, the parameters of the two sets become

\vspace{5mm}
\noindent
\begin{equation}\label{eqn:Lpcc}
\begin{array}{llllll }
\rm{Set}\;\rm{C}\rm{-}\rm{C:} & L_{\rm p} = R \cdot\theta(\xi) = \eta(\xi) r_{\rm p} f^{-1/3} , & \qquad & \qquad & \vspace{2mm} \\
& S_{\rm Dcc} = R\frac{\sqrt{3}}{2}|1-\xi|, \quad C_{\rm Vcc} = \frac{S_{\rm Dcc}}{L_{\rm pcc}} = \sqrt{3}\frac{|1-\xi|}{(3+\xi)}, & \qquad & \qquad &  \vspace{2mm} \\
& \textrm{where} \quad \theta = \frac{3+\xi}{2}, \quad \eta = \left(\frac{9}{4\xi}\right)^{1/3}\left(1+\frac{\xi}{3}\right), \quad \textrm{with} \quad \xi = H/R . & \qquad  & \qquad &
\end{array}
\end{equation}
\vspace{5mm}
\noindent
\begin{equation}\label{eqn:Lpss}
\begin{array}{llll}
\rm{Set}\;\rm{S}\rm{-}\rm{S:} & L_{\rm pss} = L_{\rm pcc}-2r_{\rm p} = R \cdot \theta_{\rm{ss}}(\xi,f) = \eta_{\rm ss}(\xi,f) r_{\rm p} f^{-1/3}, & \qquad & \qquad   \vspace{2mm} \\
& S_{\rm Dss} = S_{\rm Dcc}, \quad  C_{\rm Vss} = C_{\rm Vcc}\frac{L_{\rm pcc}}{L_{\rm pcc}-2r_{\rm p}} = C_{\rm{Vcc}} \frac{\theta}{\theta_{\rm ss}}, & \qquad & \qquad   \vspace{2mm} \\
& \textrm{where} \quad  \theta_{\rm{ss}} = \theta \cdot \left(1-\frac{4}{(3+\xi)} \left(\frac{3 \xi f}{2}\right)^{1/3}\right), \quad \eta_{\rm ss} = \eta \cdot  \left(1-2f^{1/3}\right). & \qquad  & \qquad 
\end{array}
\end{equation}
\noindent
The distribution parameters of the two sets will be investigated in the analysis.

\subsection{Strain gradient plasticity formulation and constitutive relations}

\noindent The primary kinematic variables in the higher order SGP theory of \cite{Gudmundson04} are the displacements $u_{i}$ and the plastic strains $\varepsilon_{ij}^{\rm p}$, related as
\begin{equation}\label{eqn:kinematic}
\varepsilon_{ij} = \frac{1}{2}(u_{i,j}+u_{j,i}), \quad \varepsilon_{ij} = \varepsilon_{ij}^{\rm e} + \varepsilon_{ij}^{\rm p},  \quad  \varepsilon_{kk}^{\rm p} = 0,
\end{equation}
where $\varepsilon_{ij}^{\rm e}$ are the elastic strain components, and where plastic incompressibility has been assumed.

The equilibrium relations are established by use of the principle of virtual work, where the work performed by plastic strains and the gradients of plastic strains in the unit cell volume $V$ are included. In addition, also the particle/matrix interface $S^{\Gamma}$ contributes to the internal virtual work. In absence of body forces, balance between internal and external virtual work becomes
\begin{equation}\label{eqn:VirtualWork}
\int_{V}\left[ \sigma_{ij}\delta\varepsilon_{ij} + (q_{ij}-s_{ij})\delta\varepsilon_{ij}^{\rm p} + m_{ijk}\delta\varepsilon_{ij,k}^{\rm p} \right]{\rm d}V +  \int_{S^{\Gamma}}\left[ M^{\Gamma}_{ij}\delta\varepsilon^{\rm p}_{ij}  \right]{\rm d}S = \int_{S^{\rm ext}}\left[ T_{i}\delta u_{i} + M_{ij}\delta\varepsilon_{ij}^{\rm p} \right]{\rm d}S.
\end{equation}
Here, $ \sigma_{ij} $ and $ s_{ij} = \sigma_{ij} - \delta_{ij}\sigma_{kk}/3 $ denote the components of the Cauchy stress tensor and its respective deviator, with $ \delta_{ij} $ being the Kronecker delta symbol. In addition, the micro stress $ q_{ij} $ and the moment stress $ m_{ijk} $ are introduced as work conjugates to the plastic strain tensor and its spatial gradient tensor, respectively. On the internal and external surfaces, the higher order moment tractions $M^{\Gamma}_{ij}$ and $M_{ij}$ consistently arise, both being work conjugates to $\varepsilon_{ij}^{\rm p}$. It should be noted that only the matrix material contributes to the internal virtual work at the interface, since particles are assumed to be elastic. From integration by parts of \eref{eqn:VirtualWork} the equilibrium equations, the natural boundary conditions, and the conditions at the internal interface are obtained as
\begin{equation}\label{eqn:equilibrium}
\begin{array}{lcll}
  \sigma_{ij,j} = 0        & {\rm and} & m_{ijk,k} + s_{ij} - q_{ij} = 0 & \textrm{ in $V = V_m + V_p$}, \vspace{2mm} \\
  \sigma_{ij}n_{j} = T_{i} & {\rm and} & m_{ijk}n_{k} = M_{ij}           & \textrm{ on $ S^{\rm ext} $}, \vspace{2mm} \\
  & &  M^{\Gamma}_{ij} + m_{ijk}n^{\Gamma}_{k} = 0 & \textrm{ on $ S^{\Gamma} $}.
\end{array}
\end{equation}
Here, $n_{k}$ denotes the outward normal vector to surface $S^{\rm ext}$, and $n^{\Gamma}_{k}$ denotes the normal vector to the interface surface $S^{\Gamma}$ in the direction from the matrix into the particle.

In both particles and matrix, the elastic strains are governed by 
\begin{equation}\label{eqn:Cijkl}
\sigma_{ij} = 2G\left(\frac{1}{2}(\delta_{ik}\delta_{jl}+\delta_{il}\delta_{jk}) + \frac{\nu}{1-2\nu}\delta_{ij}\delta_{kl} \right) \varepsilon_{kl}^{\rm e},
\end{equation}
where $G$ is the shear modulus and $\nu$ is Poisson's ratio. To distinguish between material parameters in the matrix and the particles, the sub-indices `m` and `p` will henceforth be used.

The inelastic deformation of the matrix material is assumed to be purely dissipative, such that the rate of plastic dissipation in the matrix material can be expressed as
\begin{equation}\label{eqn:dissipation}
\dot{D} = q_{ij} \dot{\varepsilon}_{ij}^{\rm p} + m_{ijk}\dot{\varepsilon}_{ij,k}^{\rm p} = \Sigma \dot{E^{p}} \geq 0,
\end{equation}
where $\Sigma$ and $\dot{E^{p}}$ are introduced as scalar work conjugate quantities to the stresses and the plastic strain rates, respectively. The effective quantities, which reduce to the standard $ J_{2} $-plasticity in absence of plastic strain gradients, may be defined as
\begin{equation}\label{eqn:EffectiveStress}
\Sigma = \sqrt{\frac{3}{2}\left( q_{ij}q_{ij} + \frac{m_{ijk}m_{ijk}}{\ell^{2}} \right)},
\end{equation}
\begin{equation}\label{eqn:EffectivePeeqRate}
\dot{E}^{\rm p} = \sqrt{\frac{2}{3}\left( \dot{\varepsilon}_{ij}^{\rm p}\dot{\varepsilon}_{ij}^{\rm p}   +  \ell^{2}\dot{\varepsilon}_{ij,k}^{\rm p}\dot{\varepsilon}_{ij,k}^{\rm p} \right) }.
\end{equation}
The inclusion of higher order terms necessarily brings in a length parameter $ \ell $, here viewed as a material constant. In the context of the current problem, $ \ell $ sets the scale on which plastic strain gradients may influence the strengthening due to the build-up of plastic strain gradients at the particle/matrix interface.

Rate-independent constitutive relations that satisfy \eref{eqn:dissipation}-\eref{eqn:EffectivePeeqRate} will be used to formulate the inelastic strain rates according to
\begin{equation}\label{eqn:const_qij}
\dot{\varepsilon}_{ij}^{\rm p} = \dot{E}^{\rm p} \frac{3q_{ij}}{2\Sigma},
\end{equation}
\begin{equation}\label{eqn:const_mijk}
\dot{\varepsilon}_{ij,k}^{\rm p} = \dot{E}^{\rm p}\frac{3m_{ijk}}{2\ell^{2}\Sigma},
\end{equation}
These equations are only valid if the yield condition $\Sigma = \sigma_{f}(E^{\rm p})$ is fulfilled and if continued plastic loading occurs, otherwise $\dot{E}^{\rm p}$ vanishes and the loading is elastic. In \eref{eqn:const_qij} and \eref{eqn:const_mijk}, $\sigma_{f}(E^{\rm p})$ is the flow stress which depends on the plastic work hardening of the matrix material, which was assumed to follow a power law hardening function as
\begin{equation}\label{eqn:flowfcn}
\sigma_{f} = \sigma_{0} \left( 1 + \frac{E^{\rm p}}{\varepsilon_{0}}\right)^{N}, \quad {\rm{with}} \quad  E^{\rm p} = \int{\dot{E}^{\rm p} {\rm{d}}t}, 
\end{equation}
where, $\sigma_{0}$ is the initial yield stress of a matrix material in absence of particles, $N$ is a plastic work hardening exponent and $\varepsilon_0 = \sigma_0 / (2G(1+\nu))$. 
 
\subsection{Modeling of the particle/matrix interface}
\label{sec:InterfaceDescription}
\noindent The purpose of the model for the particle/matrix interface is to account for the accumulation of GNDs around the particle in a phenomenological manner. \cite{Fredriksson05, Fredriksson07} have investigated different formulations, both of dissipative and energetic nature. They conclude that both formulations will yield the same response for proportional loading. Assuming that a surface energy $\Psi_{\Gamma}$ exists at the interface, the rate of dissipation at the interface can generally be expressed as

\begin{equation}\label{eqn:dissInt}
\dot{D}_{\Gamma } = \left( M^{\Gamma}_{ij} - \frac{\partial \Psi_{\Gamma}}{\partial \varepsilon_{ij}^{\rm p}} \right)\dot{\varepsilon}_{ij}^{\rm p} \le 0 
\end{equation}
Here, it will be assumed that the response of the interface will be purely energetic ($\dot{D}_{\Gamma } = 0$) and that the surface energy depends on a scalar measure of plastic strain as $\Psi_{\Gamma} = \Psi_{\Gamma}(\varepsilon_{\Gamma})$, where $\varepsilon_{\Gamma} = \sqrt{2 \varepsilon_{ij}^{\rm p} \varepsilon_{ij}^{\rm p} / 3}$ is the effective plastic strain at the interface. It is convenient to introduce $ \psi_{\Gamma} = {\rm d} \Psi_{\Gamma} / {\rm d} \varepsilon_{\Gamma}$. Thus, the moment tractions at the interface and $ \psi_{\Gamma} $ are given by
\begin{equation}\label{eqn:PsiInt}
M^{\Gamma}_{ij} = \frac{\partial \Psi_{\Gamma}}{\partial \varepsilon_{ij}^{\rm p}} =    \frac{{\rm d} \Psi_{\Gamma}}{{\rm d} \varepsilon_{\Gamma}} \frac{\partial \varepsilon_{\Gamma}}{\partial \varepsilon_{ij}^{\rm p}} =    \psi_{\Gamma} \frac{2}{3} \frac{\varepsilon_{ij}^{\rm p}}{\varepsilon_{\Gamma}} \quad \Rightarrow \quad \psi_{\Gamma} = \sqrt{\frac{3}{2}M^{\Gamma}_{ij}M^{\Gamma}_{ij}}
\end{equation}

\begin{figure}[!htb]
\begin{center}
\includegraphics[width=0.4\textwidth]{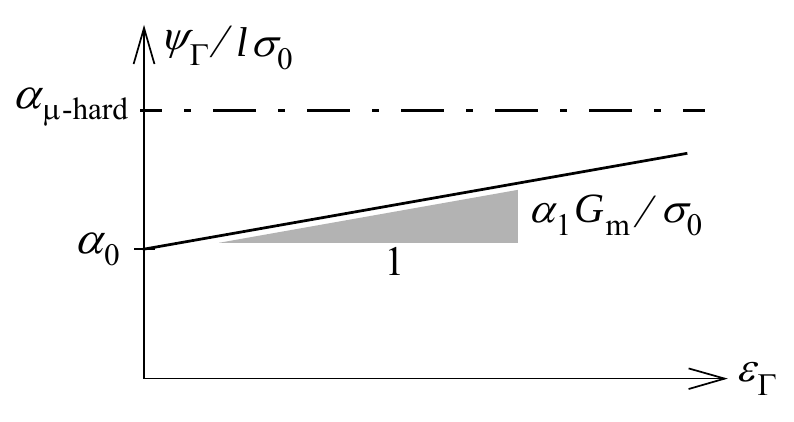}
\caption{Schematic of the behavior of introduced interface formulation, illustrating the meaning of the two key parameters $\alpha_0$ and $\alpha_1$.}
\label{fig:Interface_schematic}
\end{center}
\end{figure}

In view of \eref{eqn:PsiInt}, and to develop a suitable function for $\psi_{\Gamma}(\epsilon_{\Gamma})$, insight can be gained by exploring the limiting conditions at the interface, i.e., the so called 'micro-soft' and 'micro-hard' conditions. The micro-soft condition puts no constraint on the plastic strains at the interface, and only a minimal increase in strength would be obtained in the material. By contrast, for micro-hard condition, plastic strains at the interface are constrained to zero. For this condition ($\dot{\varepsilon}_{ij}^{\rm p} = 0$) it is possible to give a heuristic estimate as follows: from Eq. \eref{eqn:const_qij} it is seen that $q_{ij} = 0$ is valid at the interface and hence a plastic strain gradient can only develop normal to the interface; Eqs. \eref{eqn:const_mijk} and \eref{eqn:equilibrium} then gives $m_{ijk}m_{ijk} \rightarrow  M^{\Gamma}_{ij} M^{\Gamma}_{ij}$ and by virtue of \eref{eqn:EffectiveStress} that moment tractions on $S^{\Gamma}$ are limited by the yield condition which gives an upper bound for the surface energy as
\begin{equation}\label{eqn:PeakMomTrac}
\psi_{\Gamma} = \sqrt{ \frac{3}{2} M^{\Gamma}_{ij} M^{\Gamma}_{ij}} = \sigma_{\rm{0}} \ell   \quad  {\rm{for}} \quad \epsilon_{ij}^{p}=0 \quad {\rm{on}} \quad S^{\Gamma}.
\end{equation}
We will assume that the surface energy to first order is linear in the effective plastic strain at the interface with a second order correction. Hence, function $\psi_{\Gamma}(\epsilon_{\Gamma})$ may be expressed as
\begin{equation}\label{eqn:FaleskogAsgharzadehFunction}
  \psi_{\Gamma}(\epsilon_{\Gamma}) = \alpha_0 \ell \sigma_0 \cdot \left( 1 + \alpha_1 \frac{\varepsilon_{\Gamma} G_{\rm{m}} } {\alpha_0 \sigma_0} \right)  \thickspace .
\end{equation}
Here, the shear modulus of the matrix material $G_{\rm m}$ is introduced for dimensional reasons. This function contains two parameters $\alpha_0$ and $\alpha_1$ with the following interpretation. The first term defines the strength of the interface and the increase in yield stress of the material, whereas the second term affects the increase in plastic work hardening of the material. The interface response is depicted in Figure \ref{fig:Interface_schematic}, where also a micro-hard limit $\alpha_{\mu{\rm-hard}} $ is depicted, which constitutes an upper limit for the interface strength heuristically found to be one above.

\subsection{Numerical implementation and modeling}
  
\begin{figure}[htb!]
\begin{center}
    \subfigure[]{ \includegraphics[scale=0.65]{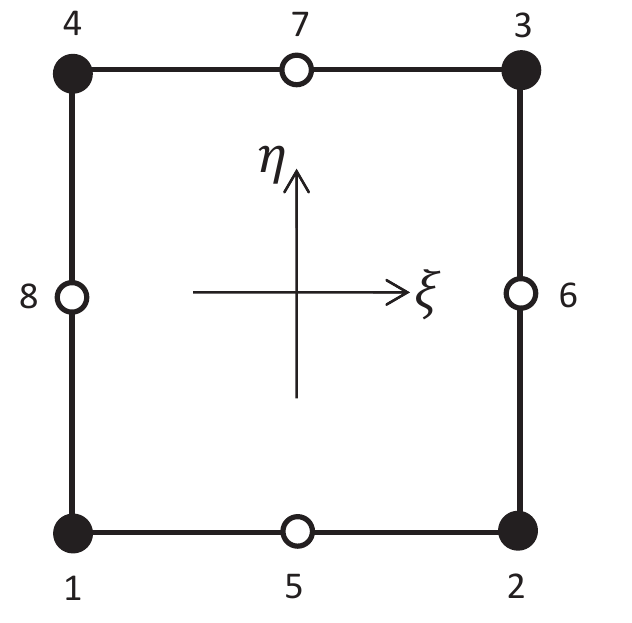} }
    \subfigure[]{ \includegraphics[scale=0.75]{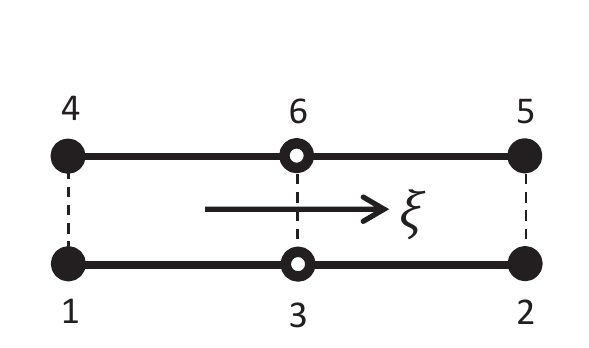} }
		\caption{The FEM elements employed; (a) 2D axi-symmetric continuum element, and (b) 1D interface element. Filled circles denote nodes with both displacement and plastic strains as degrees of freedom, while hollow circles denote nodes where only displacements degrees of freedom are described.}
	 \label{fig:FemElement}
\end{center}
\end{figure}

\noindent The micro-mechanical model introduced above was numerically analyzed by the finite element method. An axisymmetric 2D model with the mesh shown in Figure \ref{fig:AxiModel}c was employed. The mesh consists of quadrilateral 8-node elements for the bulk material (particle in color blue and matrix in color black) and 6-node interface elements to account for the special description of the particle/matrix interface (color red). In both types of elements, a quadratic interpolation is used for the displacement fields, and a linear interpolation is used for the plastic strain fields. Hence, all nodes contain displacement degrees of freedom (DOFs), and only the vertex nodes contain plastic strain DOFs, as illustrated in Figure \ref{fig:FemElement}. The primary variables at the nodes in the 2D continuum element are then described by the nodal displacement vector
\begin{equation}
\vect{d}_{\rm u}^{\rm T} = [u_{r}^{1} \quad u_{z}^{1} \ldots \quad u_{r}^{8} \quad u_{z}^{8}]_{16 \times 1},
\end{equation}
and the nodal plastic strain vector (only vertex nodes)
\begin{equation}
\vect{d}_{\rm p}^{\rm T} = [\varepsilon_{rr}^{{\rm p}~1} ~ \varepsilon_{zz}^{{\rm p}~1} ~ \gamma_{rz}^{{\rm p}~1} ~ \ldots ~ \varepsilon_{rr}^{{\rm p}~4} ~ \varepsilon_{zz}^{{\rm p}~4} ~ \gamma_{rz}^{{\rm p}~4}]_{12 \times 1}.
\end{equation}
Note that $ \gamma_{rz}^{\rm p} = 2\varepsilon_{rz}^{\rm p} $, and that plastic strain in the $ \phi $-direction is missing as it is given by plastic incompressibility ($ \varepsilon_{\phi\phi}^{{\rm p}} = - (\varepsilon_{rr}^{{\rm p}} + \varepsilon_{zz}^{{\rm p}})$). The DOFs for the primary variables in the interface element are described accordingly. A detailed account of the finite element discretization and implementation of this theory is presented in Appendix \ref{appendA} and \ref{appendB}. This element technology is used in \cite{Fredriksson09} and \cite{Dahlberg13b} to study problems based on a similar higher order SGP theory.

To study the strengthening effects on the macroscopic level (on the scale of the unit cell), two different load cases may be analyzed, uniaxial tension and biaxial tension. The conventional boundary conditions for these load cases are described as follows.

\vspace{2mm}
\noindent
Uniaxial tension:
\begin{equation}
\begin{array}{lcl}
  \dot{u}_{z} = \dot{u}_{z0} = {\rm constant} &    & \textrm{ on } z = H  \vspace{2mm} \\
  \dfrac{\partial u_{r}}{\partial z} = 0, \quad 2 \pi R \int T_{r}{\rm d}z = 0  &  & \textrm{ on } r = R.
\end{array}
\end{equation}
Biaxial tension:
\begin{equation}
\begin{array}{lcl}
  \dot{u}_{r} = \dot{u}_{r0}  = {\rm constant}  &   & \textrm{ on } r = R \vspace{2mm} \\
  \dfrac{\partial u_{z}}{\partial r} = 0, \quad  2\pi \int T_{z}r{\rm d}r = 0 & & \textrm{ on } z = H. 
\end{array}
\end{equation}

The above requirement of keeping the boundaries straight during loading was accomplished by use of Lagrange multipliers. In addition, symmetry boundary conditions were applied according to
\begin{equation}
 u_{z} = 0 \quad {\rm on} \quad z = 0, \quad {\rm and} \quad u_{r} = 0  \quad  {\rm on} \quad r = 0.
\end{equation}

\noindent
For both load cases, the higher order boundary conditions are described as
\begin{equation}
\begin{array}{ll}
M_{rr} = M_{zz} = 0, \thickspace \dot{\gamma}_{rz}^{\rm p} = 0  & {\rm on} \quad S^{\rm ext} \quad {\rm and} \quad z = 0, r = 0 \vspace{2mm} \\
\dot{\varepsilon}^{\rm p}_{ij} = 0  &  {\rm for} \quad 0 \le \sqrt{r^2+z^2} \le r_{\rm p},
\end{array}
\end{equation}
where the latter constraint is to ensure that the particle will remain elastic during loading. This was achieved by prescribing $\vect{d}_{\rm p} = 0$ for elements belonging to the particle (color blue in Fig. \ref{fig:AxiModel}c). 

It is not straight forward to establish a numerical procedure for the higher order SGP theory employed in the current study. For instance, how to define the onset of plastic yielding based on $q_{ij}$ and $m_{ijk}$ that are indeterminate in the purely elastic regime is not unambiguous (reminiscent of the situation in rigid plastic material). In recognition of this conundrum, \cite{Fleck09} put forward two minimum principles by which unique rate-independent solutions can be obtained. Utilizing these minimum principles, \cite{Nielsen09} propose a numerical model formulation for the present type of SGP theory. An alternative method is to use a visco-plastic formulation that in the limit reduces to the rate-independent formulation in \eref{eqn:const_qij} and\eref{eqn:const_mijk}. This approach was used in the present investigation as described in \cite{Dahlberg13a}. Equations \eref{eqn:const_qij} and\eref{eqn:const_mijk} are then altered such that the effective plastic strain rate (or plastic multiplyer) is replaced by a viscosity law according to
\begin{equation}\label{eqn:viscosity_law}
   \dot{E}^{\rm p} = \dot{\varepsilon}_{0} \Phi(\Sigma,\sigma_{\rm f}),
\end{equation}
where $ \dot{\varepsilon}_{0} $ is a reference strain rate, and $ \Phi(\Sigma,\sigma_{\rm f}) =\dot E^p/\dot\varepsilon_0$ represents a visco-plastic response function that depends on the effective stress $\Sigma$ and the flow stress $ \sigma_{\rm f}$ in the inviscid limit. The response function was chosen as (\cite{Dahlberg13a})
\begin{equation}\label{eqn:viscofcn}
\Phi(\Sigma,\sigma_{\rm f}) = \kappa\frac{\Sigma}{\sigma_{\rm f}} + \left(\frac{\Sigma}{\sigma_{\rm f}}\right)^{n}.
\end{equation}
This Ramberg-Osgood type of function remedies the indeterminacy of $q_{ij}$ and $m_{ijk}$ in the purely elastic regime, since a small but insignificant inelastic deformation will be present already upon initial loading. However, the outcome of an analysis will reduce to a rate-independent elastic-plastic response, provided that a sufficiently small value of $\kappa$ and a high enough value of the exponent $n$ are chosen. In addition, the overall applied rate of loading should preferably be chosen such that the resulting plastic strain rate is of order $ \dot{\varepsilon}_{0} $ or less. Thus, a possible rate dependency was neglected in the current study (cf. \cite{Anand05}) and accomplished by the choices $\kappa = 0.005 \varepsilon_{0}$ and $n = 2000$. A limited study was conducted to establish the latter, see Appendix \ref{appendC}.

Also, the equation for the interface strength \eref{eqn:FaleskogAsgharzadehFunction} is not amenable to numerical implementation as it is. The $\psi_{\Gamma}$-$\epsilon_{\Gamma}$ response must be altered to have an initial slope at $\epsilon_{\Gamma} = 0$, which was achieved by the modification
\begin{equation}\label{eqn:Mod_interface}
  \psi_{\Gamma}(\epsilon_{\Gamma}) \rightarrow \psi_{\Gamma}(\epsilon_{\Gamma}) \cdot \frac{ (\varepsilon_{\Gamma} / \kappa_{\Gamma})} {\left(1 + (\varepsilon_{\Gamma} / \kappa_{\Gamma})^{p}  \right)^{1/p}}  \thickspace ,
\end{equation}
where the additional function (essentially a continuous Heaviside function) introduces an initial slope of $\alpha_0 \ell \sigma_0 / \kappa_{\Gamma}$ as well as provides a smooth transition to  \eref{eqn:FaleskogAsgharzadehFunction}. If $\kappa_{\Gamma} \ll \varepsilon_{0}$ and $p \gg 1$, \eref{eqn:FaleskogAsgharzadehFunction} is retained and the interface response depicted in Figure \ref{fig:Interface_schematic} obtained.  In the present study, this was fulfilled by the values $\kappa_{\Gamma} = 0.1\varepsilon_{0}$ and $p = 5$.

The axisymmetric element and the new interface model were implemented in the SGP-FEM code developed by \cite{Dahlberg13a}, where a sparse solver (Pardiso) and OpenMP parallelization were utilized to speed up the computations. A fully backward Euler method was used to update stresses and to calculate a consistent tangent stiffness for the elements. To obtain accurate solutions for a case with a micro-hard interface, a mesh with a high resolution of elements in the matrix near the interface is needed. Here, continuum elements with a radial length of $0.002 r_{\rm p}$ near the interface proved to be sufficient as discussed in Appendix C.

Macroscopic stresses and strains were obtained as volume average values by homogenization. 
Homogenization involving higher order SGP theories is discussed in \cite{Dahlberg13b}, and for the boundary conditions employed in the current work, standard equations may be used. The macroscopic strains may then be evaluated as
\begin{equation}\label{eqn:EijDefinition}
E_{ij} = \frac{1}{V}\int_{S^{\rm ext}} \frac{1}{2}\left( u_{i}n_{j} + u_{j}n_{i} \right)~{\rm d}S,  
\end{equation}
which in the present problem simplifies to
\begin{equation}\label{eqn:EijDefByU}
   E_{rr} = E_{\varphi\varphi} = \frac{\dot{u}_{r0} t}{R}, \quad E_{zz} = \frac{\dot{u}_{z0} t}{H},
\end{equation}
where the other $E_{ij} = 0$, and $t$ represents a pseudo time, which is zero at the start of an analysis and unity at the end. 
\noindent
The macroscopic stresses can be evaluated as
\begin{equation}\label{eqn:SigmaijDef}
\Sigma_{ij} = \frac{1}{V} \int_{S^{\rm ext}} \frac{1}{2} ( \sigma_{ik}x_{j}n_{k} + \sigma_{jk}x_{i}n_{k} ) ~{\rm d}S  ~ \Rightarrow ~ \Sigma_{rr} = \Sigma_{\varphi\varphi} = 2 \pi R \int T_{r}{\rm d}z, ~ \Sigma_{zz} = 2 \pi \int T_{z}r{\rm d}z,
\end{equation}
where the other $\Sigma_{ij} = 0$, and $x_k$ is components of the position vector. Below, stress-strain curves will be presented in terms of effective measures. From \eref{eqn:SigmaijDef}, the effective macroscopic stress is given by
\begin{equation}\label{eqn:EffStressDef}
\Sigma_{\rm e} = \sqrt{\frac{3}{2} S_{ij}S_{ij}} ~, ~ S_{ij} = \Sigma_{ij} - \frac{1}{3}\Sigma_{kk}\delta_{ij},
\end{equation}
and from \eref{eqn:EijDefByU} effective macroscopic strain is defined as
\begin{equation}\label{eqn:EffStrainDef}
E_{\rm e} = \sqrt{\frac{2}{3} E^{\rm d}_{ij}E^{\rm d}_{ij} } ~, ~  E^{\rm d}_{ij} = E_{ij} - \frac{1}{3}E_{kk}\delta_{ij}.
\end{equation}

\section{Results and discussion}
\label{sec:3}

\begin{figure}[htb!]
\begin{center}
  \includegraphics[scale=0.9]{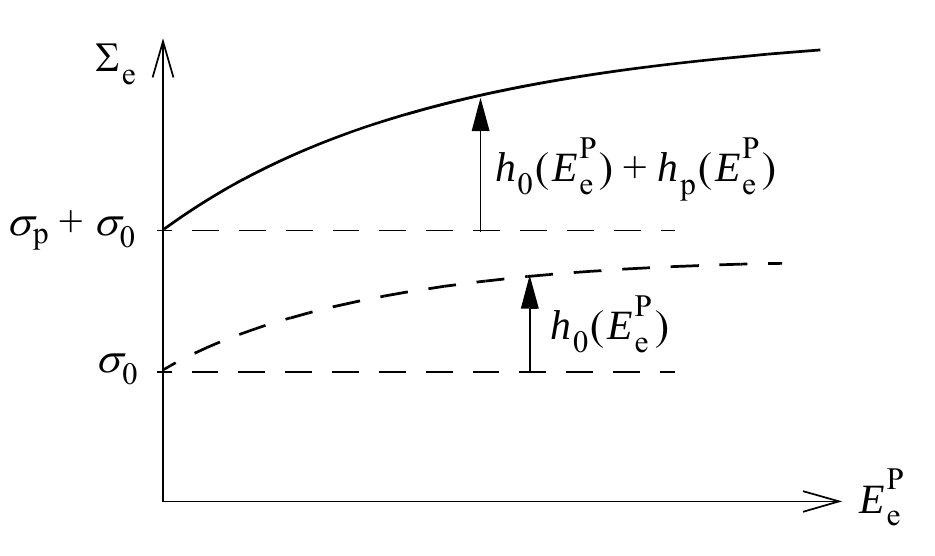}
  \caption{Definitions of yield and hardening for pure matrix material (dashed line) and precipitation-hardened material (solid line).}
	\label{fig:Def_of_Sp}
\end{center}  
\end{figure}

\noindent To qualitatively and quantitatively investigate how the yield strength and the hardening are influenced by model parameters, an extensive parametric study  will now be presented in terms of changes in the yield strength $\sigma_{\rm p}$ and hardening $h_{\rm p}$. A schematic graph of the macroscopic effective stress $\Sigma_{\rm e}$ versus the macroscopic effective plastic strain $E^{\rm p}_{\rm e}$ is shown in Figure \ref{fig:Def_of_Sp}, where $\sigma_{\rm p}$ and  $h_{\rm p}$ are indicated. The macroscopic effective plastic strain was calculated as $ E^{\rm p}_{\rm e} = E_{\rm e} - \Sigma_{\rm e}/(3\bar{G})$, where $\bar{G}$ was numerically evaluated from the initial elastic slope. The macroscopic yield stress, $\Sigma_{\rm e}(E^{\rm p}_{\rm e}=0)$, was then determined at the onset of overall plastic yielding as the intersection point between the initial elastic slope and a linear curve fit to the initial part of the post yield response obtained by linear regression.

The matrix material length parameter $\ell$ arising in the constitutive formulation (\sref{sec:2}) defines the physical length scale in the problem, and the characteristic dimensions of the micro-structure must scale with $\ell$. The micro-structurally associated parameters to be investigated are: $L_{\rm p}/ \ell$, $f$, $G_{\rm p}/G_{\rm m}$ (the mismatch in particle/matrix elastic modulus), and $H/R$ ($=\xi$) or alternatively the coefficient of variance $C_{\rm V}$ (representing the inhomogeneity of the particle spacing distribution). The particle radius $r_{\rm p}$ is determined by the values of $L_{\rm p}$ and $f$ from the relations in \eref{eqn:Lpcc} or \eref{eqn:Lpss}. The relative influence of these parameters will be governed by the interface parameters $\alpha_{0}$ and $\alpha_{1}$. With this in mind, a non-dimensional function $F_{\sigma}$ is assumed to govern the increase in yield stress according to
\begin{equation}\label{eqn:Sp_fcn}
  \sigma_{\rm p} = \sigma_{0} F_{\sigma} \left( \alpha_0, \alpha_1; \frac{L_{\rm p}}{\ell}, f,\frac{G_{\rm p}}{G_{\rm m}}, \rm{`Geomtry`} \right),
\end{equation}
and the change in hardening as a function on the form
\begin{equation}\label{eqn:hp_fcn}
  h_{\rm p} = G_{\rm m} F_{\rm h} \left( \varepsilon_{\rm e}^{\rm p}; N; \alpha_0, \alpha_1; \frac{L_{\rm p}}{\ell}, f,\frac{G_{\rm p}}{G_{\rm m}}, \rm{`Geomtry`} \right).
\end{equation}

The material length parameter of the matrix $\ell$ is defined in the following manner: for each analysis a value of the ratio $\lambda = L_{\rm p}/\ell$ was chosen. The scale of the discretized FEM model is set by the outer cell radius $R$. Then, from \eref{eqn:Lpcc} or \eref{eqn:Lpss}, the material length parameter is given by $\ell = R \theta / \lambda$, where $\theta$ is either $\theta_{\rm cc}$ or $\theta_{\rm ss}$ depending on the measure used to define the average particle spacing (C-C or S-S).

All computations were carried out for a matrix material with $\nu = 0.3$ and $\varepsilon_0 = \sigma_0 / [2G_{\rm m}(1+\nu)] = 0.002$. In most analyses, a homogeneous particle spacing ($H = R \Rightarrow C_{\rm V} = 0$) was considered, except in one set of analyses where effects of a less homogeneous particle distribution were studied. Furthermore, in the first part of this investigation, an elastic-perfectly plastic matrix material ($N = 0$) and the interface parameter $\alpha_1 = 0$ were used to exclude effects from matrix strain hardening. In the second part, the combined effect of matrix hardening and $\alpha_1 \ge 0$ on the macroscopic hardening $h_{\rm p}$ were explored.

By solving several problems in both uniaxial and biaxial tension, including a large variety of $H/R$ ratios, it was found that the evaluation of $\sigma_{\rm p}$ was insensitive to the type of loading. Therefore, only results from unit cells subjected to uniaxial tension will be presented below.

\subsection{General model features and influence of particle spacing}
\label{sec:3.1}

\begin{figure}[htb!]
\begin{center}
    \subfigure[]{ \includegraphics[width=0.44\textwidth]{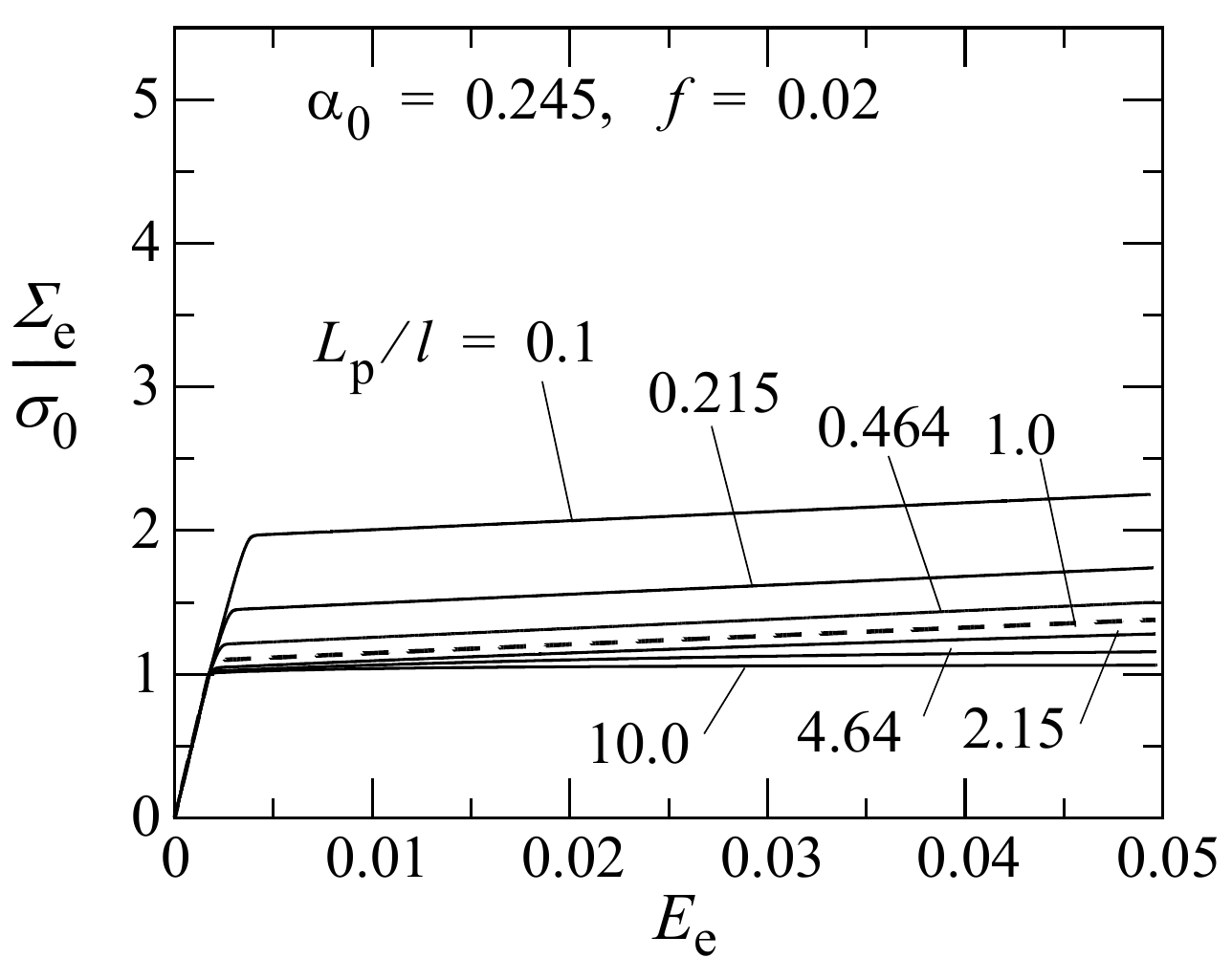} }
    \subfigure[]{ \includegraphics[width=0.46\textwidth]{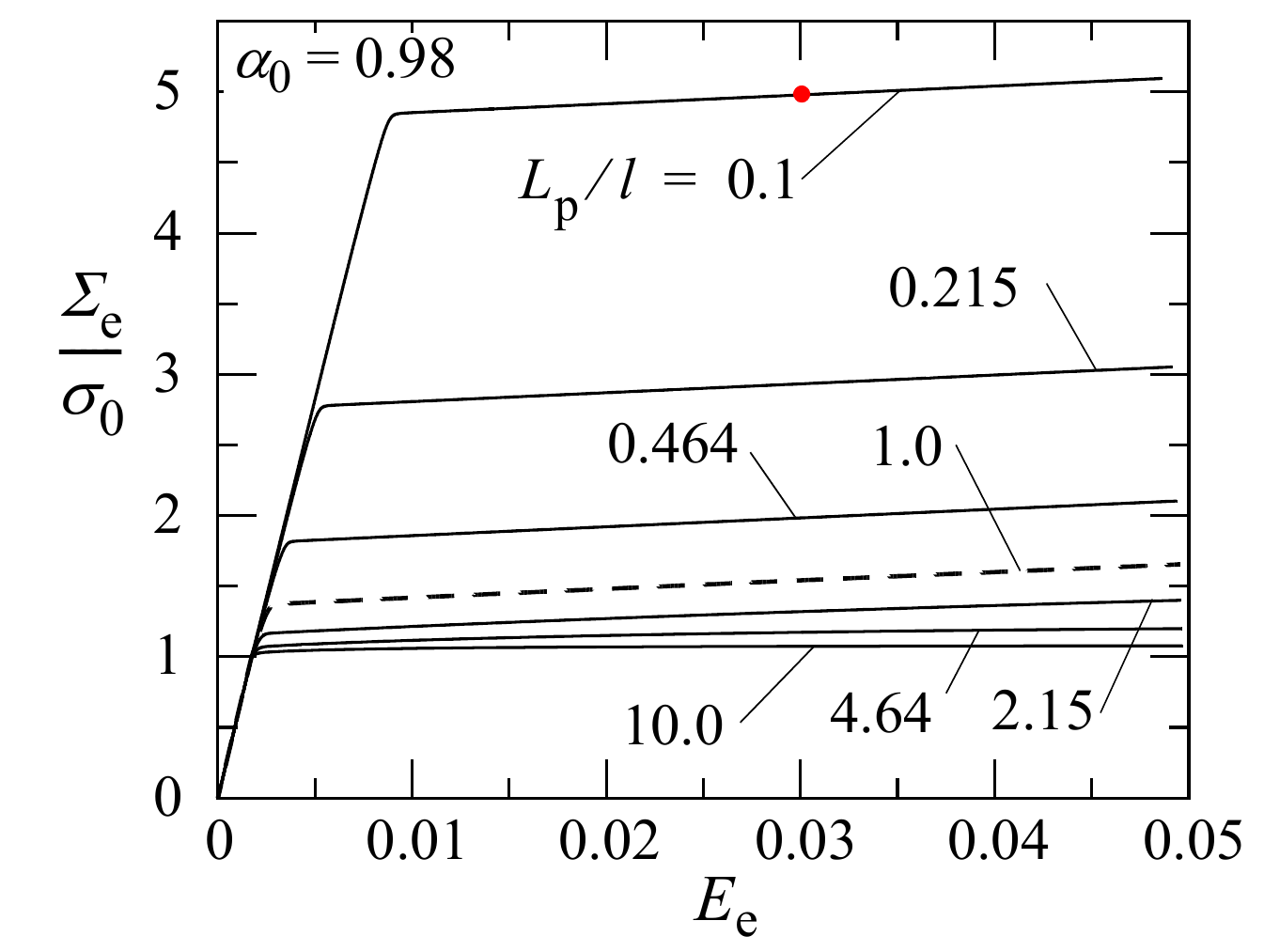} }
    \caption{Stress-strain curves for $\alpha_0 = 0.245$ shown in (a), $\alpha_0 = 0.98$ shown in (b). The matrix material is elastic-perfectly plastic ($N = 0$). The red point in (b) marks the load level at which the effective plastic strain field is examined in Figure \ref{fig:GND}.}
    \label{fig:Res1}
\end{center}
\end{figure}

The general features of the model will be introduced for a set of micro-structures with matched elastic constants ($G_{\rm p} = G_{\rm m}$) that have the same volume fraction of particles $f$ but differ in the strength of particle/matrix interface $\alpha_0$ and particle spacing set by the center-to-center definition. If all micro-structures have the same matrix material with a fixed length parameter $\ell$, equation \eref{eqn:Lpcc} gives $L_{\rm p}/\ell = 2R = 1.75 f^{1/3} r_{\rm p}/\ell$. From \eref{eqn:Sp_fcn} it is expected that the change in yield stress will depend on $L_{\rm p}/\ell$ and $\alpha_0$. Figure \ref{fig:Res1} illustrates this influence, where macroscopic stress-strain curves are plotted for materials containing a particle volume fraction of 2\%. In graph (a) the materials are equipped with a weak particle/matrix interface, $\alpha_0 = 0.245$, whereas the curves in (b) are generated with a strong interface, $\alpha_0 = 0.98$. The particle spacing $L_{\rm p}$ in each graph ranges from $10\ell$ down to $0.1\ell$, and as can be seen, substantial strengthening requires that $L_{\rm p} < \ell$. Also, note that the strength of the interface significantly amplifies the effect of spacing, and to connect with physics of inelastic deformation, this behavior mimics the fact that a decreasing spacing makes it more difficult for dislocations to bypass particles, and hence there is more strengthening.

\begin{figure}[htb!]
\begin{center}
    \subfigure[]{ \includegraphics[width=0.44\textwidth]{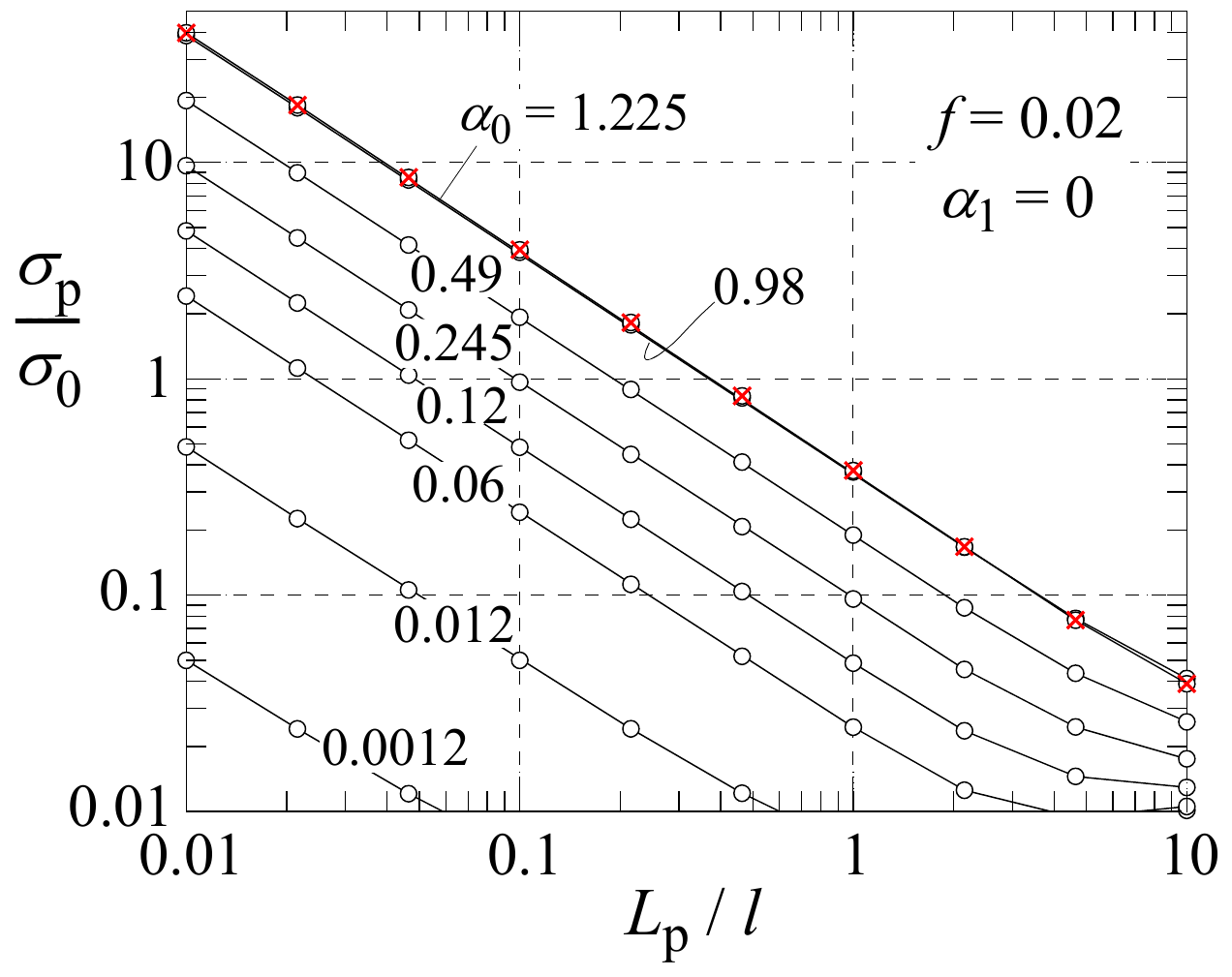} }
    \subfigure[]{ \includegraphics[width=0.46\textwidth]{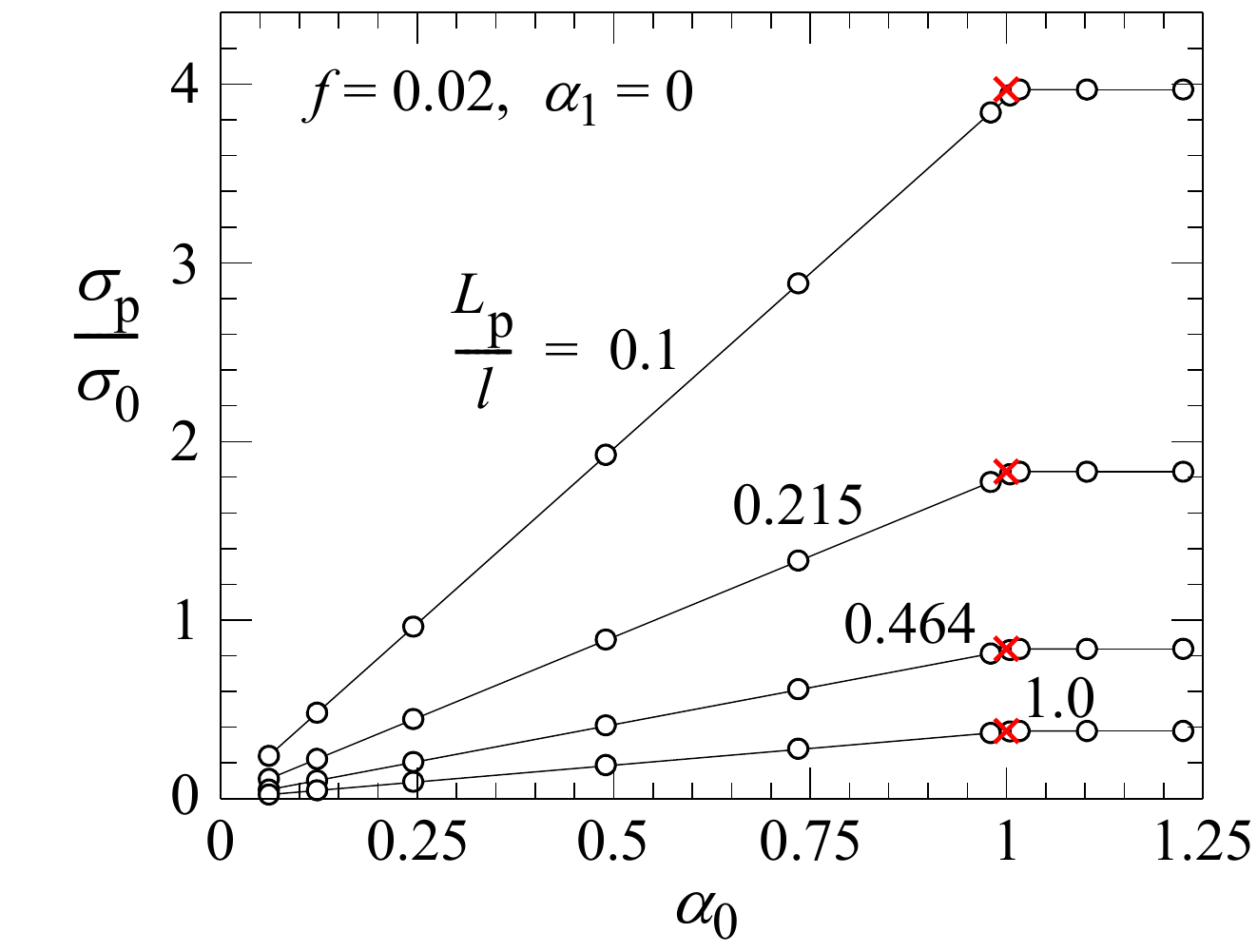} }
    \caption{Influence of particle spacing and interface strength on the increase in yield strength of the material. In (a) $\sigma_{\rm p}/\sigma_{\rm 0}$ versus $L_p/\ell$ is shown for a large set of different $\alpha_0$ values (solid lines), red symbols); and in (b) $\sigma_{\rm p}/\sigma_{\rm 0}$ versus $\alpha_0$ is shown for four different values of $L_p/\ell$. The red crosses in (a) and (b) corresponds to solutions generated with a purely micro-hard interface, i.e., where $ \varepsilon_{ij}^{\rm p} = 0 $ at the interface.}
    \label{fig:Res2}
\end{center}
\end{figure}

The effects of the parameters $L_{\rm p}/\ell$ and $\alpha_0$ on $\sigma_{\rm p}$ observed in Figure \ref{fig:Res1} will be quantified next. To isolate the specific influence of spacing, a set of analyses was carried out where $\alpha_0$ was kept constant and the spacing varied in the range $0.01 \le L_{\rm p}/\ell \le 10$. The result is shown in Figure \ref{fig:Res2}a, where $\sigma_{\rm p} / \sigma_0$ is plotted versus $L_{\rm p}/\ell$ in a log-log diagram. Several observations can be made; when the strength of the interface decreases towards zero, here represented by $\alpha_0 < 0.001$, $\sigma_{\rm p}$ vanishes and the interface essentially responds in a `micro-soft` manner where effects of spacing become insignificant on the macro-scale. When the strength of the interface increases, there is an upper limit for its influence. From Figure \ref{fig:Res2}a it can be observed that the upper limit confirms the micro-hard limit discussed in Section \ref{sec:InterfaceDescription}. Note that the curves pertaining to $\alpha_0 = 1.225$ and to the one with a purely micro-hard interface (red symbols) collapse on top of each other. These two curves also yield results close to the result of $\alpha_0 = 0.98$. Thus, when $\alpha_0$ exceeds the micro-hard limit, the interface displays a `micro-hard` response and the development of plastic strains is completely blocked as discussed in Section \ref{sec:InterfaceDescription} and depicted in Figure \ref{fig:Interface_schematic}. Recall that the `micro-hard` limit for $\alpha_0$ was deduced from \eref{eqn:PeakMomTrac} to be equal to one, which is supported by the result found here. Moreover, for combinations of spacing and interface strength that give a noticeable increase in yield strength, say $\sigma_{\rm p} > 0.05 \sigma_0$, the slopes of the curves in the log-log diagram are constant and essentially equal to $-1$. This, suggests that $\sigma_{\rm p}/\sigma_0 \propto \ell/L_{\rm p}$, as would be predicted by the Orowan strengthening mechanism for dislocations that loops around hard particles (\cite{Orowan48}). For $\sigma_{\rm p} < 0.05 \sigma_0$ the solutions are less affected by size scale and a transition towards solutions given by local $J_2$ plasticity is anticipated.

It should be noted that the slope of $-1$ found in Figure \ref{fig:Res2}a would also be obtained if the results had been plotted versus $L_{\rm pss} / \ell$ instead of $L_{\rm pcc} / \ell$, since $f$ was held constant in this analysis, cf. \eref{eqn:Lpss}.

To quantify the isolated effect of the interface strength on the overall yield strength, an analysis was performed for the full range of $\alpha_0$ while keeping $L_{\rm p}/\ell$ constant. The results are shown in Figure \ref{fig:Res2}b. The effect of the interface strength exhibits a clear trend, i.e., the increase in yield stress is proportional to $\alpha_0$ if $L_{\rm p}/\ell$ and $f$ is kept fixed. Note also that the influence of $\alpha_0$ saturates for $\alpha_0 \ge 1$, independent on the value of $L_{\rm p}/\ell$, which agrees the results shown in Figure \ref{fig:Res2}a and supports the estimate of the upper limit for $\alpha_0$ based on reaching `micro-hard` conditions at $\alpha_0 = 1$.

To further connect to the physics of the considered strengthening mechanisms in the context of a SGP continuum, it is of interest to explore the development of a strong plastic strain gradient near the interface, as it would correspond to an accumulation of GNDs. This can be done by investigating the gradient of the conventional effective plastic strain $\varepsilon^{\rm p}_{\rm e} = \sqrt{2 \varepsilon^{\rm p}_{ij}\varepsilon^{\rm p}_{ij} /3}$ in the normal direction to the particle/matrix interface. By introducing the spherical radius $\rho = \sqrt{r^2+z^2}$, the gradient of $\varepsilon^{\rm p}_{\rm e}$ in the direction perpendicular to the interface can be expressed as (here denoted as $\bigtriangledown_{\rho}\varepsilon^{\rm p}_{\rm e}$)
\begin{equation}\label{eqn:grad_peeq}
\bigtriangledown_{\rho}\varepsilon^{\rm p}_{\rm e} = \frac{\partial \varepsilon^{\rm p}_{\rm e} }{\partial \rho} = \frac{2}{3}\frac{\varepsilon^{\rm p}_{ij}}{\varepsilon^{\rm p}_{\rm e}} \varepsilon^{\rm p}_{ij,k} \bar{n}_k \quad {\rm with} \quad \bar{n}_r = \frac{r}{\sqrt{r^2+z^2}}, \quad \bar{n}_z = \frac{z}{\sqrt{r^2+z^2}}.
\end{equation}

\begin{figure}[htb!]
\begin{center}
    \subfigure[]{ \includegraphics[width=0.45\textwidth]{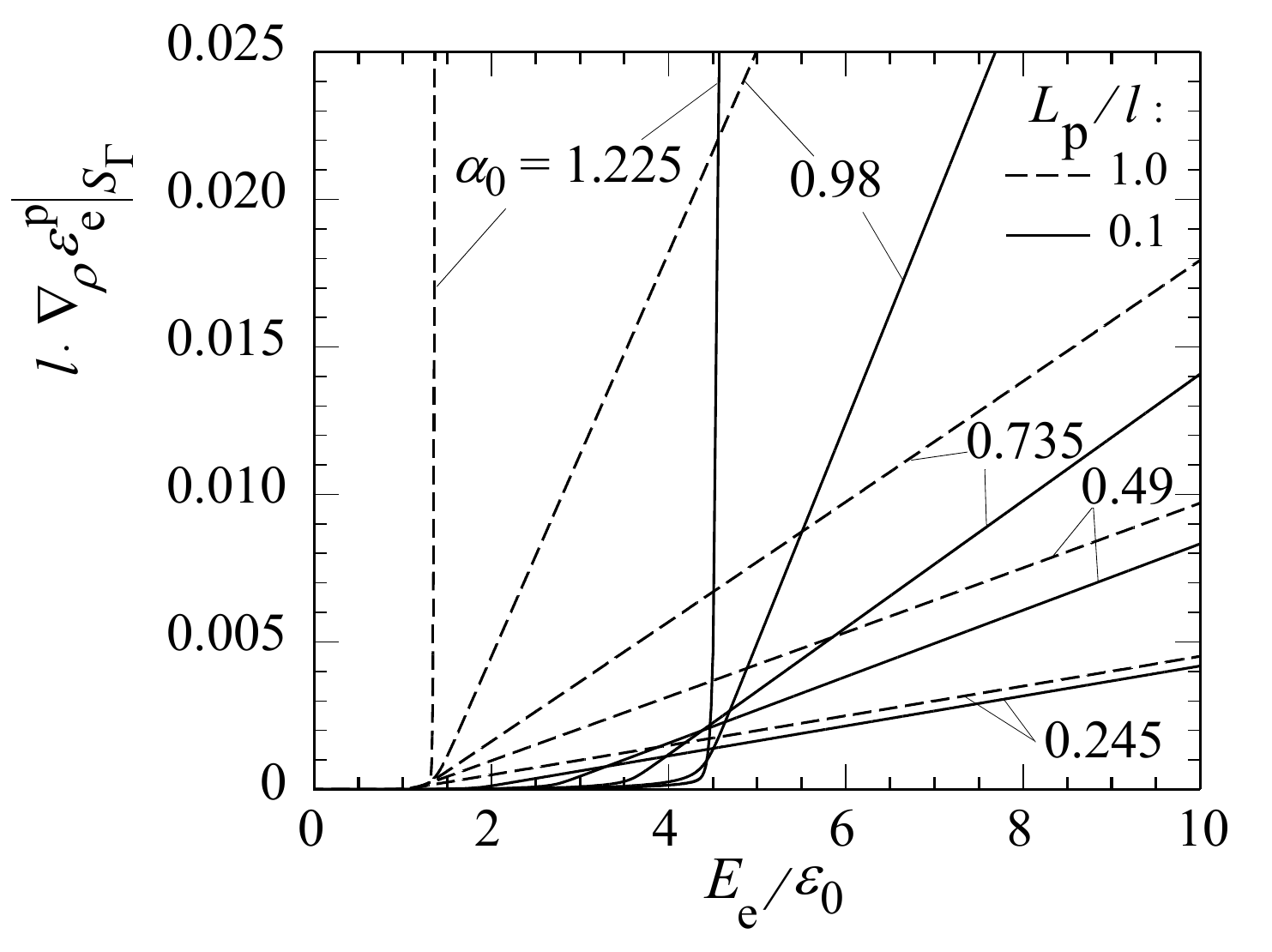} }
    \subfigure[]{ \includegraphics[width=0.45\textwidth]{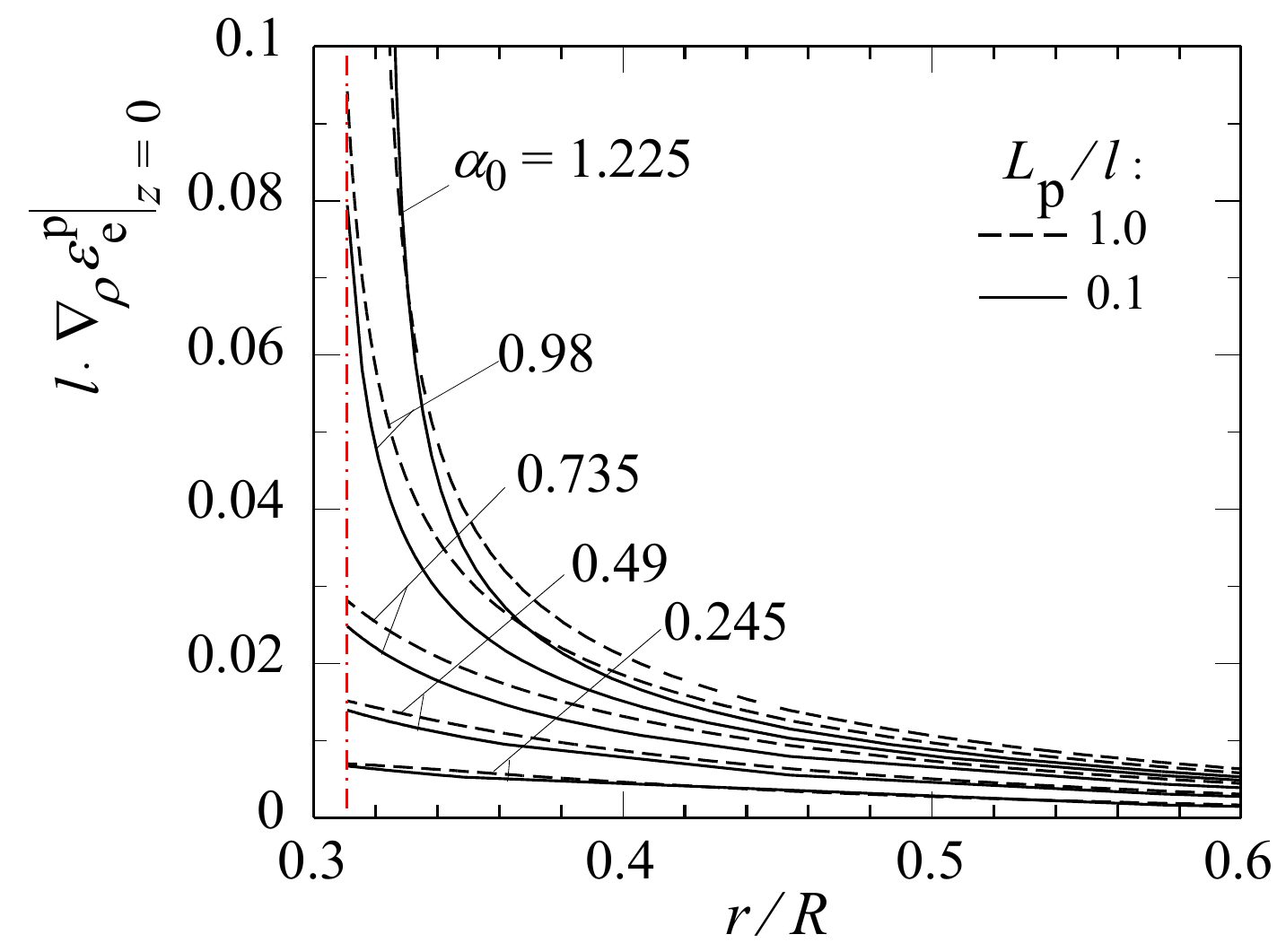} }
    \subfigure[]{ \includegraphics[width=0.45\textwidth]{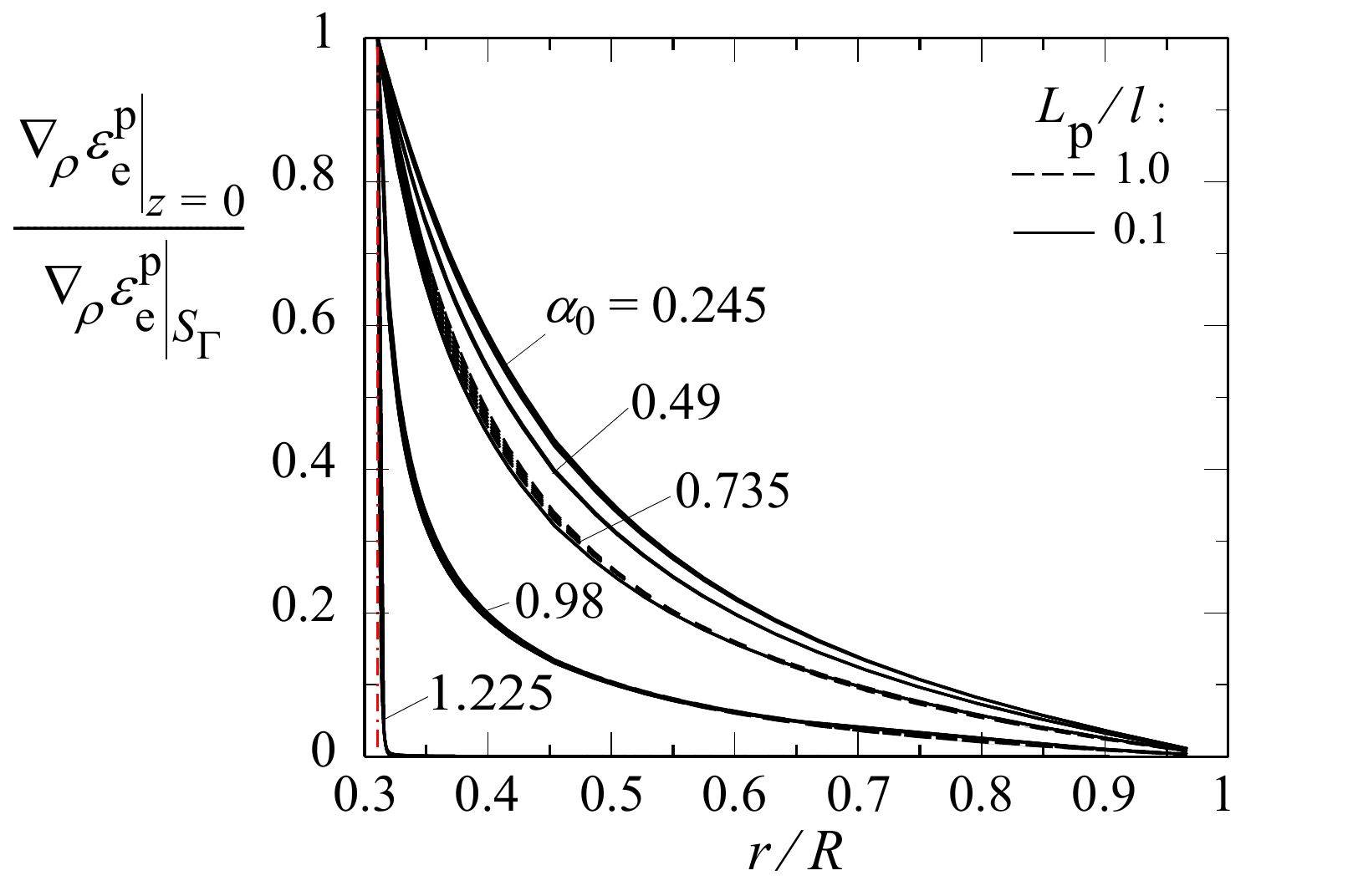} }
    \subfigure[]{ \includegraphics[width=0.32\textwidth]{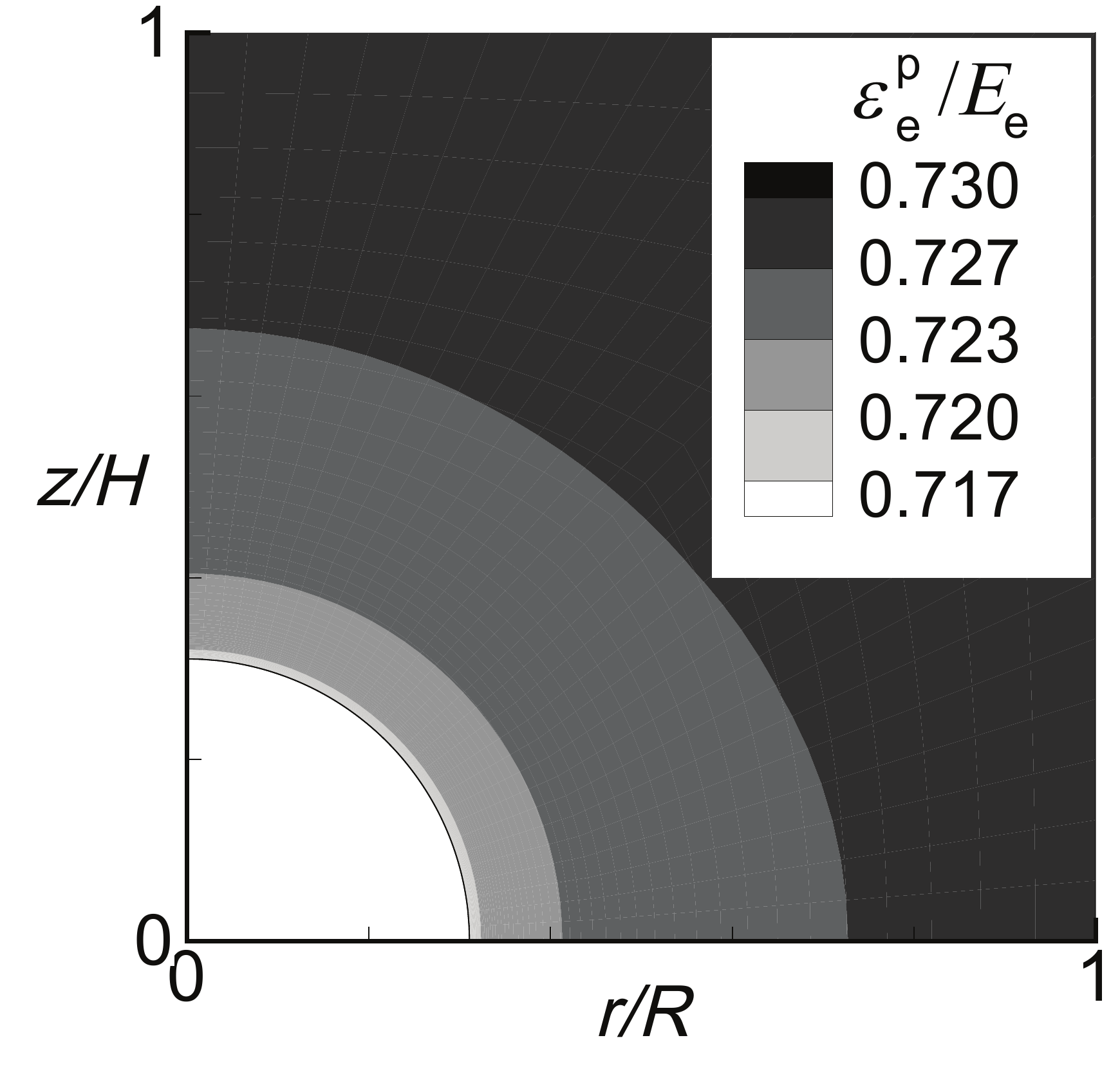} }
  \caption{(a) Development of GND at the particle/matrix interface with overall strain, where the GND is represented by the derivative of the effective plastic strain in the direction normal to the interface. (b) Distribution of GNDs along radius of the matrix at the overall load level $E_{\rm e}/\epsilon_{0} = 15$, where the red dashed line is the particle/matrix border. (c) Normalized GND levels with respect to the GND level at the particle surface plotted versus $r/R$ for combination of $\alpha_{0}$ and $L_{\rm p}/l$ at six six different load levels, $E_{\rm e}/\epsilon_{0} = \{6, 10, 14, 18, 22, 26\}$. (d) The distribution of plastic strain within the RVE at $E_{\rm e}/\epsilon_{0} = 15$ (red point on Figure \ref{fig:Res1}b). $f = 0.02, N = 0, \alpha_{1} = 0$ in all graphs.}
	\label{fig:GND}
\end{center}
\end{figure}

Calculation of the gradient $\bigtriangledown_{\rho}\varepsilon^{\rm p}_{\rm e}$ revealed that it is essentially constant along the interface $S_{\Gamma}$ for any combination of parameters and applied loads. For practical purposes, $\bigtriangledown_{\rho}\varepsilon^{\rm p}_{\rm e}$ was then evaluated as the average value along $S_{\Gamma}$, below referred to as $\bigtriangledown_{\rho}\varepsilon^{\rm p}_{\rm e} |_{S_{\Gamma}}$. The scale over which the gradients develop is set by the matrix material length $\ell$. Hence, to facilitate a comparison of gradients from different micro-structures, it must be multiplied with $\ell$. The evolution of $\ell \bigtriangledown_{\rho}\varepsilon^{\rm p}_{\rm e} |_{S_{\Gamma}}$ with the normalized macroscopic effective strain $E_{\rm e}/\varepsilon_0$ is shown in Figure \ref{fig:GND}a for ten combinations of $L_p/\ell$ and $\alpha_0$ as indicated in the figure. Note that $\ell \bigtriangledown_{\rho}\varepsilon^{\rm p}_{\rm e} |_{S_{\Gamma}}$ is zero prior to the onset of macroscopic plastic deformation, which manifests itself in the delayed response of the solid lines ($L_p/\ell = 0.1$, high $\sigma_{\rm p}$) if compared to the dashed lines ($L_p/\ell = 1.0$, low $\sigma_{\rm p}$) for the same value of $\alpha_0$. Also, note that the rate of increase (the slope of a curve) is governed by $\alpha_0$, and becomes infinity when `micro-hard` conditions prevail at the interface (here represented by $\alpha_0 = 1.225$). In the numerical framework used, the latter manifests itself by development of a very sharp boundary layer solution for the plastic strains at the interface. The rate of increase also seems to be insensitive to $L_p/\ell$ for the same interface strength.

The distribution of $\ell \bigtriangledown_{\rho}\varepsilon^{\rm p}_{\rm e}$ in the radial direction for $z = 0$ is plotted in Figure \ref{fig:GND}b at the load level $E_{\rm e} = 15\varepsilon_0$ for the same combinations of parameters as used in Figure \ref{fig:GND}a. The values of $\ell \bigtriangledown_{\rho}\varepsilon^{\rm p}_{\rm e}$ at the particle/matrix interface ($r=r_{\rm p} = 0.31R$ in this case, red dash-dotted line) comply with those in graph (a). All gradients vanish at $r = R$ (although not shown in the graph) and increase at different rates when approaching the interface. A sharp increase would be reminiscent of a pile up of dislocation loops, i.e. GNDs, at the interface. It should be mentioned that $\varepsilon^{\rm p}_{\rm e}$ at the interface is finite and increases with the applied load for $\alpha_0$ values below the `micro-hard` limit. Above this limit, $\varepsilon^{\rm p}_{\rm e} = 0$ at the interface and jumps to a finite value just outside the interface.

A generalized picture of this behavior can be obtained by scaling the gradient distributions presented in Figure \ref{fig:GND}b with their respective value at the interface. This is presented in Figure \ref{fig:GND}c, where $\bigtriangledown_{\rho}\varepsilon^{\rm p}_{\rm e} |_{z=0} /\bigtriangledown_{\rho}\varepsilon^{\rm p}_{\rm e} |_{S_{\Gamma}}$ is plotted versus $r/R$. This graph contains results generated with the same ten combinations of parameters as above, and for the six load levels $E_{\rm e}/\varepsilon_0 = \{6, 10, 14, 18, 22, 26\}$. Interestingly, all curves belonging to the same $\alpha_0$ value collapse on top of each other, independent of the values of $L_p/\ell$ and the level of applied load. This self-similar behavior suggests that the development of sharp gradients is entirely governed by the interface strength parameter $\alpha_0$, which then controls the ability to accommodate an accumulation of GNDs at the interface. 

To give a perspective of the overall distribution of plastic strains in the unit cell, iso-contours of $\varepsilon^{\rm p}_{\rm e}$ are plotted in Figure
\ref{fig:GND}d, for the parameter combination and load level indicated by a red solid circle in Figure \ref{fig:Res1}b. The overall strengthening in this case is $\sigma_{\rm p} = 3.86 \sigma_0$. It can be observed that, despite the fact that a relatively sharp gradient develops in this case (cf. Figure \ref{fig:GND}c with $\alpha_0 = 0.98$), the variation of $\varepsilon^{\rm p}_{\rm e}$ is less than 1\% over a majority of the domain, except close to the interface, where a mild decrease of $\varepsilon^{\rm p}_{\rm e}$ is seen. Thus, the plastic strain field seems to be essentially constant for sufficiently small values of $L_{\rm p} / \ell$.

\subsection{Influence of volume fraction of particles}
\label{sec:VolumeFractionEffects}

\begin{figure}[htb!]
\begin{center}
    \subfigure[]{ \includegraphics[width=0.445\textwidth]{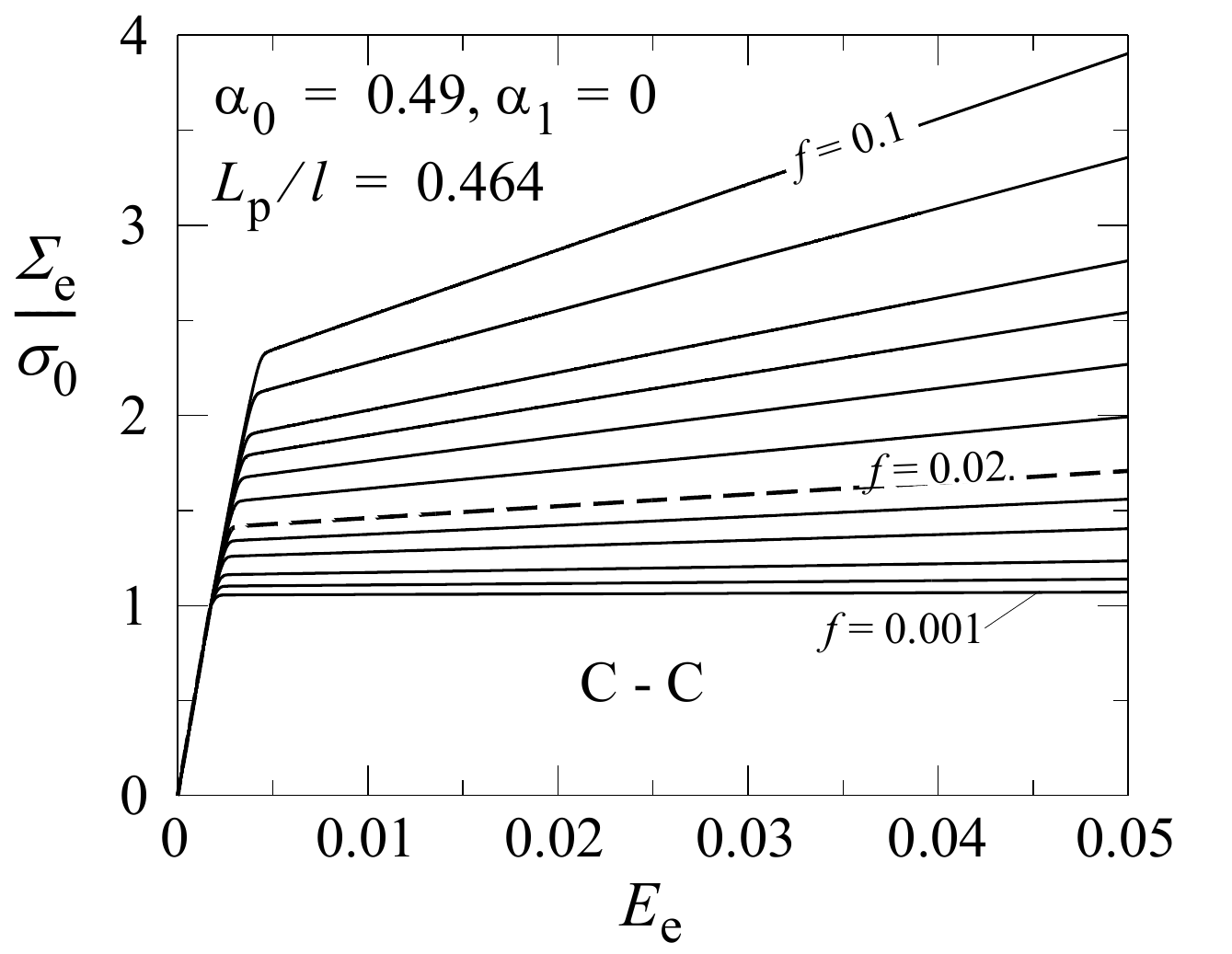}}
    \subfigure[]{ \includegraphics[width=0.44\textwidth]{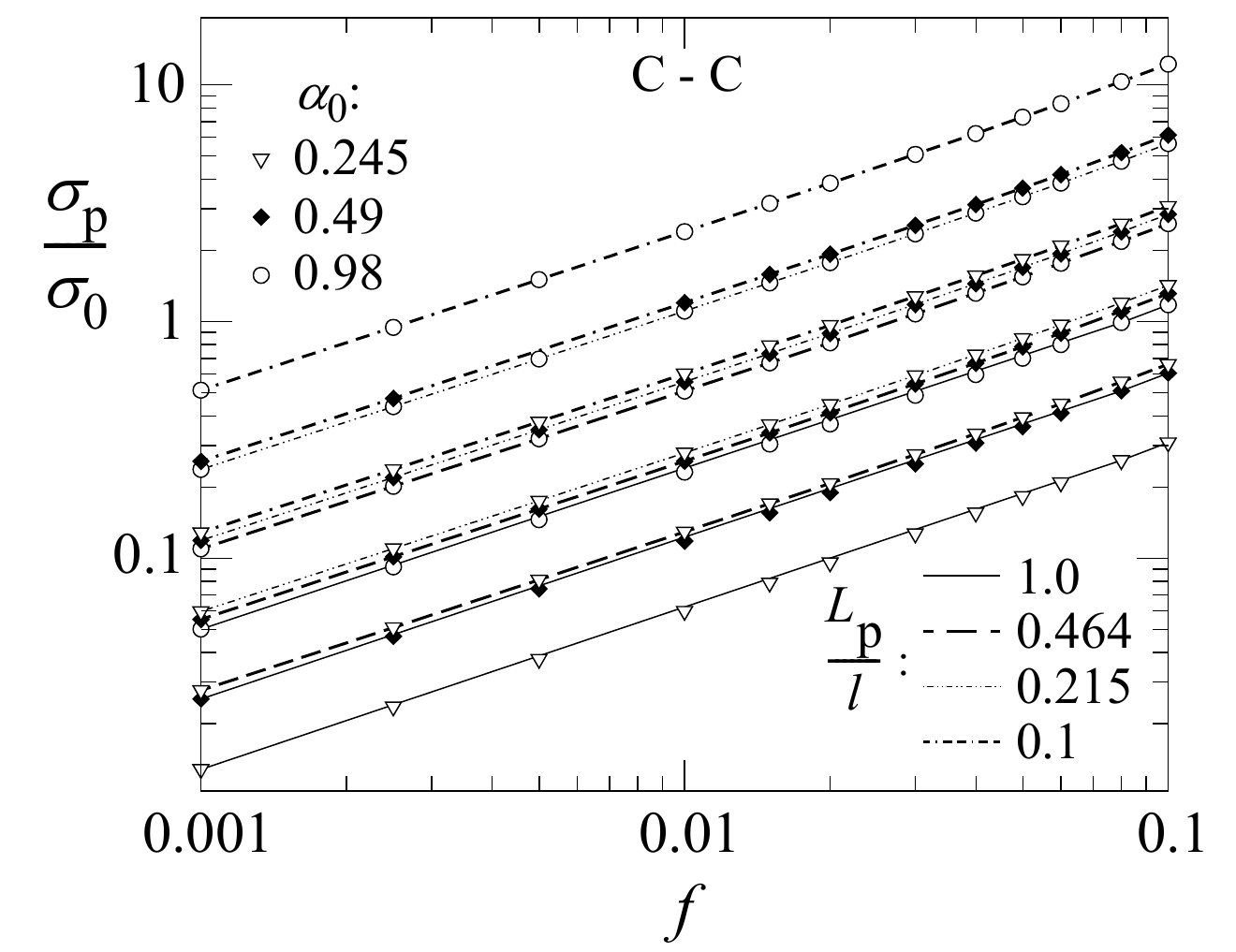} }
    \subfigure[]{ \includegraphics[width=0.44\textwidth]{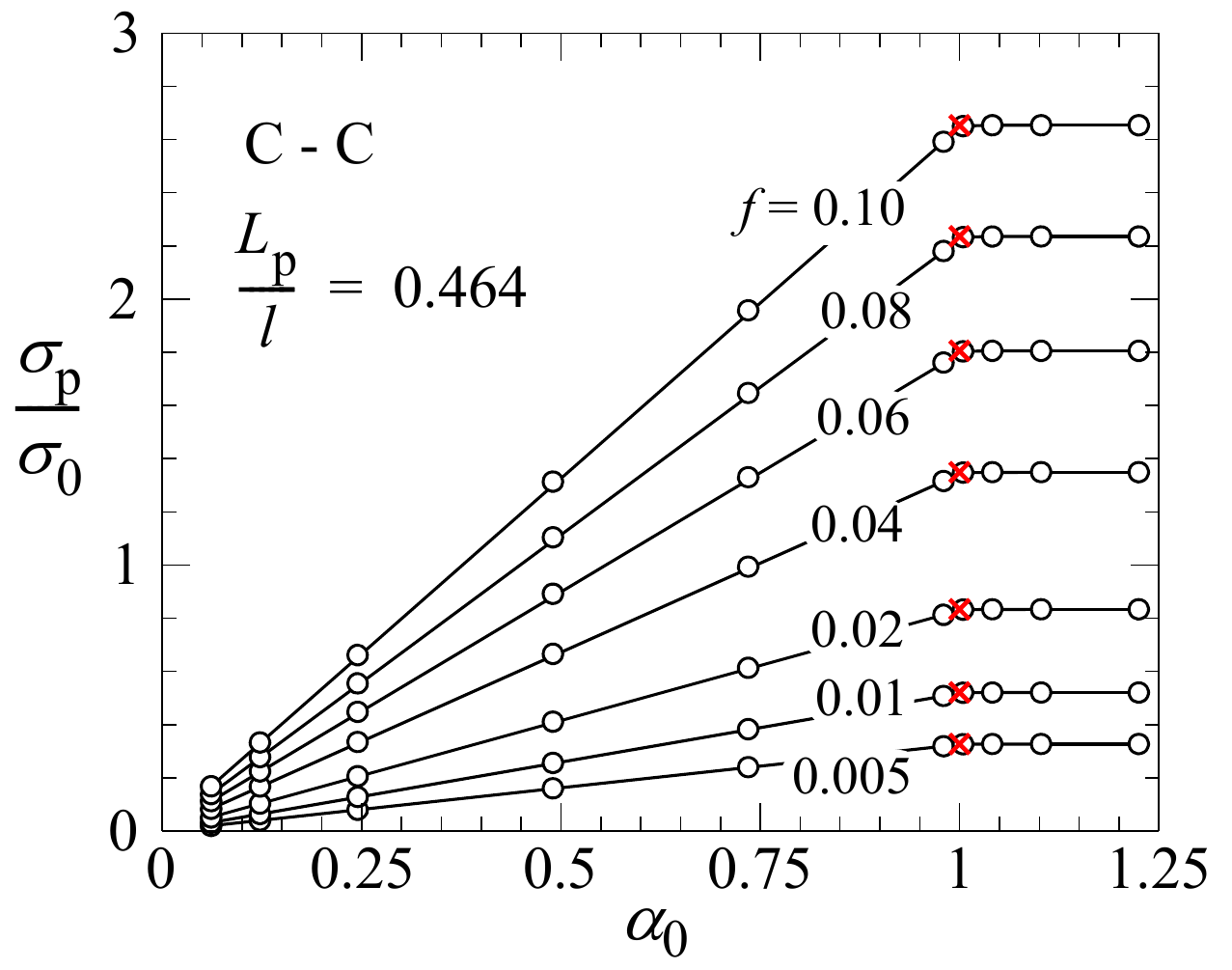} }
\caption{Influence of the volume fraction of particles on the stress-strain response, based on the center-to-center measure of particle spacing, i.e., $L_{p} = L_{pcc}$ in this case. (a) The macroscopic stress-strain response, (b) increase strength versus volume fraction for twelve combinations of parameters, (c) increase in strength versus the strength of the interface for seven different values of $f$, where the red cross symbols pertain to solutions with a purely micro-hard interface. All parameters used to generate the curves are indicated in the respective graph.}
\label{fig:fig9}
\end{center}
\end{figure}

\noindent Having established that strengthening seems to scale with $\alpha_0 \cdot L_{\rm p}/ \ell$, the influence of $f$ on $\sigma_{\rm p}$ in \eref{eqn:Sp_fcn} will now be examined. Since $L_{\rm p}/ \ell$ and $f$ are interrelated as seen from \eref{eqn:Lpcc} and \eref{eqn:Lpss}, the effect of volume fraction of particles on strengthening depends on the measure used to define the mean particle spacing. The centre-to-centre measure will be investigated first. Consider a set of micro-structures that have the same particle spacing in terms of $\lambda = L_{\rm pcc}/\ell$, but that differ in particle volume fraction. The particle size is then obtained by \eref{eqn:Lpcc} as $r_{\rm p}/\ell =  0.57\lambda f^{1/3} $ ($\xi=1$). The surface-to-surface distance between the particles will then change with $f$ as $\lambda\ell (1-1.14f^{1/3})$. Stress-strain curves are shown in Figure \ref{fig:fig9}a for $f$ in the range 0.001 to 0.1 for $\lambda = 0.464$ and $\alpha_0 = 0.49$. A clear strengthening is noted in materials with larger volume fractions. Figure \ref{fig:fig9}a also reveals the existence of a particle size dependent hardening $h_{\rm p}$. A polynomial fit to the post yield response resulted in $h^{'}_{\rm p}/G_{\rm m} = 1.3f + 3.2f^2$, which seems to hold independent of the values of $\alpha_0$ and $\lambda$.

In Figure \ref{fig:fig9}b, the relative increase in yield stress $\sigma_{\rm p}/\sigma_0$ is plotted versus $f$ in a log-log diagram for twelve combinations of $\alpha_0$ and $\lambda$. Note that all curves are virtually linear with a slope of approximately $2/3$. Hence, with a center-to-center measure of mean particle distance, $\sigma_{\rm p}/\sigma_0 \propto f^{2/3}$ appears to govern the influence of volume fraction on strengthening.

In Figure \ref{fig:fig9}c, $\sigma_{\rm p}/\sigma_0$ is plotted versus $\alpha_0$ for seven different values of $f$ in the range 0.005 to 0.1, with $\lambda = 0.464$. And again, as in Figure \ref{fig:Res2}b, it can be observed that $\sigma_{\rm p}/\sigma_0$ exhibits a linear dependence of $\alpha_0$, but in this case for a range of volume fractions. Not shown, but this was also obtained for $\lambda$ values in the full range 0.1 to 1.0. This confirms the conclusion made from Figure \ref{fig:Res2}b about linearity between $\sigma_{\rm p}/\sigma_0$ and $\alpha_0$. Furthermore, the transition to `micro hard' condition occurs at $\alpha_0 \approx 1$ independent of $f$. 

\begin{figure}[htb!]
\begin{center}
    \subfigure[]{ \includegraphics[width=0.44\textwidth]{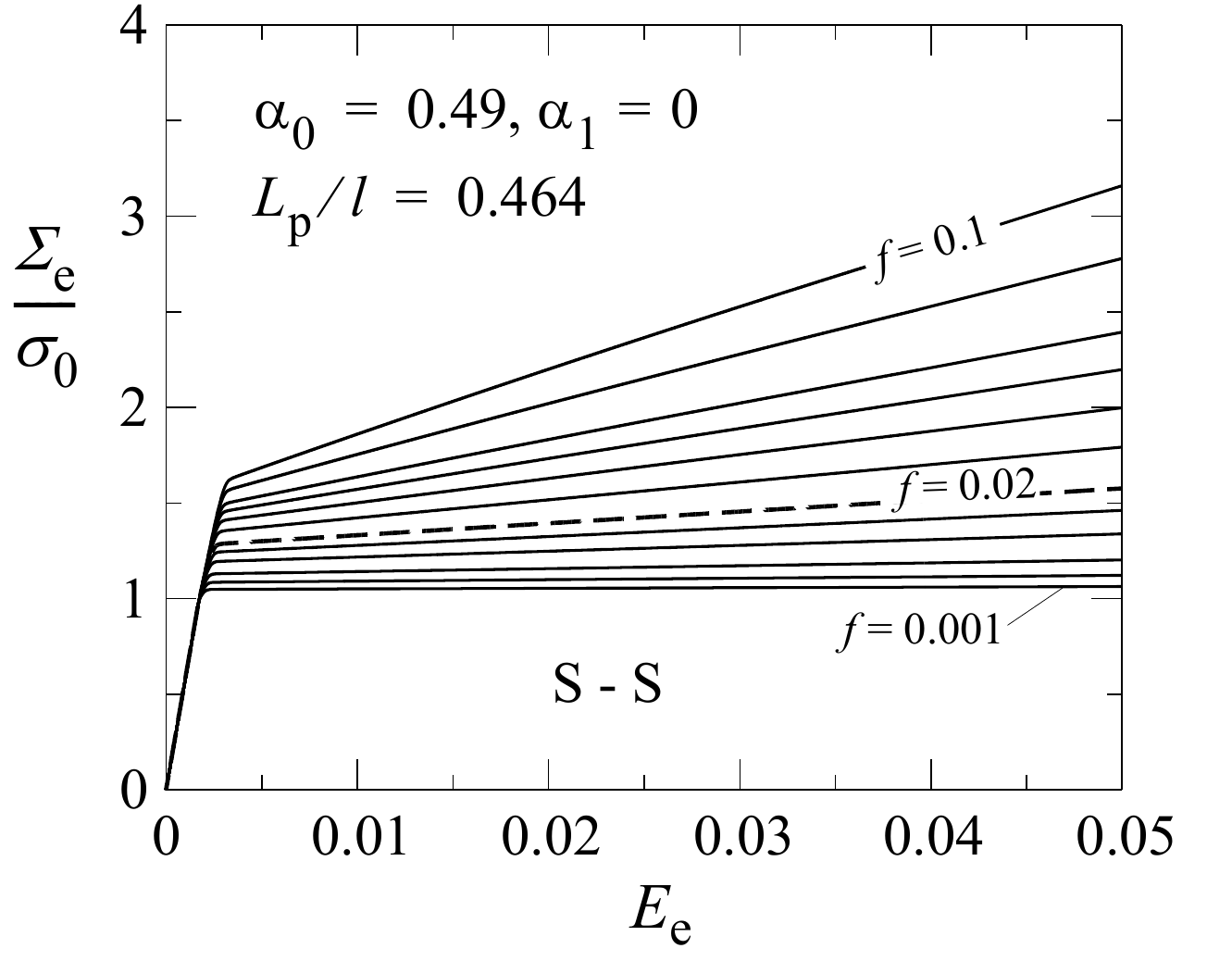} }
    \subfigure[]{ \includegraphics[width=0.44\textwidth]{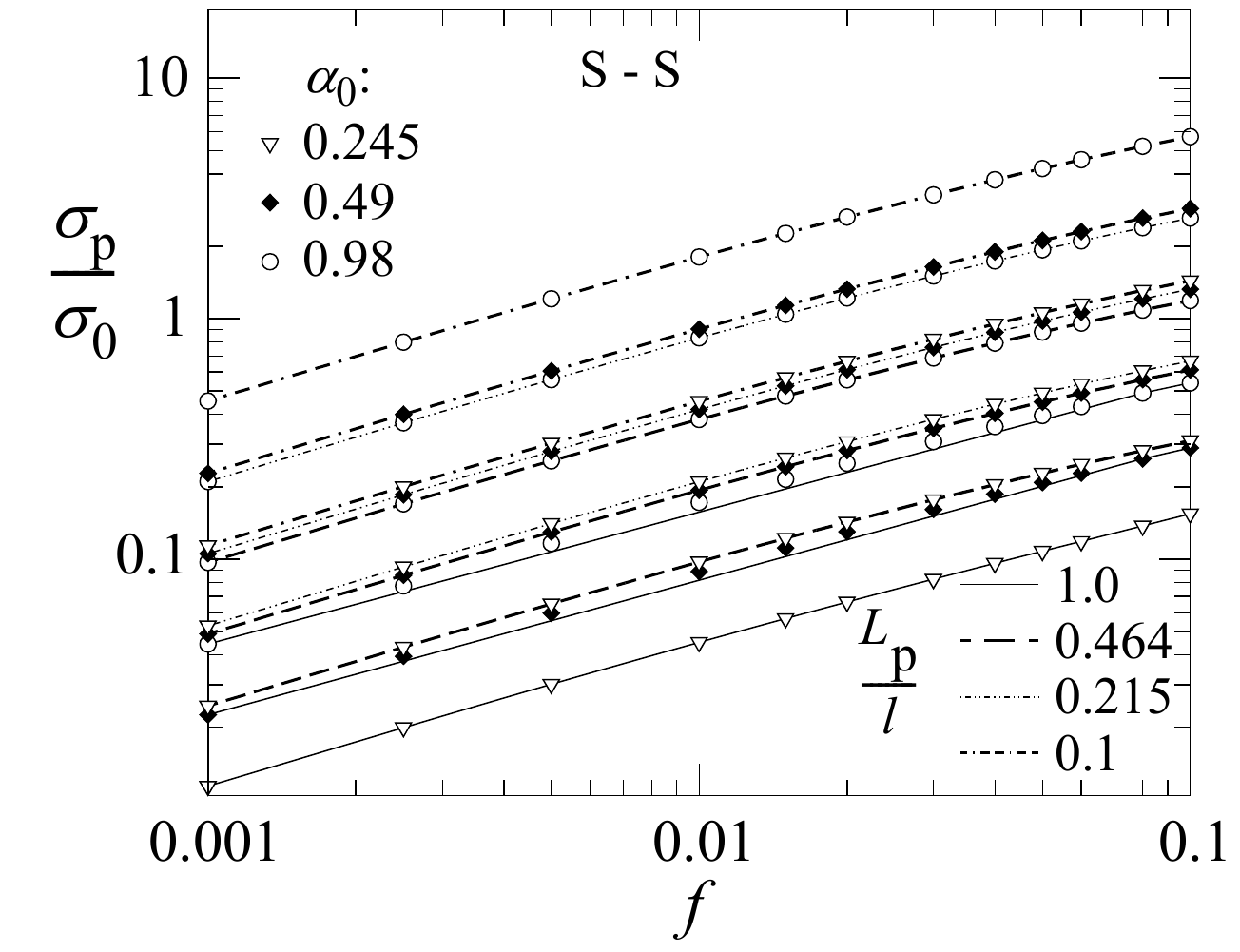} }
    \subfigure[]{ \includegraphics[width=0.44\textwidth]{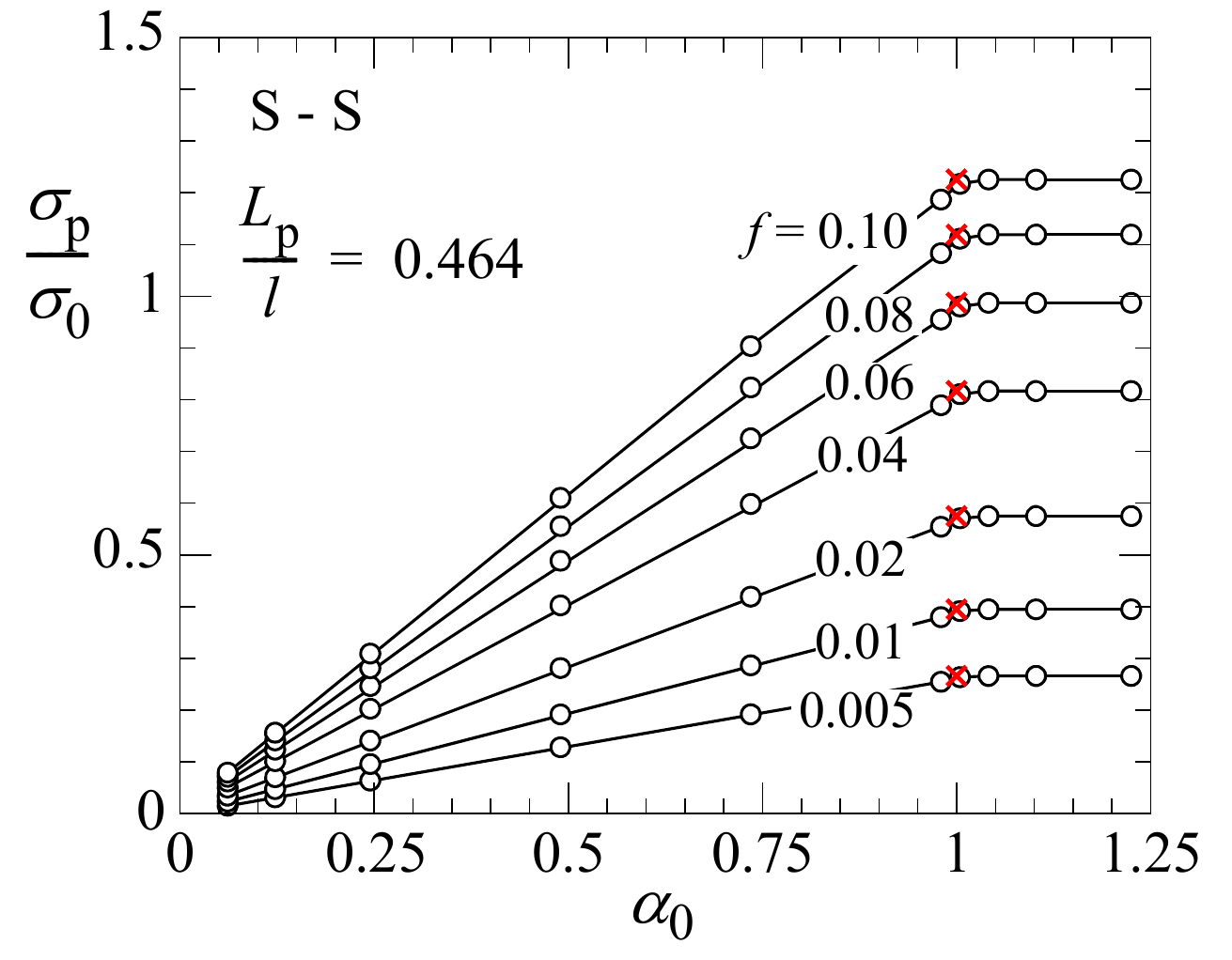} }
\caption{Influence of the volume fraction of particles on the stress-strain response, based on the surface-to-surface measure of particle spacing, i.e., $L_{p} = L_{pss}$ in this case. (a) The macroscopic stress-strain response, (b) increase strength versus volume fraction for twelve combinations of parameters, (c) increase in strength versus the strength of the interface for seven different values of $f$, where the red cross symbols pertain to solutions with a purely micro-hard interface. All parameters used to generate the curves are indicated in the respective graph.}
\label{fig:fig10}
\end{center}
\end{figure}

We now switch to examine the surface-to-surface measure of mean particle distance on strengthening. Again, a set of micro-structures that have the same particle spacing and that only differ in the particle volume fractions will be considered. But, this time, $\lambda = L_{\rm pss}/\ell$ will be held constant. The relative particle size in this case is given from \eref{eqn:Lpss} as $r_{\rm p}/\ell =  0.57 \lambda f^{1/3} / (1-2f^{1/3})$ with $\xi=1$. Here, the surface-to-surface distance between particles is independent of $f$, whereas the center-to-center distance will increase with $f$ according to $\lambda\ell (1-0.86f^{1/3})/(1-2f^{1/3})$. Macroscopic stress-strain curves for volume fractions in the range $0.001 \le f \le 0.1$ are shown in Figure \ref{fig:fig10}a for $\lambda = 0.464$. Also, in this case, an increasing volume fraction leads to significant strengthening, although less pronounced if compared to the case when $L_{\rm pcc}/\ell$ was held constant. Furthermore, a particle size dependent hardening is noted in this case as well, and a closer examination of the hardening slopes revealed essentially the same relation as found in Figure \ref{fig:fig9}a.

The influence of $f$ on strengthening based on the surface-to-surface measure of mean particle distance is shown in Figure \ref{fig:fig10}b. Here, $\sigma_{\rm p}/\sigma_0$ is plotted versus $f$ in a log-log diagram for the same twelve combinations of $\alpha_0$ and $\lambda$ as used to generate the curves in Figure \ref{fig:fig9}b. In contrast to the case (C-C) above, the curves are not entirely linear. A continuous decrease in slope can be observed as the volume fraction increases. The reason for the difference in slopes between cases (C-C) and (S-S) will be addressed below.

In Figure \ref{fig:fig10}c, $\sigma_{\rm p}/\sigma_0$ is plotted versus $\alpha_0$ for the same combinations of parameters as used in Figure \ref{fig:fig9}c. The same trend is seen in Figure \ref{fig:fig10}c as noted in Figure \ref{fig:fig9}c, confirming that $\sigma_{\rm p}/\sigma_0$ depends linearly on $\alpha_0$. 

\begin{figure}[htb!]
\begin{center}
    \subfigure[]{ \includegraphics[width=0.43\textwidth]{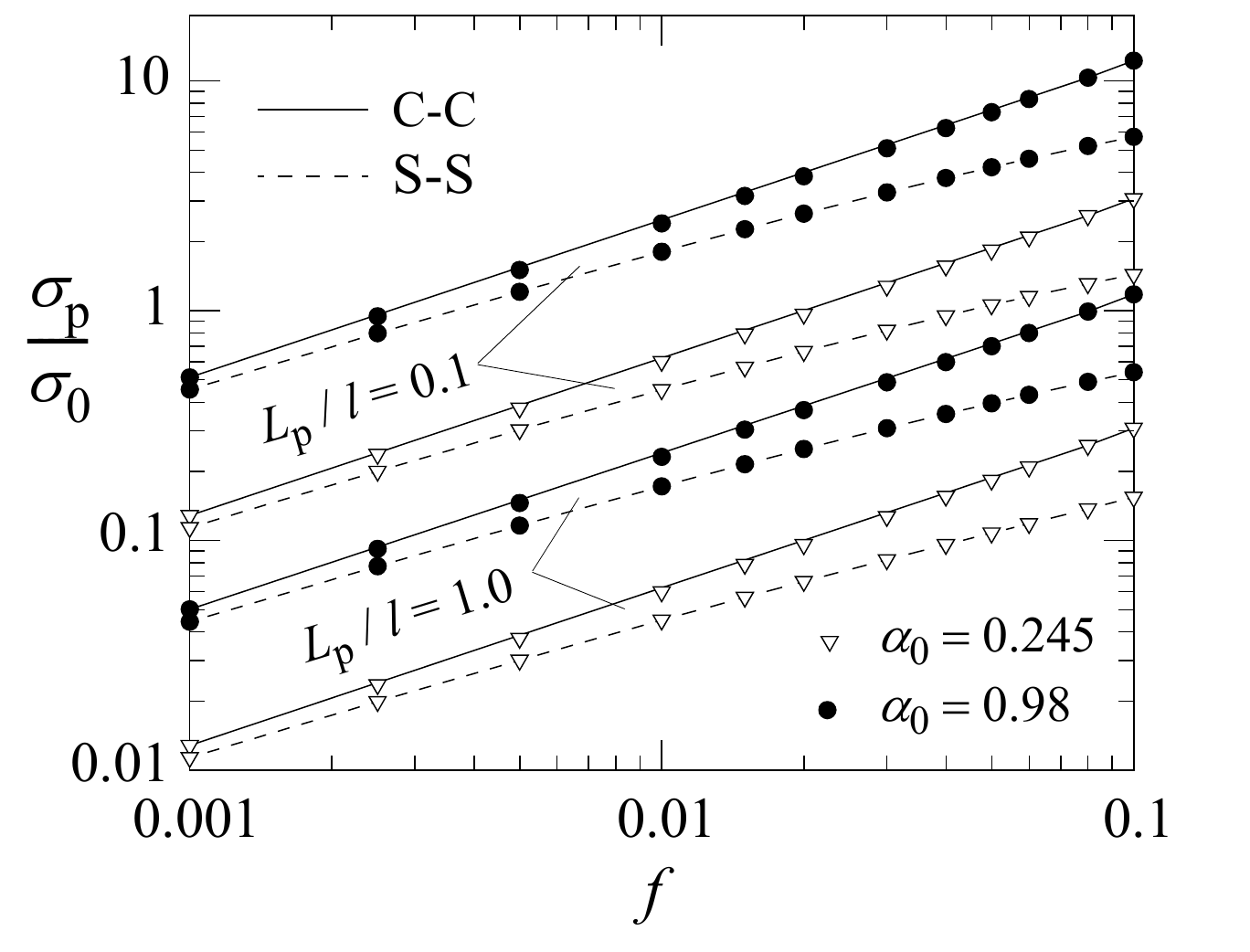}	}
    \subfigure[]{ \includegraphics[width=0.40\textwidth]{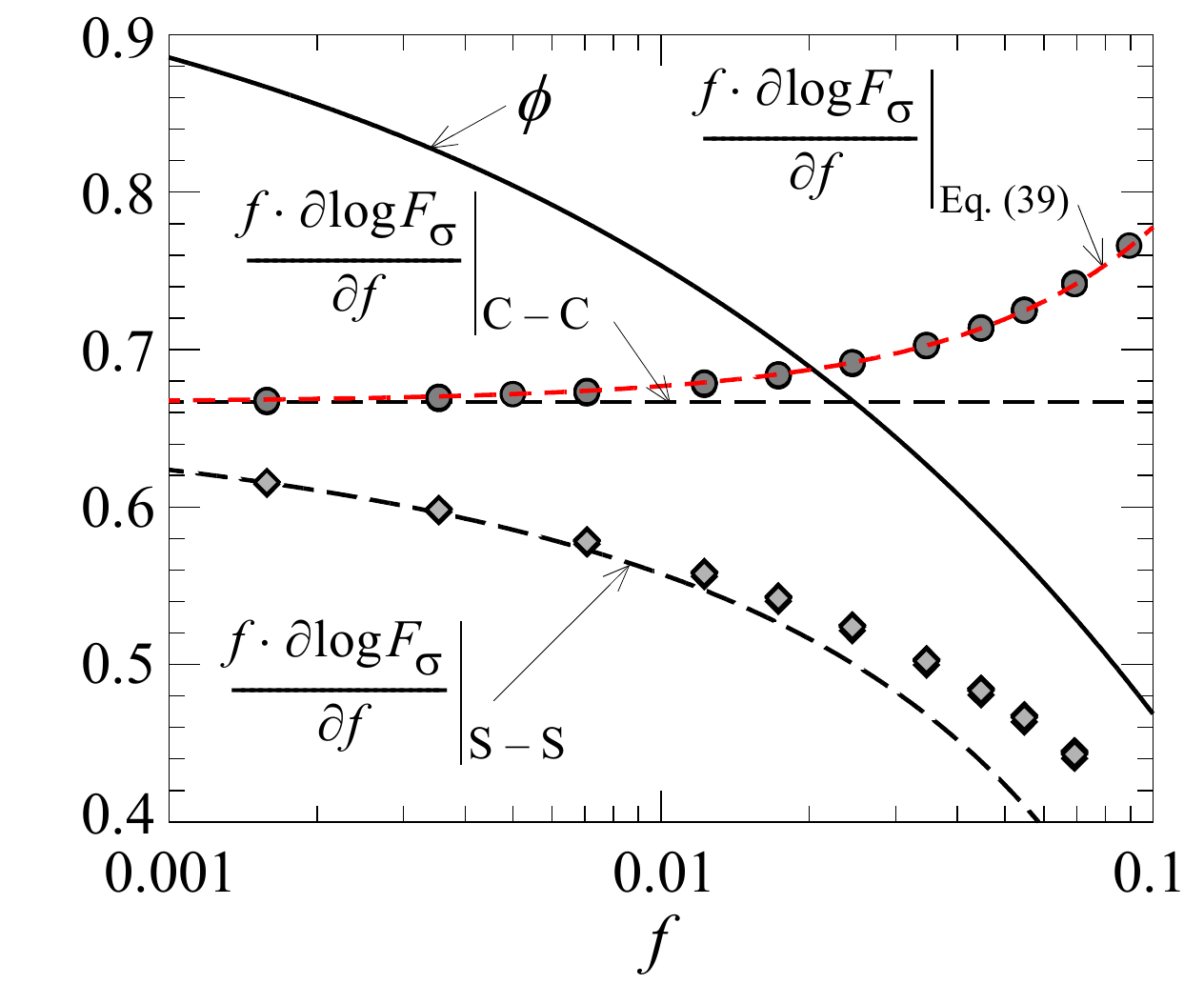} }
\caption{(a) The effects of $f$ on $\sigma_{\rm p}$ in a comparison between results from the (C-C) estimate of mean particle spacing (solid line), and the corresponding (S-S) estimate. Four sets of curves are compared with the parameters shown in the graph. (b) Function $\phi$ and estimated slopes of the assumed $\log(F_{\sigma}) - \log f$ response for case (C-C) and case (S-S), respectively. The red dashed curve represent the modified $f$ dependency introduced in Section 3.6. The symbols represent numerical results obtained from Figures \ref{fig:fig9}b and \ref{fig:fig10}b.}
	\label{fig:fig11}
\end{center}  
\end{figure}

An attempt to explain the different strengthening response with respect to $f$ between case (C-C) and case (S-S) as seen by comparing Figure \ref{fig:fig9}b with Figure \ref{fig:fig10}b will now be made. A representative selection of results from these figures are replotted in Figure \ref{fig:fig11}a, showing $\sigma_{\rm p}/\sigma_0 $ versus $f$ on log-log scale. By comparison, it is clear that the curves belonging to case (C-C) are linear, whereas the curves pertaining to case (S-S) are not. To resolve the cause of this difference, it is assumed that $F_{\rm \sigma}$ in \eref{eqn:Sp_fcn} exhibits the following dependency
\begin{equation}\label{eqn:FpropLpcc}
  F_{\rm \sigma} \thickspace \propto \thickspace \ell/L_{\rm pcc} \cdot f^{2/3}
\end{equation}
By use of relations \eref{eqn:Lpcc} and \eref{eqn:Lpss} we note that $\ell/L_{\rm pcc} = \ell/L_{\rm pss} \cdot \phi$, where $\phi(\xi,f) = \theta_{\rm ss}(\xi,f)/\theta_{\rm cc}(\xi)$. Utilizing this, \eref{eqn:FpropLpcc} can be recast in terms of $L_{\rm pss}$ as
\begin{equation}\label{eqn:FpropLpss}
  F_{\rm \sigma} \thickspace \propto \thickspace \ell/L_{\rm pss} \cdot \phi \cdot f^{2/3}.
\end{equation}
The tangent slope in a log-log diagram of a function, say $h(x)$, is given by the operation $x\partial \log(h) / \partial x$. Thus, the tangent slope of the curves plotted in Figure \ref{fig:fig10}b and \ref{fig:fig11}a pertaining to (S-S) would by keeping $L_{\rm pss}$ then be represented by
\begin{equation}\label{eqn:dlogFdf}
   f \cdot \frac{\partial\log(F_{\sigma})}{\partial f} = f \cdot \frac{\partial\log(\phi f^{2/3})}{\partial f} = \frac{2}{3} - \frac{1}{3} \left(\frac{4}{3+\xi}\right) \left(\frac{3\xi f}{2}\right)^{1/3} \left(1-\frac{4}{(3+\xi)} \left(\frac{3 \xi f}{2}\right)^{1/3}\right)^{-1}.
\end{equation}
The function $\phi$, the slope expressed in \eref{eqn:dlogFdf} and the slope $2/3$ representing case (C-C) are plotted in Figure \ref{fig:fig11}b for $\xi=1$. The symbols included in the graph are data obtained from numerical derivation of the results in Figures \ref{fig:fig9}b and Figure \ref{fig:fig10}b. As can be observed, the theoretical curves agree completely with the numerical results at low $f$ values and then starts to deviate somewhat when the volume fraction of particles becomes larger than approximately 5 $\%$.

To summarize the observations made so far, the numerical results based on the proposed model suggests that the increase in yield stress $\sigma_{\rm p}/\sigma_0 \propto f^{2/3}\alpha_0 \ell/L_{\rm pcc}$.

\subsection{Influence of mismatch in particles/matrix modulus}

\begin{figure}[htb!]
\begin{center}
    \subfigure[]{ \includegraphics[width=0.47\textwidth]{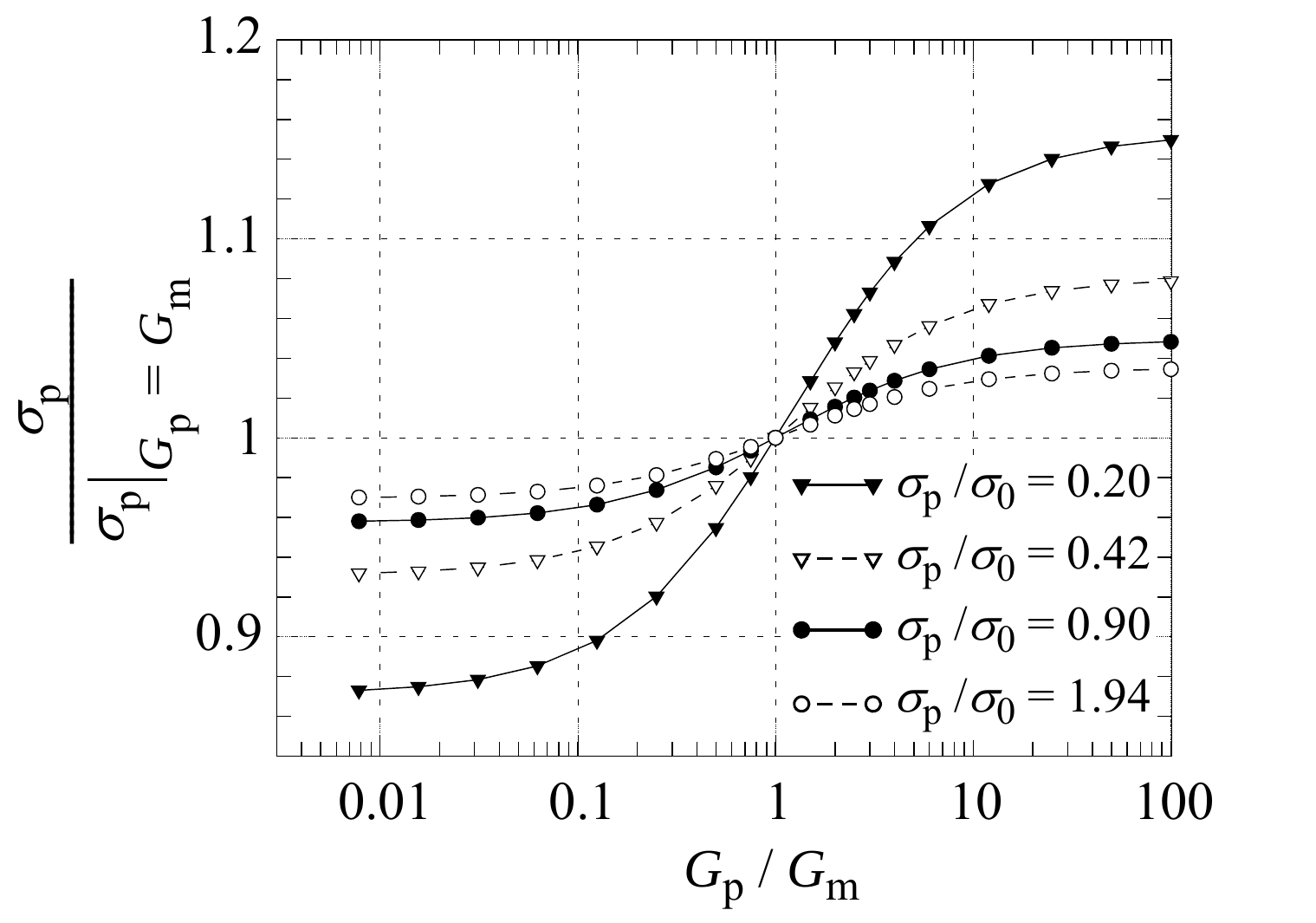} }
    \subfigure[]{ \includegraphics[width=0.46\textwidth]{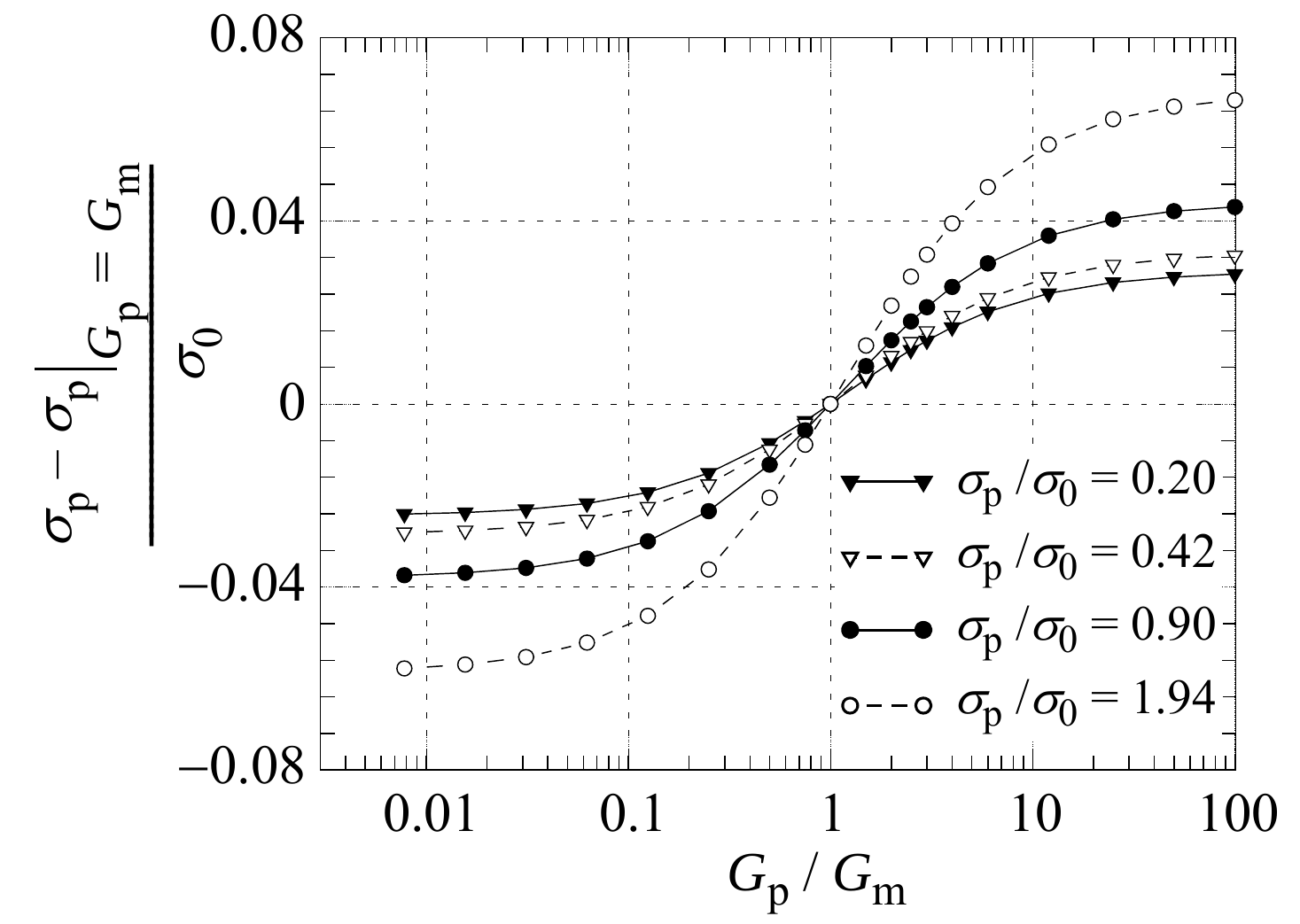} }
    \caption{Influence of the mismatch in between elastic modulus in particle and matrix. The curves pertain to $\alpha_{0} = 0.49$ and $\alpha_{1} = 0$. The legends represent the change in $\sigma_{p}$ for the matched case ($E_{\rm p}=E_{\rm m}$), obtained with $L_{\rm p}/l = 0.1, 0.215, 0.464, 1.0$. In (a) the relative change in $\sigma_{\rm p}$ is shown, and in (b) the absolute change in $\sigma_{\rm p}$ is shown.}
    \label{fig:fig12}
\end{center}
\end{figure}

\noindent The effects of mismatch in shear modulus between particle and matrix are shown in Figure \ref{fig:fig12} for $f = 0.02$ and demonstrated for particles with vanishing stiffness to particles that are virtually rigid. In both graphs, the effect on strengthening is plotted versus the ratio $G_{\rm p}/G_{\rm m}$ with $G_{\rm m} = 192.3\sigma_0$ held constant. In Figure \ref{fig:fig12}a, the relative change in yield stress is shown, and in Figure \ref{fig:fig12}b, the absolute change in yield strength is displayed. The four curves in each graph are labelled according to the increase in yield stress obtained for the matched case ($G_{\rm p} = G_{\rm m}$), as indicated by the legends. In relative terms, the change in yield stress is most pronounced for the curve belonging to the lowest $\sigma_{\rm p}/\sigma_0$ value, whereas in absolute terms the change in yield stress is most pronounced for the curve belonging to the highest $\sigma_{\rm p}/\sigma_0$ value. Other combinations of $\alpha_0 \ell$ resulting in the same change in $\sigma_{p}$ as considered in Figure \ref{fig:fig12} were also examined, but that did not alter the results in Figure \ref{fig:fig12}. Hence, the influence of $G_{\rm p}/G_{\rm m}$ on $\sigma_{\rm p}/\sigma_0$ appears to be independent of $\alpha_0 \ell$. In addition, simulations were carried out with higher and lower $f$ values yielding the same $\sigma_{\rm p}/\sigma_0$ values with  $G_{\rm p} = G_{\rm m}$ as considered in Figure \ref{fig:fig12}. Those results indicates that the effect of modulus mismatch is amplified when $f$ increases.

\subsection{Effects of a heterogeneous distribution of particles}
\label{Clustering}

\begin{figure}[htb!]
\begin{center}
    \subfigure[]{ \includegraphics[width=0.465\textwidth]{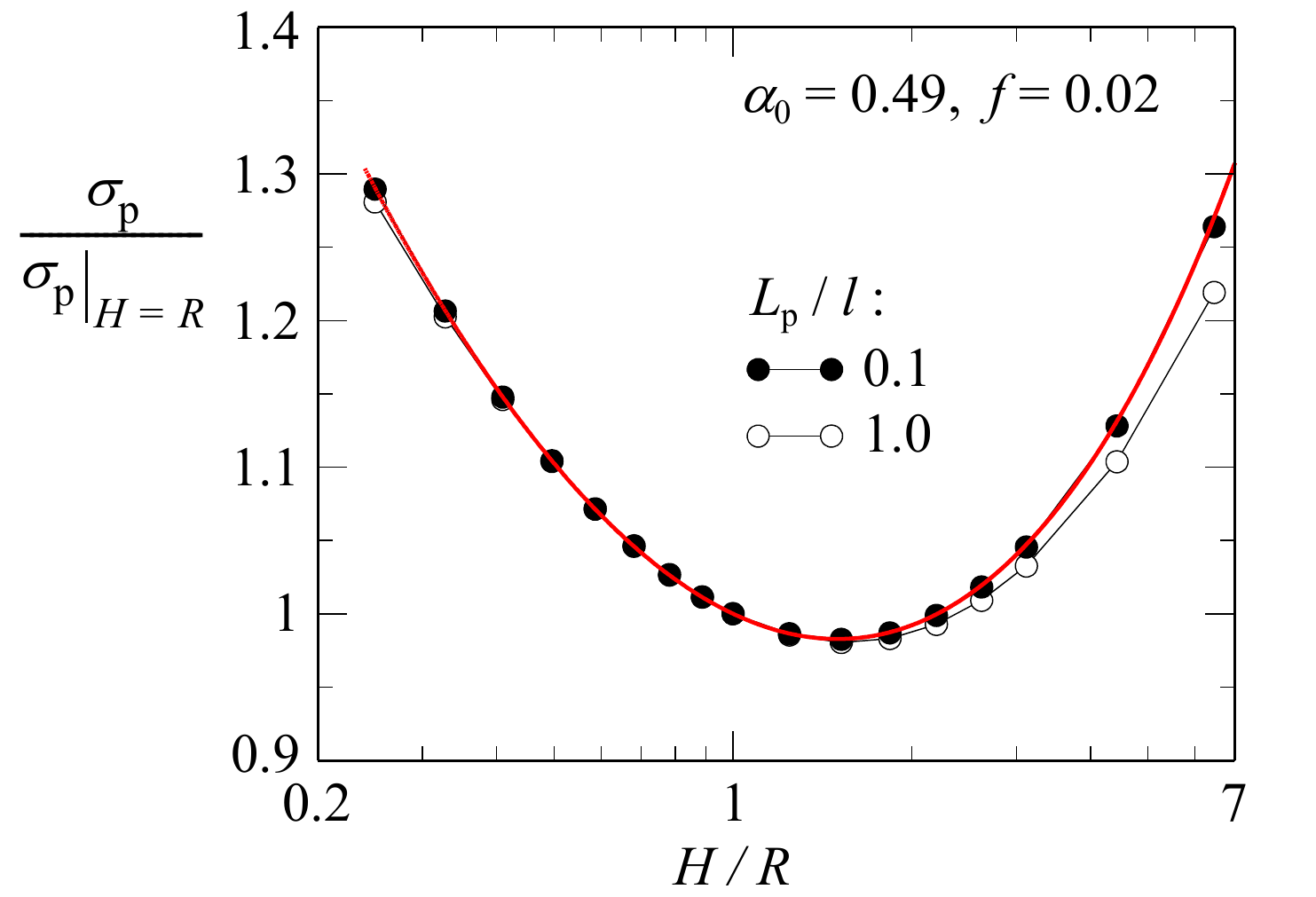} }
    \subfigure[]{ \includegraphics[width=0.45\textwidth]{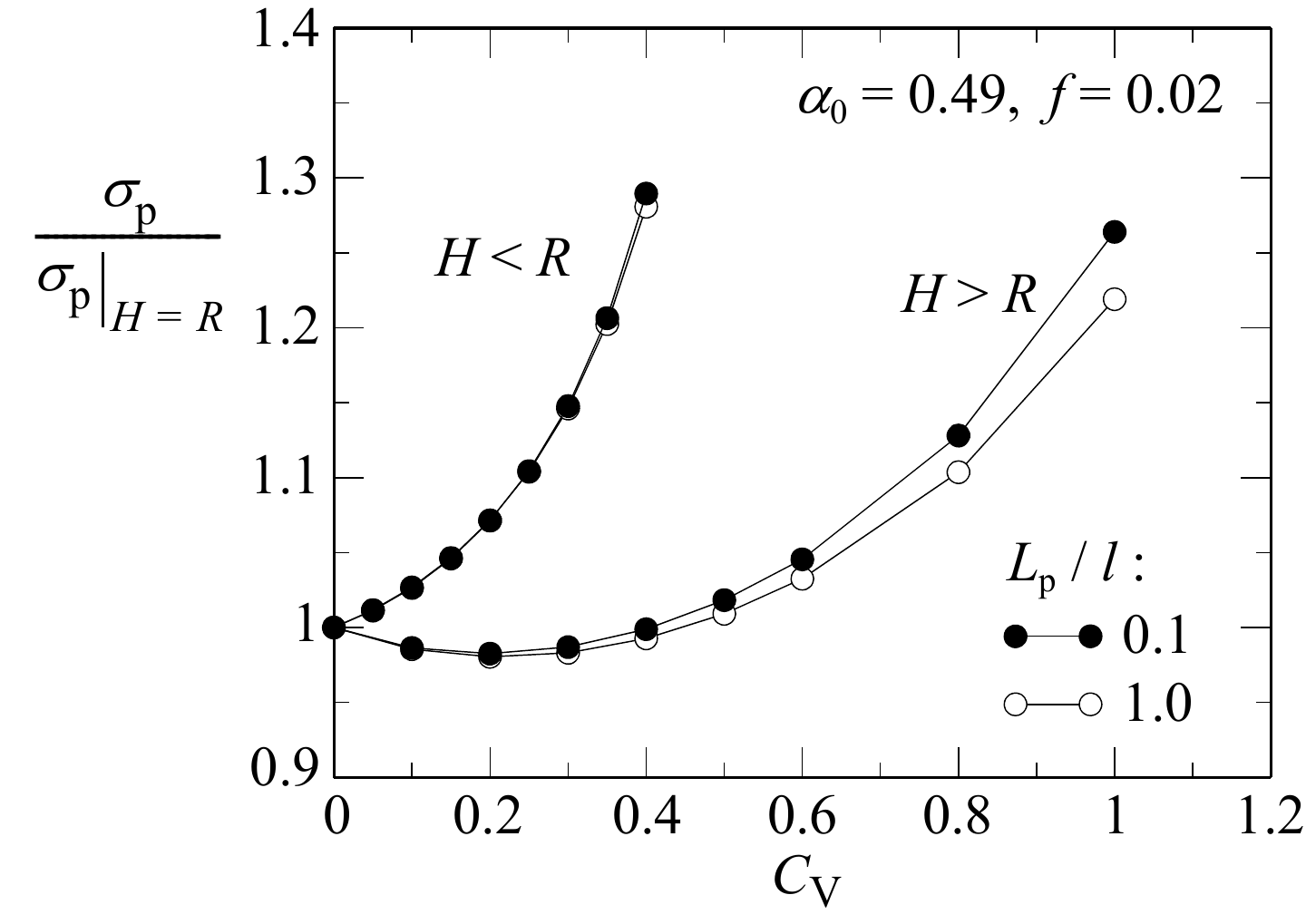} }
    \caption{Influence of an inhomogeneous particle distribution on strengthening. The solid lines pertain to $L_{\rm p}$ defined by the center-to-center distance, and the dashed lines pertain to $L_{\rm p}$ defined by the surface-to-surface distance. In (a) inhomogeneity is represented by the variation of $H/R$, and in (b) by the coefficient of variance $C_{\rm V}$. The parameters used to generate the curves are indicated in (a). The red dash-dot curve in (a) is a prediction based on $\chi(\xi)/\chi(1)$ as discussed in Subsection 3.6.}
    \label{fig:fig13}
\end{center}
\end{figure}

\noindent The present unit cell model allows for a limited investigation of how an inhomogeneous particle distribution may affect strengthening. By varying the ratio between height and radius of the unit cell $H/R$, the particle spacing distribution will be inhomogeneous with $C_{\rm V} > 0$. A decreasing $H/R$ will lead to a distribution with columns of tightly packed  particles, whereas an increasing $H/R$ will lead to a distribution with planes of tightly packed particles. Two curves were generated with $f=0.02$ and $\alpha_0=0.49$ for the case (C-C), one with $L_{\rm p}/\ell=0.1$ and one with $L_{\rm p}/\ell=1$. The result is presented as the increase in yield stress $\sigma_{\rm p}$ normalized by the value obtained for a homogeneous distribution ($H=R$). In Figure \ref{fig:fig13}a, the relative change in yield stress is plotted versus $H/R$ on log-scale. Strengthening is minimum for $H/R \approx 1.5$, it increases considerably for $H/R$ smaller or larger than this value and appears to be insensitive to the particle spacing $L_{\rm p} = 0.1\ell$.

The result in Figure \ref{fig:fig13}a is replotted in Figure \ref{fig:fig13}b, but now as a function of the coefficient of variance $C_{\rm V}$ instead. As can be observed, the two branches $H<R$ and $H>R$, respectively, deviate significantly from each other and do not collapse. Thus, the coefficient of variance does not characterize the influence of an inhomogeneous particle spacing on strengthening within the current modelling framework. Not shown here, but exactly the same results were obtained by applying biaxial tension on the unit cell and for several other values of $f$ and $\alpha_0$.

\subsection{Hardening of the macrosopic stress-strain curve}

\noindent From Figures \ref{fig:fig9}a and \ref{fig:fig10}a it was observed that particles embedded in an ideal plastic matrix give rise to a plastic work hardening that depends on the volume fraction of particles. But if also the matrix is a strain hardening material and/or the interface parameter $ \alpha_1 > 0 $, additional hardening will develop. This was examined for materials containing a particle volume fraction of 2\%, with matrix hardening exponents $ N = \{0, 0.1, 0.2\} $ and interfaces with $ 0 \le \alpha_1 \le 2$. Stress-strain curves were generated such that the combination of interface parameter $ \alpha_0 $ and the particle center-to-center spacing $ L_{\rm p}/\ell $ gave $ \sigma_{\rm p} = \sigma_{0} $ for the case $ N = \alpha_1 = 0 $.

\begin{figure}[htb!]
	\begin{center}
		\subfigure[]{ \includegraphics[width=0.3\textwidth]{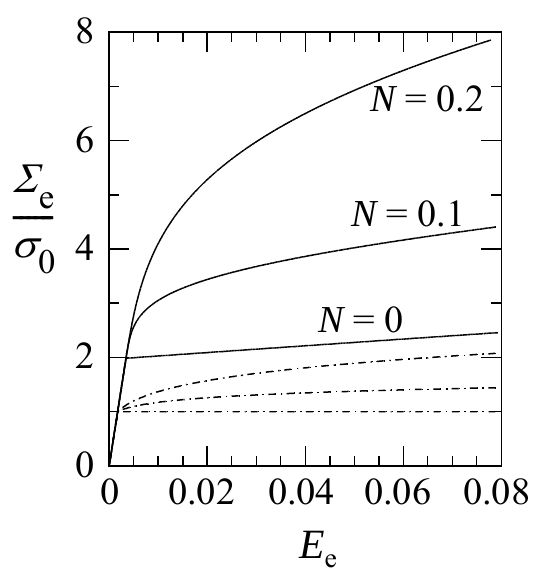} }
		\subfigure[]{ \includegraphics[width=0.80\textwidth]{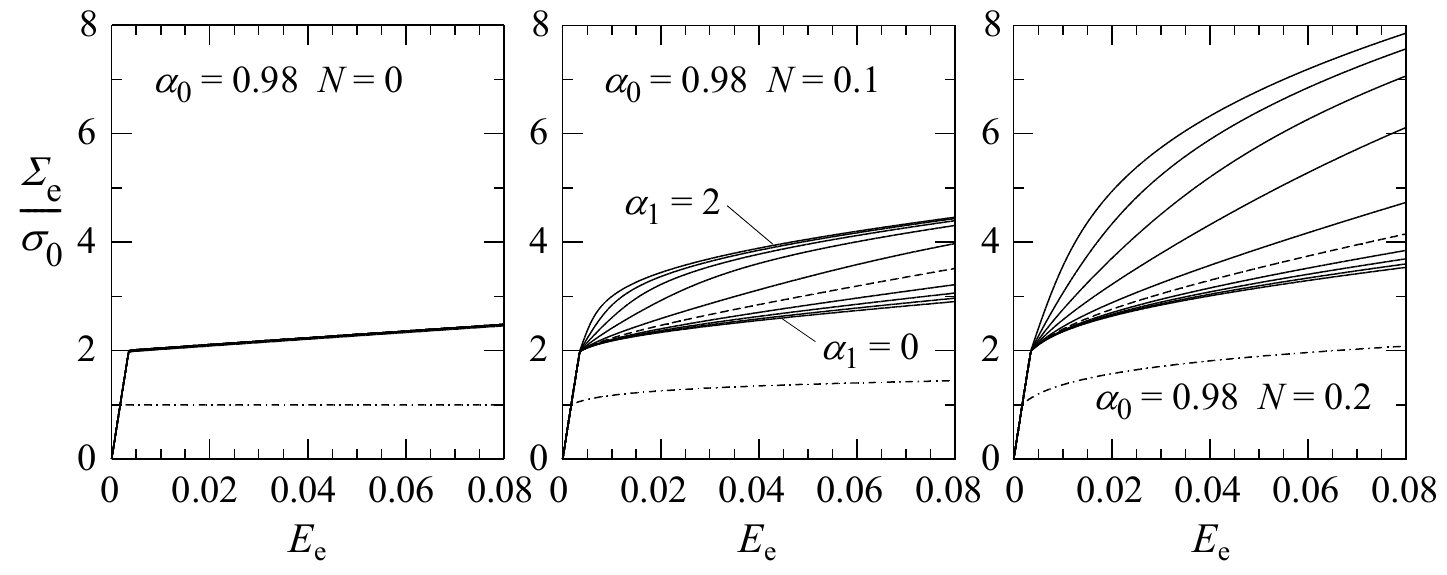} }
		\subfigure[]{ \includegraphics[width=0.80\textwidth]{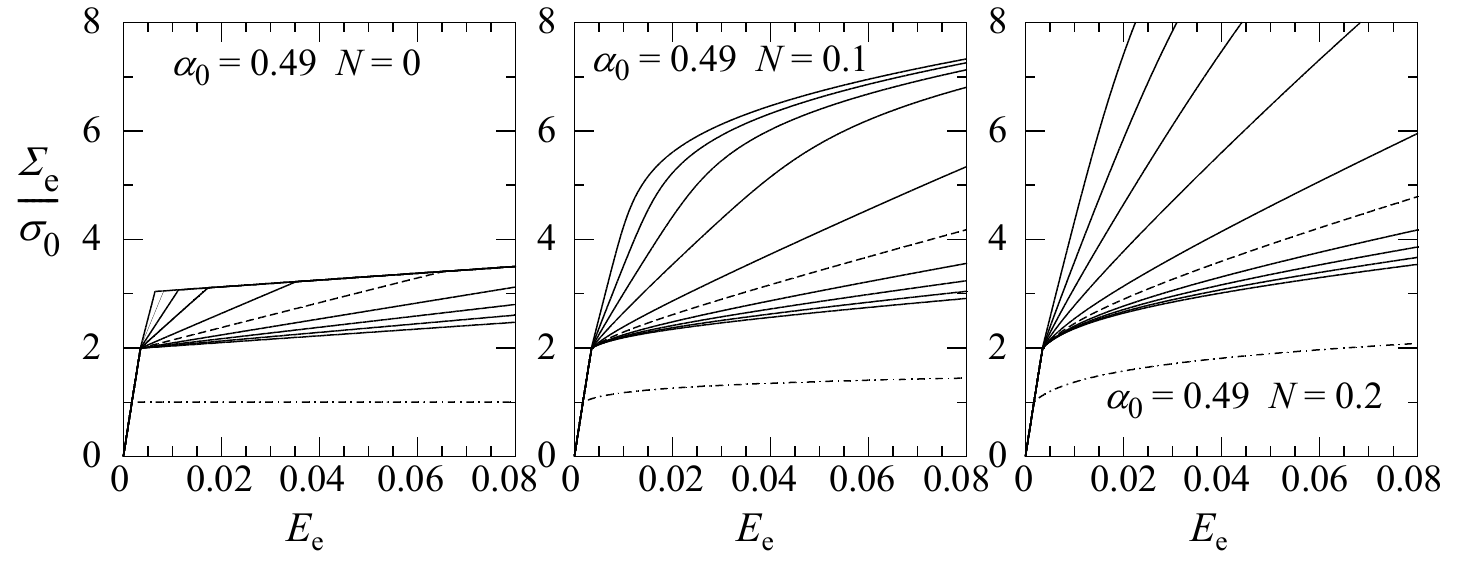} }
		\caption{Influence of $ N $ and $ \alpha_1 $ on the macroscopic stress-strain response. The results in (a) corresponds to a micro-hard interface with $ L_{\rm p}/\ell = 0.3941 $, and in (b) to an interface with $ \alpha_0 = 0.98 $ and $ L_{\rm p}/\ell = 0.3861 $, and in (c) to an interface with $ \alpha_0 = 0.49 $ and $ L_{\rm p}/\ell = 0.1931 $, respectively. In (b) and (c) the curves are generated with $\alpha_1 = \{0, 0.004, 0.01, 0.02, 0.04, 0.08, 0.2, 0.04, 0.08, 2.0\}$, where the stress level increases with increasing values of $ \alpha_1 $. The dash-dotted curves in all graphs represent the matrix response in absence of particles.}
		\label{fig:fig14}
	\end{center}
\end{figure}

The maximum attainable strength of an interface and its effect on the overall strengthening is limited by the yield condition of the matrix material, as argued in Section \ref{sec:InterfaceDescription}. Thus, a micro-hard interface would yield an upper limit of hardening in this respect. Figure \ref{fig:fig14}a shows the stress-strain curves for such an interface for three different values of $ N $. By comparing the solid curves (composite) with their corresponding dash-dotted curves (matrix), it can be observed that the additional hardening caused by the presence of particles ($ h_{\rm p} $ defined in Figure \ref{fig:Def_of_Sp}) increases with $ N $ of the matrix. This may be understood from the fact that the flow stress is allowed to significantly surpass $ \sigma_0 $ after the onset of plastic deformation when $ N > 0 $, thus effectively strengthening the interface and leaving room for additional hardening.

Macroscopic stress-strain curves corresponding to materials with an $ \alpha_0 $ value of the particle/matrix interface below the micro-hard limit ($ = 1 $) are shown in Figures \ref{fig:fig14}b and \ref{fig:fig14}c. First, consider a case with $ \alpha_1 = 0 $ (the lowest solid curve in all graphs). For this case the `interface strength' will not evolve with plastic deformation, it will be constant and equal to $ \psi_{\Gamma}(\epsilon_{\Gamma}) = \alpha_0 \ell \sigma_0 $, and it can be observed that the increase in hardening due to particles, $ h_{\rm p} $, is essentially independent of $ N $. By contrast, when $ \alpha_1 > 0 $, the additional hardening observed is a strong function of both $ N $ and $ \alpha_1 $. For instance, when $ \alpha_0 $ is close to the micro-hard limit as in Figure \ref{fig:fig14}b ($ \alpha_0 = 0.98 $) and $ \alpha_1 $ is large enough, the macroscopic stress-strain curves rapidly approach the response obtained with the micro-hard interface shown in Figure \ref{fig:fig14}a. 

In Figure \ref{fig:fig14}c results are generated with $ \alpha_0 = 0.49 $, which is approximately half the micro-hard limit. Thus, in this case a large $\alpha_1$ value can lead to a substantial additional hardening. Recall that the results in Figure \ref{fig:fig14} were generated with pairs of $ \{\alpha_0, L_{\rm p}/\ell\} $ that give $\sigma_{\rm p} = \sigma_{0}$ for $ N = \alpha_1 = 0 $. Thus, an $ \alpha_0 $ value significantly lower than the micro-hard limit opens up for both additional strengthening and hardening if $ \alpha_1 > 0 $. This is especially apparent in the left sub-graph in Figure \ref{fig:fig14}c ($ N = 0 $), where it is clear that the character of the hardening curves is directly correlated to the development of plastic straining at the interface. For instance, consider the dashed curve, which pertains to $ \alpha_1 = 0.04 $. Here, a considerable hardening is seen as plastic straining starts to develop at the interface at the onset of macroscopic yielding and continues to evolve until a knee is reached, where the slope of the macroscopic stress-strain curve exhibits a drastic decrease. The knee corresponds to a stage in loading where plastic straining at the interface cease to since micro-hard conditions has been reached. Furthermore, if $\alpha_1$ is sufficiently high, $\sigma_{\rm p}$ is strongly affected, leaving little room for post-yield hardening, since `micro-hard' conditions practically has been attained at the onset of macroscopic yielding.

Sufficiently large values of $\alpha_1$ will effectively change the character of the surface energy $\Psi_\Gamma$ from a linear dependence of the effective plastic strain to a quadratic dependency. If combined with an $\alpha_0$ value significant below the `micro-hard' limit, an unrealistic high strain hardening will result if the composite contains a strain hardening matrix material. Furthermore, if the matrix material is strain hardening, a micro-hard interface also seems to result in an unrealistically high post yield strain hardening.

In conclusion, an interface with $\alpha_0 \le 1$ and $\alpha_1 = 0$ seems to be more realistic choice for capturing post yield strain hardening in engineering materials with the model under investigation.

\subsection{Discussion and a function proposed for $\sigma_{\rm p}$}
\noindent Effects on strengthening of particle spacing and volume fraction, mismatch in particle/matrix modulus, and irregularity in particle distribution have so far been examined within the proposed nonlocal continuum framework. The systematic trend in the obtained numerical results suggests that the increase in yield stress \eref{eqn:Sp_fcn} may be split up into two terms as
\begin{equation}\label{eqn:Fs_function}
	\sigma_{\rm p} = \sigma_{\rm pl} +  \sigma_{\rm pnl}.
\end{equation}
where the first term is independent of particle spacing (a local term), whereas the second term (a non-local term) depends on particle spacing that vanishes when the spacing becomes sufficiently large and/or the interface strength is negligible. Following \cite{Bao91}, the local term may be written on the form
\begin{equation}\label{eqn:Sloc}
	\sigma_{\rm pl} = g_{\rm l} \, \sigma_0 \, f,
\end{equation}
where $ g_{\rm l} $ is dimensionless function that depends on  modulus mismatch, $ G_{\rm p}/G_{\rm m} $, and $ f $ to cope with larger volume fractions of particles. If $G_{\rm p} = G_{\rm m}$, $ g_{\rm l} $ must be zero. For sufficiently small volume fractions of rigid spherical particles ($ G_{\rm p}/G_{\rm m} \rightarrow \infty $), \cite{Bao91} find that $g_{\rm l} = 0.375$ is suitable for the type of particle distributions considered here. Regarding the non-local term, the numerical results suggest that a multiplicative form of the influence of particle size and spacing may capture the observed trend in essence, which can be formulated as
\begin{equation}\label{eqn:Snloc}
 \sigma_{\rm pnl} = g_{\rm nl} \cdot C_{\rm nl} \frac{\psi_{\Gamma}\:f^{2/3}}{(1-f) L_{\rm p}},  \quad {\rm where} \quad \psi_{\Gamma} = \alpha_0 \ell \sigma_0,  \quad  0 \le \alpha_0 \le 1.
\end{equation}
Here, $\psi_{\Gamma}$ defines the initial interface strength, $\alpha_0$ is limited by the micro-hard condition $\alpha_0 \le \alpha_{\mu \rm{-hard}} \approx 1$, and $C_{\rm nl}$ is a dimensionless scaling parameter that also carry a possible dependence on the character of the particle distribution. The power $2/3$ of $f$ is the outcome from Figure \ref{fig:fig11}b in Section \ref{sec:VolumeFractionEffects} for $L_{\rm p} = L_{\rm pcc}$. Linked to this, the term $1/(1-f)$ has been included as a first order correction to capture the numerical results observed in Figure \ref{fig:fig11}b for an increasing $f$. The red dashed curve represents the slope of \eref{eqn:Snloc} in the log-log graph, i.e., $f \cdot \partial \log( f^{2/3}/(1-f) ) = 2/3 + f/(1-f)$. This term clearly brings \eref{eqn:Snloc} in closer agreement with the numerical results in Figure \ref{fig:fig11}b. Finally, a possible mismatch in particle/matrix modulus is accounted for by function $g_{\rm nl}$, which depends on $ G_{\rm p}/G_{\rm m}$ in a manner displayed in Figure \ref{fig:fig12}a.

\begin{figure}[htb!]
\begin{center}
    \subfigure[]{ \includegraphics[width=0.46\textwidth]{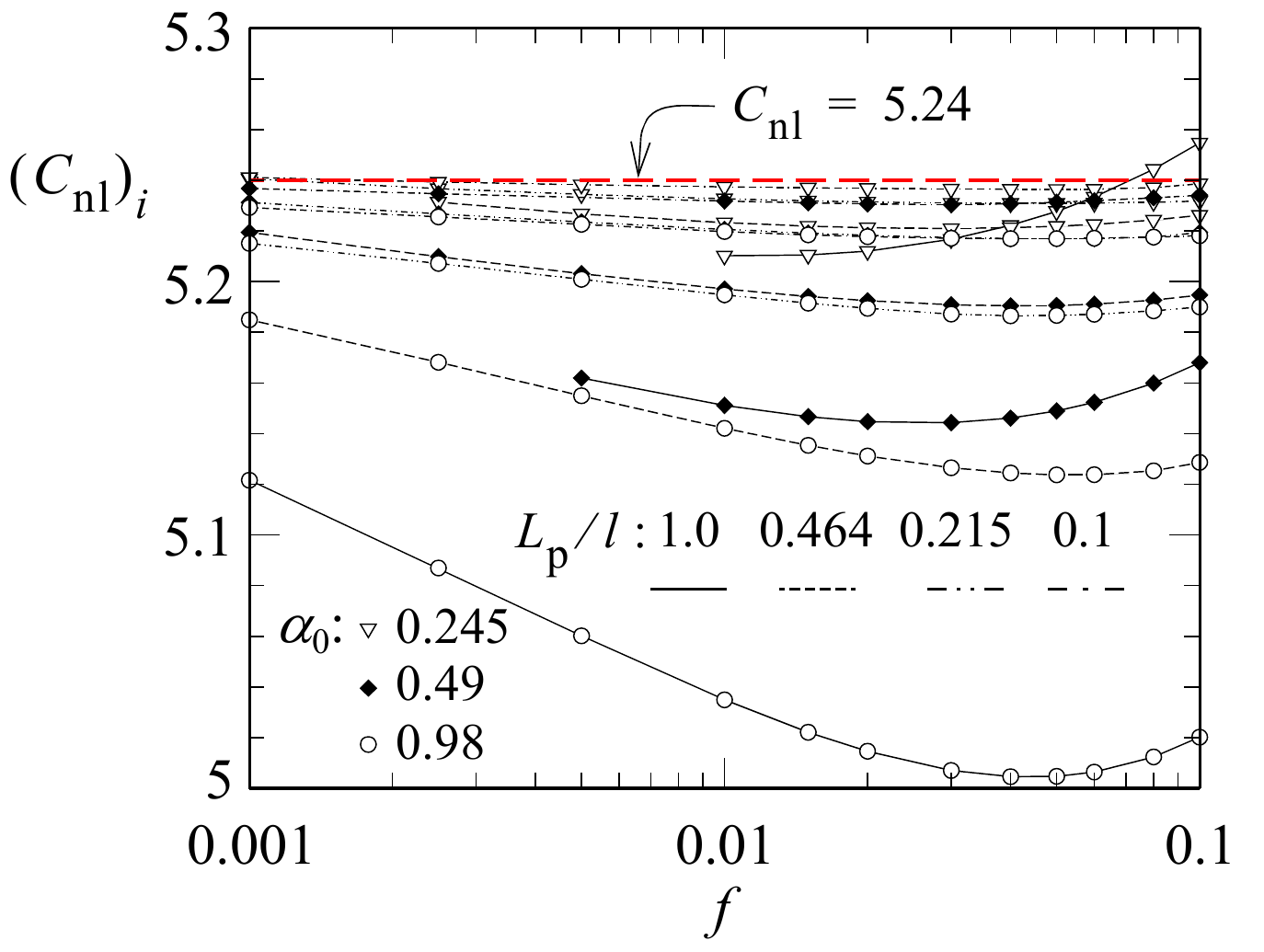} }
    \subfigure[]{ \includegraphics[width=0.46\textwidth]{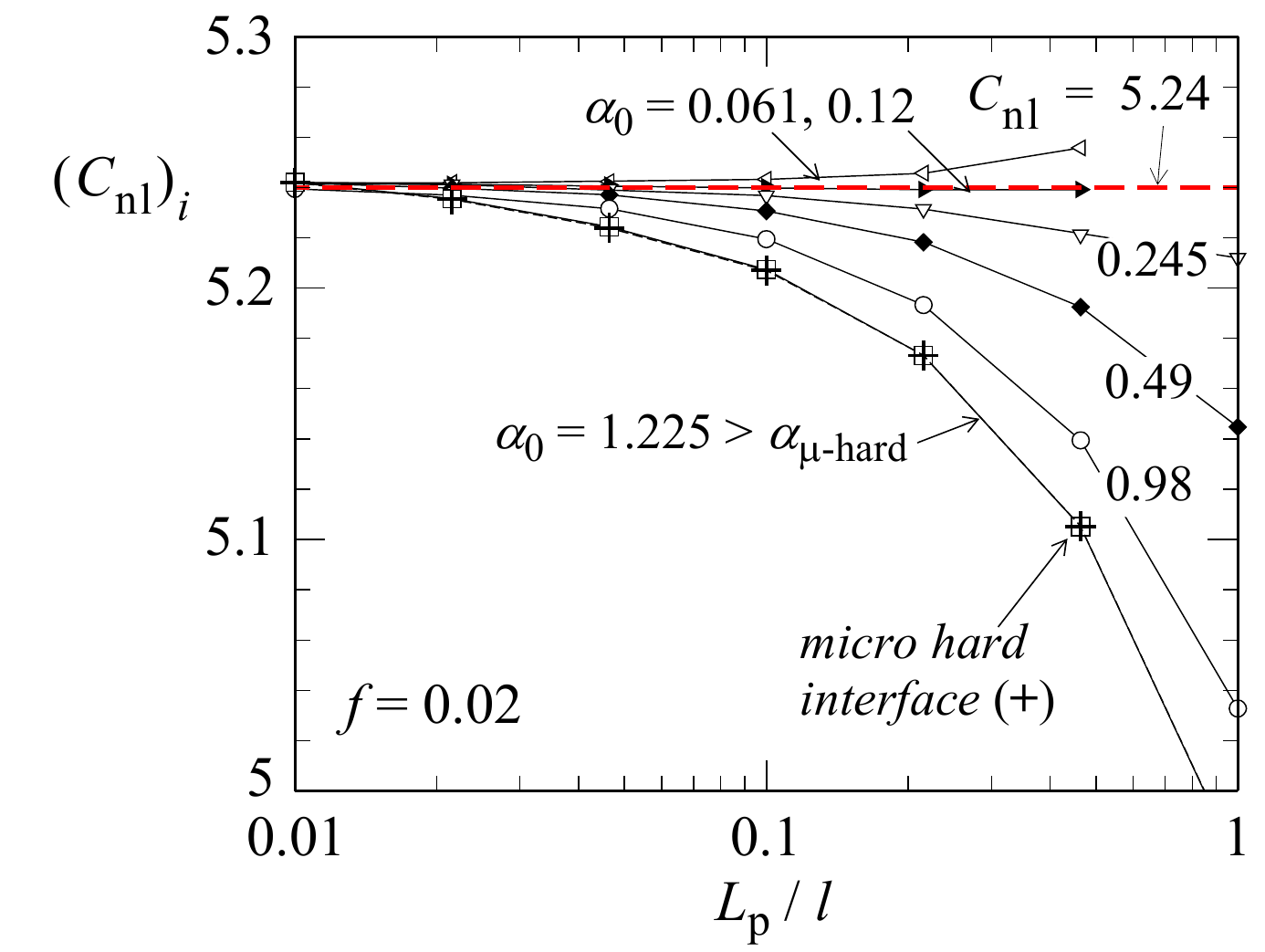} }
    \caption{Estimate of coefficient $(C_{\rm nl})_i$ for each data point $i$ plotted versus particle fraction $f$ in graph (a), and versus relative particle spacing in graph (b). Data in (b) pertain to $f=0.02$.}
    \label{fig:fig15} 
\end{center}
\end{figure}

To give an estimate of the coefficient $C_{\rm nl}$, the data points for the parameter combinations: $\alpha_0 = \{0.245, 0.49, 0.98\}$, $L_{\rm p}/\ell = \{0.1, 10^{-2/3}, 10^{-1/3}, 1\}$, and $f = \{0.001, ..., 0.1\}$ were considered (data points for combinations pertaining to $\alpha_0 = 0.49$ are displayed in Figure \ref{fig:fig9}b). In the light of Figure \ref{fig:Res2}a, only data fulfilling $\sigma_{\rm p}/\sigma_0 \ge 0.05$ were used to alleviate an improper scatter. For each data point $i$, $(C_{\rm nl})_i$ was calculated by use of equations \eref{eqn:Fs_function}-\eref{eqn:Snloc} and plotted versus $f$ and $L_{\rm p}/\ell$ as shown in Figure \ref{fig:fig15}. In Figure \ref{fig:fig15}a the value of $C_{\rm nl}$ is not unambiguous, but the curves show convergence when $f$ becomes small. However, Figure \ref{fig:fig15}b reveals that all curves converge towards the same value of $C_{\rm nl}$ equal to approximately 5.24 when $L_{\rm p} \ll \ell$. The red dash-dotted horizontal line included in both graphs of Figure \ref{fig:fig15} represents $ C_{\rm nl} = 5.24 $.

\begin{figure}[htb!]
\begin{center}
    \subfigure[]{ \includegraphics[width=0.46\textwidth]{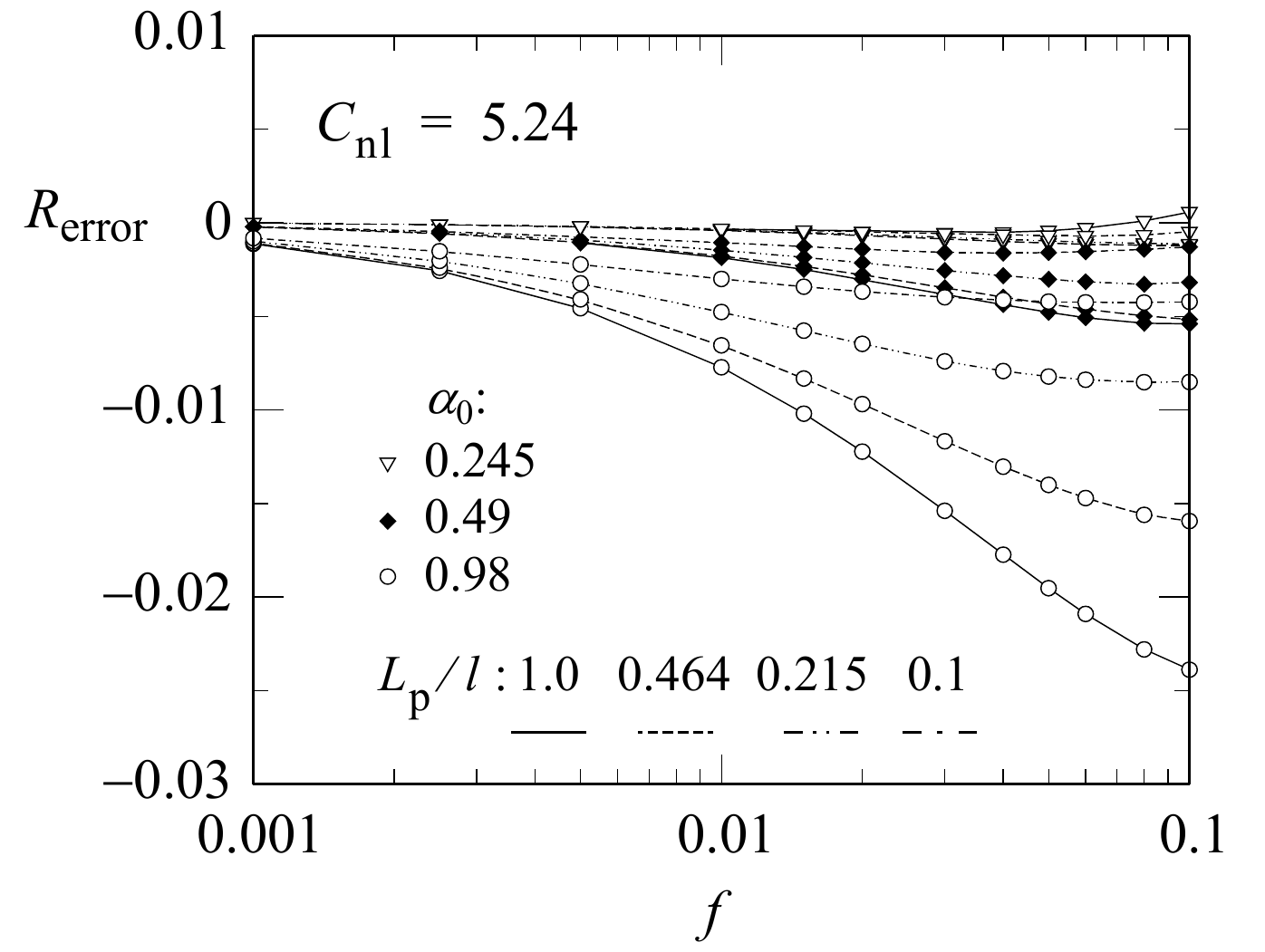} }
    \subfigure[]{ \includegraphics[width=0.46\textwidth]{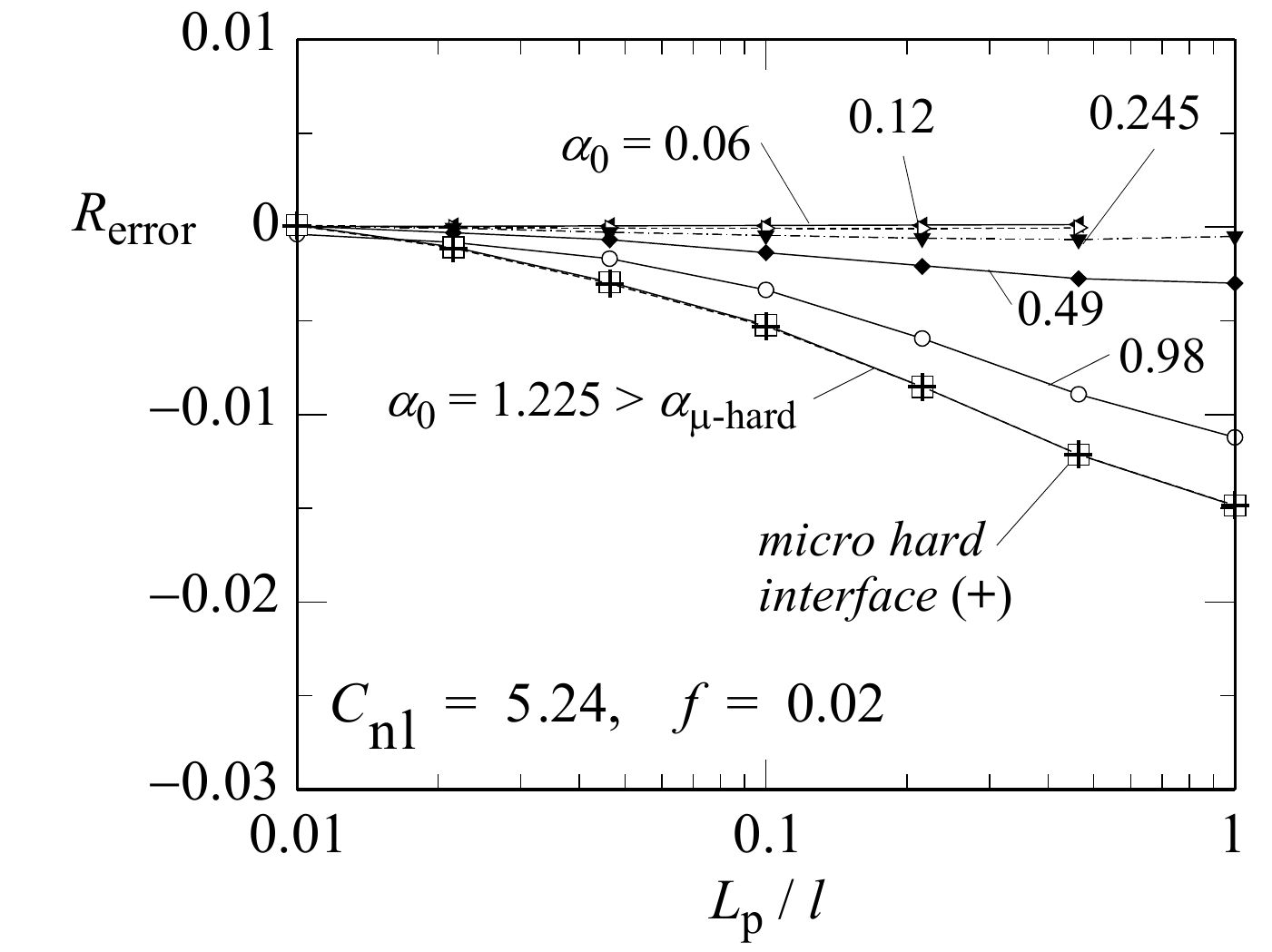} }
    \caption{Relative error in the yield stress predicted by equation \eref{eqn:Fs_function} plotted versus volume fraction of particles in (a) and plotted versus mean particle spacing in (b). In both graphs, $R_{\rm error}$ is calculated based on data from Figure \ref{fig:fig15}.}
    \label{fig:fig16} 
\end{center}
\end{figure}

To further scrutinize the validity of \eref{eqn:Fs_function}-\eref{eqn:Snloc}, a measure of relative error of the predicted macroscopic yield stress is introduced as
\begin{equation}\label{eqn:RelErr}
  R_{\rm error} = 1 - \frac{\sigma_0 + \sigma_{\rm p}|_{\rm \eref{eqn:Fs_function}-\eref{eqn:Snloc}}}{\Sigma_{\rm e}(E_{\rm e}^{\rm p} = 0)}.
\end{equation}
This error measure was evaluated for all data points in Figure \ref{fig:fig15}. $R_{\rm error}$ is plotted versus $f$ in Figure \ref{fig:fig16}a, and versus $L_{\rm p}/\ell$ in Figure \ref{fig:fig16}b, respectively. As can be observed, the relative error is small and less than 0.5\% for $\alpha_0 \le 0.49$. Furthermore, the proposed relation in \eref{eqn:Snloc} with $ C_{\rm nl} = 5.24 $ appears to give an upper bound for the relative error. 

The term $f^{2/3}$ in \eref{eqn:Snloc} will be rationalized as follows. The strength of the interface plays a key role for the increase in yield stress and enters the problem by contributing to the internal virtual work. Its contribution in the variational principle \eref{eqn:VirtualWork} can by use of \eref{eqn:dissInt} and \eref{eqn:PsiInt} for spherical particles of radius $r_{\rm p}$ becomes
\begin{equation}\label{eqn:ivw_int}
\int_{S^{\Gamma}}\left[ M^{\Gamma}_{ij}\delta\varepsilon^{\rm p}_{ij}  \right]{\rm d}S = \int_{S^{\Gamma}} \psi_{\Gamma} \delta\varepsilon_{\Gamma}  {\rm d}S  = 4 \pi r_{\rm p}^2 \psi_{\Gamma} \delta\varepsilon_{\Gamma},
\end{equation}
where $ \delta\varepsilon_{\Gamma} $ is chosen to be constant in order to obtain the right-hand side. The relative contribution of this term in volume $V=2\pi R^3 \xi$ can by use of \eref{eqn:Lpcc} and some manipulations be expressed as
\begin{equation}\label{eqn:ivw_int_byV}
\frac{4 \pi r_{\rm p}^2}{2\pi R^3 \xi} \; \psi_{\Gamma} \delta\varepsilon_{\Gamma} = \chi(\xi) \: \alpha_0 \ell \sigma_0
\: \frac{f^{2/3}}{L_{\rm p}} \: \delta\varepsilon_{\Gamma}, \quad {\rm where} \quad \chi(\xi) = \left( \frac{3+\xi}{\xi} \right) \left( \frac{3\xi}{2} \right)^{2/3}.
\end{equation}
This expression brings out the response proposed in \eref{eqn:Fs_function}. Moreover, $\chi(\xi)/\chi(1)$ introduced in \eref{eqn:ivw_int_byV} represents the relative change in the area of surface $S^{\Gamma}$ as a function of $\xi$ for a fixed combination of $f$ and $L_{\rm p}$. This ratio would then be expected to account for the change in yield stress observed when $\xi$ deviates from one, as is  illustrated in Figure \ref{fig:fig13}a. The red dash-dotted curve in this figure represents function $\chi(\xi)/\chi(1)$ plotted versus $\xi$ ($=H/R$), and indeed, it falls on top of the curve pertaining to $L_{\rm pcc}=0.1\ell$ with $\alpha_0 = 0.49$. 
Thus, the results obtained in this study and elucidated by the heuristic argument above, supports that the increase in yield stress is determined by the total surface area of particles per unit volume of the material times the strength of the particle/matrix interface. Furthermore, the non-dimensional scaling parameter can be identified as $C_{\rm nl} = \chi(1) = 5.24$, since the numerical results analysed in Figures \ref{fig:fig15} and \ref{fig:fig16} belonged to $\xi=1$. The length parameter is brought out by the ratio total surface area of particles per unit volume, which by \eref{eqn:Lpcc} and \eref{eqn:ivw_int_byV} can be expressed as $S^{\Gamma} / V = \chi(\xi) f^{2/3} / L_{\rm pcc} = 3 f / r_{\rm p}$. Hence, the non-local contribution to the increase in yield strength \eref{eqn:Snloc} can be expressed as
\begin{equation}\label{eqn:SnlocGen}
 \sigma_{\rm pnl} = g_{\rm nl} \cdot \frac{1}{(1-f)} \frac{1}{V} \int_{S^{\Gamma}} \psi_{\Gamma} {\rm d}S =
 g_{\rm nl} \cdot \frac{1}{(1-f)} \: \frac{1}{V} \sum_{i=k}^{N_{\rm p}} ( \psi_{\Gamma} S^{\Gamma} )_k ,
\end{equation}
where $f$ is the volume fraction of particles in volume $V$, and $ S^{\Gamma} $ is the total surface area of all particles in $ V $. In the most right-hand side of \eref{eqn:SnlocGen}, it is assumed that volume $V$ contains $N_{\rm p}$ particles, where $( \psi_{\Gamma} S^{\Gamma} )_k$ represents the strengthening contribution of particle $k$.

A commonly used relation for describing precipitation hardening by dislocation looping is the Orowan-Ashby equation (\cite{Orowan48}, \cite{Ashby66}, \cite{Hirsch69}, \cite{Brown71}). If recast in a form that applies to a polycrystalline material and expressed in terms of the present notation, it may be written as
\begin{equation}\label{eqn:Orowan_Ashby}
\sigma_{\rm p}|_{\rm{O-A}} = A \: \frac{G b}{L_{\rm p}} \: \ln(\frac{2r_{\rm p}}{b}).
\end{equation}
Here, $b$ is the magnitude of Burgers vector of the mobile dislocations, $A$ is a parameter that depends on the character of the dislocations (a value 0.4 will be used here), and $L_{\rm p}$ will be interpreted as $L_{\rm pcc}$ to simplify matters. The term $\ln(2r_{\rm p}/b)$ in the Orowan-Ashby equation accounts for changes in the energy of dislocations that bow around particles which cannot be regarded as point obstacles (see discussions in \cite{Ardell85} and \cite{Gladman99} and references therein). For a fixed mean distance between particles, this term indirectly depends on $f$. Furthermore, by setting equation \eref{eqn:Snloc} equal to \eref{eqn:Orowan_Ashby}, an estimate of the material length parameter can be obtained in the context of the present model as
\begin{equation}\label{eqn:ell}
 \alpha_0 \ell = \frac{b}{\varepsilon_0} \cdot B , \quad {\rm where} \quad B = \frac{A}{2(1+\nu)C_{\rm nl}} \frac{(1-f)\ln(2r_{\rm p}/b)}{f^{2/3}}.
\end{equation}
E.g., if $\nu=0.3$, $f=0.04$, $2r_{\rm p}/b = 63.4$ with $C_{nl} = 5.24$, $B$ becomes one, and with $\varepsilon_0=0.002$ as used in this study, we obtain $\alpha_0 \ell$ equal to  $500b$. With a matrix material of pure iron, this gives $\alpha_0 \ell \approx 125$ nanometers, which is reasonable.

\section{Comparison with experimental results}
\label{sec:4}

\noindent To expedite a comparison between model predictions and experimental results it will be assumed that the material systems to be considered contain spherical particles with the same elastic constants as the matrix material, and that $\alpha_0$ is independent of particle radius. Then, by use of $\psi_{\Gamma} / V = 3 f / r_{\rm p}$ (see discussion above), relation \eref{eqn:SnlocGen} simplifies to

\begin{equation}\label{eqn:SnlocExp}
 \sigma_{\rm pnl} = \left[ \alpha_0 \ell \right] \frac{3 f}{(1-f)} \frac{1}{\bar{r}_{\rm p}}
\end{equation}

The value of $ \bar{r}_{\rm p} $ will differ somewhat from the mean value of the distribution, since the representative particle radius is evaluated as $ \bar{r}_{\rm p} = 3fV/S^{\rm \Gamma} = \sum_{i=1}^{N_{\rm p}} r^3_{\rm p i} / \sum_{i=1}^{N_{\rm p}} r^2_{\rm p i} $. Below it will be assumed that the difference between the representative radius and the mean radius, respectively, of a particle distribution can be neglected. Then, $ f $ and $\bar{r}_{\rm p}$ in \eref{eqn:SnlocExp} are quantities that typically are reported for metals reinforced by particles or precipitates, whereas the bracket term must be determined, e.g., by uniaxial tensile test data.

\begin{figure}[htb!]
	\begin{center}
		\subfigure[]{ \includegraphics[width=0.46\textwidth]{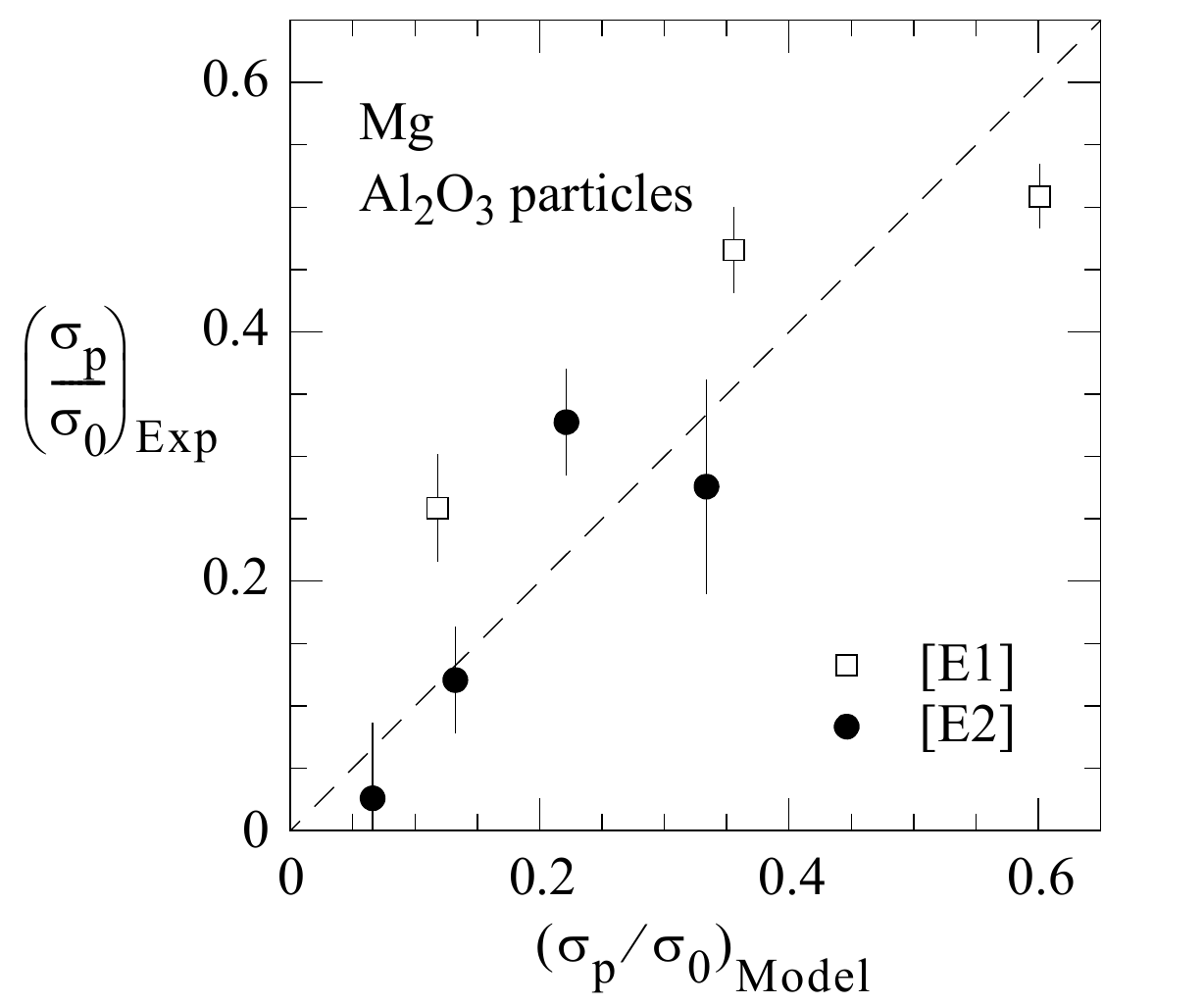} }
		\subfigure[]{ \includegraphics[width=0.46\textwidth]{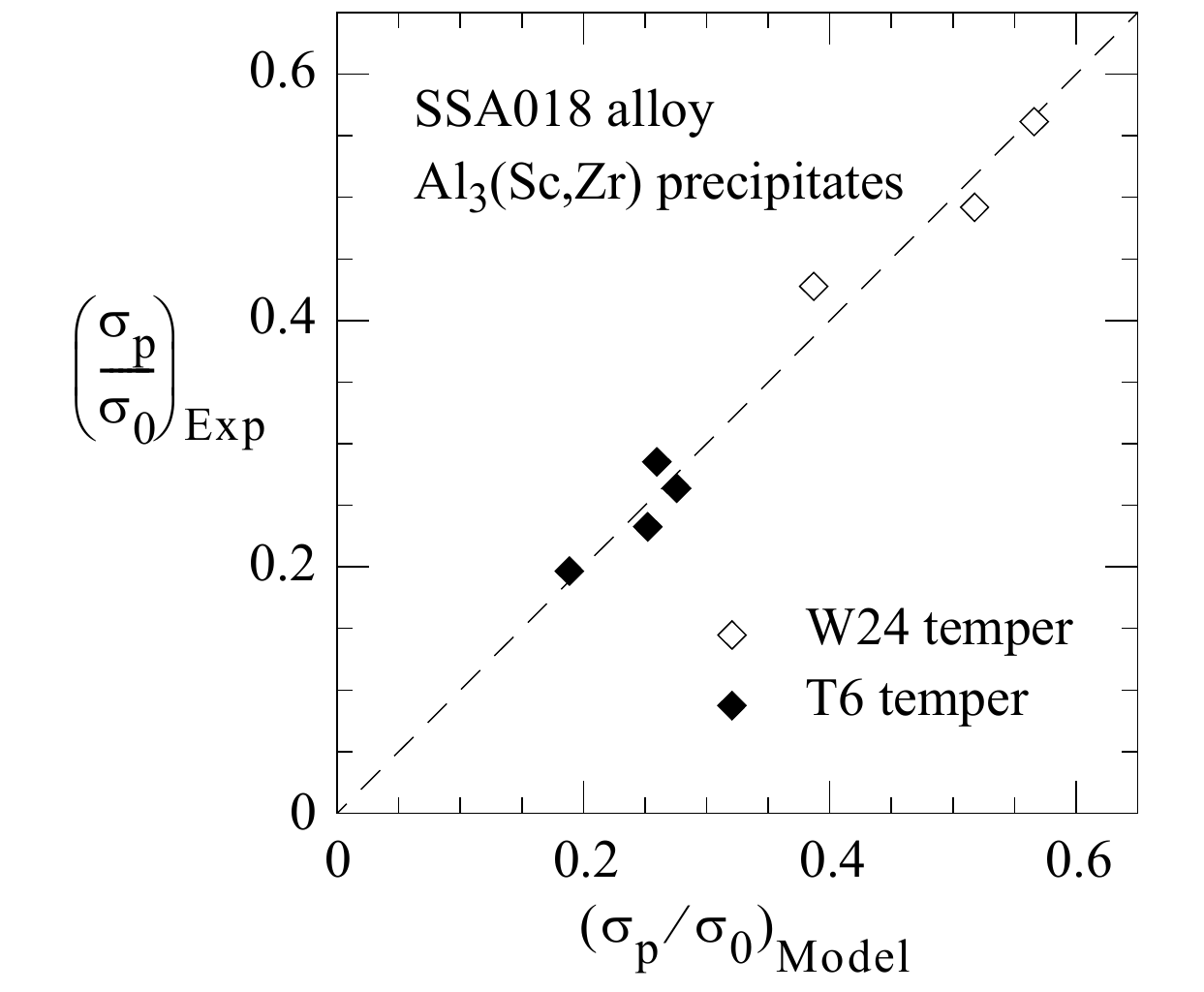} }
		\caption{Experimentally measured increase in yield stress versus model predictions for two different material systems: (a) ${\rm Mg/Al}_2{\rm O}_3$ nanocomposite, experiments by \cite{Exp1} [E1] and \cite{Exp2} [E2]; (b) Aluminum 7000 series Al-Zn-Mg-Cu-Sc-Zr alloy with ${\rm Al}_3{\rm (Sc,Zr)}$ particles, experiments by \cite{Exp3}.}
		\label{fig:fig17} 
	\end{center}
\end{figure}

The comparison to experiments will be limited to results found in the literature belonging to metal matrix composites (MMC) and alloys. From MMCs, magnesium nanocomposites reinforced with nanosized ${\rm Al}_2{\rm O}_3$ particulates with an average diameter of 50 nm and volume fractions in the range 0.22\% to 1.5 \% will be checked. The ${\rm Mg/Al}_2{\rm O}_3$ nanocomposite was produced either by a so called `disintegrated melt decomposition' (\cite{Exp1}, referred to as [E1]) or by use of a microwave sintering technique (\cite{Exp2}, referred to as [E2]). The initial yield stress of the matrix is reported as 116 MPa. Applying the current model to the experimental results, the term $\alpha_0 \ell$ in \eref{eqn:SnlocExp} was estimated to 0.73 microns for set [E1] and to 0.30 microns for set [E2]. Assuming that $\ell$ would be a constant for the Mg matrix, the discord in estimates may be related to differences in the interface property ($\alpha_0$) due to disparate manufacturing processes. The increase in yield stress as measured in the experiments are plotted versus model predictions in Figure \ref{fig:fig17}a, and as can be seen, the model captures the strengthening trend of an increasing volume fraction of particles well.

Alloys are represented by experiments of \cite{Exp3} on an aluminum 7000 series Al-Zn-Mg-Cu-Sc-Zr alloy reinforced with ${\rm Al}_3{\rm (Sc,Zr)}$ particles. The aluminum alloy, designated as SSA018, was subjected to a solution treatment (ST) temperature of $480^{\circ}C$, with a heat rate of $20^{\circ}C/h$ and different soaking times (1, 4, 20, and 48 h). After ST and water quenching, some of the material was subjected to artificial aging at $120^{\circ}C$ during 19 h (T6 temper) and some material was not (W24 temper). The treatment resulted in spherical ${\rm Al}_3{\rm (Sc,Zr)}$ particles with diameters reported to be in the range 7.9 nm to 27.3 nm, with a volume fraction that increased with the diameter from 0.16\% to 0.42\%. The initial yield stress of the matrix is reported to be 187 MPa for the W24 temper and 417 MPa for the T6 temper, respectively. By applying relation \eref{eqn:SnlocExp} to experimental data, it was found that $\alpha_0 \ell$ equal to 0.44 microns and 0.22 microns well captured the  W24 temper and T6 temper, respectively, as can be seen in Figure \ref{fig:fig17}b. A possible explanation for the difference in the estimated values of $\alpha_0 \ell$ may be that precipitates get more diffused at higher temperatures and become more coherent with the matrix, which in turn may lead to a weaker interface with a lower value of $\alpha_0$ as a consequence. Finally, by applying the proposed model to the experiments considered here gives suggestions for a lower bound (micro-hard interface) for the material length parameter, i.e. $ \ell \ge 0.89$ microns for Mg and $ \ell \ge 0.54$ microns for Aluminium 7000 series. This is in accordance with values reported in the literature, e.g., in \cite{Dunstan16}.

\section{Conclusions}
\label{sec:5}

\noindent A micro-mechanical model for strengthening and hardening due to a distribution of small particles embedded in an otherwise homogeneous matrix is developed based on the higher order strain gradient plasticity theory by \cite{Gudmundson04}. Within this modelling framework it was found that strengthening is controlled by the formulation of the particle/matrix interface, which is a key element of the model. A closed form expression for the overall increase in yield stress was deduced from the systematic parametric study carried. The proposed expression is described in equations \eref{eqn:Fs_function}, \eref{eqn:Sloc}, and \eref{eqn:SnlocGen}. An outcome of \eref{eqn:SnlocGen} is that each particle contributes to the overall increase in yield stress by its surface area times its interface strength, where the influence of volume fraction and size scale is inherent in the ratio between the overall surface area of particles and the total volume of the material.

The validity of the proposed model is limited to small volume fractions, $f < 0.1$, and particle spacings over material length scale, $L_{\rm p} / \ell < 1$. If these conditions are met, the plastic strain field developing in the matrix appears to be essentially constant, except in the close vicinity of a particle where a decrease in plastic strain levels occur which is governed by the interface strength parameter $\alpha_0$. The proposed model is able to capture the experimentally observed strengthening in some metal matrix composites and alloys published in the literature. 

The mismatch in the particle/matrix modulus affects strengthening such that particles with a higher elastic modulus amplify the increase in yield stress and vice versa. No attempt was made to describe this in detail.

Even though the 2D-axisymmetric model presently employed is an approximation of a micro-structure with particles distributed in a hexagonal pattern, it can in fact in the light of the proposed closed form relation \eref{eqn:SnlocGen} be viewed as a material system of its own. Nevertheless, particle distributions in three dimensions have been investigated within the proposed modeling framework in a forthcoming paper.

\section*{Acknowledgments}
\noindent This work was performed within the VINN Excellence Center Hero-m, financed by VINNOVA, the Swedish Governmental Agency for Innovation Systems, Swedish industry, and KTH Royal Institute of Technology. The authors are grateful to Dr. Carl Dahlberg, KTH, for providing the plane strain SGP-FEM code on which the version used in the present work was developed. The authors also appreciate fruitful discussions with Prof. Peter Gudmundson.

\begin{appendices}
\renewcommand{\theequation}{\Alph{section}.\arabic{equation}}
\counterwithin*{equation}{section}

\section{Finite element discretization and implementation}
\label{appendA}

\noindent Displacements and plastic strains within an element are interpolated by use of nodal values and their corresponding shape functions in a standard manner as
\begin{equation}
\left[ \begin{array}{c} \vect{u} \\ \vecsym{\varepsilon}^{\rm p} \end{array} \right]_{(2+4)\times1} = \left[ \begin{array}{cc} \mathds{N}_{\rm u} & \mathds{O} \\ \mathds{O} & \mathds{N}_{\rm p} \end{array} \right] \left[ \begin{array}{c} \vect{d}_{\rm u} \\ \vect{d}_{\rm p} \end{array} \right],
\end{equation}
where $\vect{u}^{T} = [u_{r} \quad u_{z} ]$, and $(\vecsym{\varepsilon}^{\rm p})^{T} = [{\epsilon_{rr}^{{\rm p}}} \quad {\epsilon_{zz}^{{\rm p}}} \quad {\epsilon_{\phi\phi}^{{\rm p}}} \quad {\gamma_{rz}^{{\rm p}}}]$, $ \mathds{O} $ is the zero matrix of appropriate size, $\mathds{N}_{\rm u}$ is the shape function matrix for displacements, and $\mathds{N}_{\rm p}$ is the shape function matrix for plastic degrees of freedom:
\begin{equation}
\mathds{N}_{\rm u} = \left[ \begin{array}{ccccccc}
N_{\rm u}^{1} & 0 & N_{\rm u}^{2} & 0 & \ldots & N_{\rm u}^{8} & 0 \\
0 & N_{\rm u}^{1} & 0 & N_{\rm u}^{2} & \ldots & 0 & N_{\rm u}^{8}
\end{array} \right]_{2 \times 16}
\end{equation}
\begin{equation}
\mathds{N}_{\rm p} = \left[ \begin{array}{ccccccc}
N_{\rm p}^{1}  & 0              & 0             & \ldots & N_{\rm p}^{4}  & 0              & 0             \\
0              & N_{\rm p}^{1}  & 0             & \ldots & 0              & N_{\rm p}^{4}  & 0             \\
-N_{\rm p}^{1} & -N_{\rm p}^{1} & 0             & \ldots & -N_{\rm p}^{4} & -N_{\rm p}^{4} & 0             \\
0              & 0              & N_{\rm p}^{1} & \ldots & 0              & 0              & N_{\rm p}^{4}
\end{array} \right]_{4 \times 12}.
\end{equation}
The third row in $\mathds{N}_{\rm p}$ enforces plastic incompressibility. The relevant strain-like fields
\begin{align}
(\vecsym{\varepsilon}^{\rm e})^{\rm T} & = [ \varepsilon^{\rm e}_{rr} ~ \varepsilon^{\rm e}_{zz} ~ \varepsilon^{\rm e}_{\phi\phi} ~ \gamma^{\rm e}_{rz}], \notag \\
(\vecsym{\varepsilon}^{\rm p})^{\rm T} & = [ \varepsilon^{\rm p}_{rr} ~ \varepsilon^{\rm p}_{zz} ~ \varepsilon^{\rm p}_{\phi\phi} ~ \gamma^{\rm p}_{rz}], \\
\vecsym{\eta} = (\nabla\vecsym{\varepsilon}^{\rm p})^{\rm T} & = [ \varepsilon^{\rm p}_{rr,r} ~ \varepsilon^{\rm p}_{zz,r} ~ \varepsilon^{\rm p}_{\phi\phi,r} ~ \gamma^{\rm p}_{rz,r} ~ \varepsilon^{\rm p}_{rr,z} ~ \varepsilon^{\rm p}_{zz,z} ~ \varepsilon^{\rm p}_{\phi\phi,z} ~ \gamma^{\rm p}_{rz,z} ~ \gamma^{\rm p}_{r\phi,\phi} ~ \gamma^{\rm p}_{z\phi,\phi}], \notag
\end{align}
can also be expressed using the nodal DOFs as
\begin{equation}
\left[ \begin{array}{c} \vecsym{\varepsilon}^{\rm e} \\ \vecsym{\varepsilon}^{\rm p} \\ \vecsym{\eta} \end{array} \right]_{(4+4+10)\times1} = \left[ \begin{array}{cc} \mathds{B}_{\rm u} & -\mathds{N}_{\rm p} \\ \mathds{O} & \mathds{N}_{\rm p} \\ \mathds{O} & \mathds{B}_{\rm p} \end{array} \right] \left[ \begin{array}{c} \vect{d}_{\rm u} \\ \vect{d}_{\rm p} \end{array} \right] = \mathds{B} \left[ \begin{array}{c} \vect{d}_{\rm u} \\ \vect{d}_{\rm p} \end{array} \right],
\end{equation}
where $\vecsym{\eta}$ is the gradient of plastic strain field (see Appendix \ref{appendB}) and $ \mathds{B} $ is the gradient operator matrix of size $ 18 \times 28 $. The geometry is interpolated using the quadratic interpolation so that the description of displacements is isoparametric. Specifically, matrix $ \mathds{B}_{\rm u} $ is
\begin{equation}
\mathds{B}_{\rm u} = \left[ \begin{array}{ccccc}
(B_{\rm u}^{1})_{r} & 0                   & \ldots & (B_{\rm u}^{8})_{r} & 0                   \\
0                   & (B_{\rm u}^{1})_{z} & \ldots & 0                   & (B_{\rm u}^{8})_{z} \\
N_{u}^{1}/r         & 0                   & \ldots & N_{u}^{8}/r         & 0                   \\
(B_{\rm u}^{1})_{z} & (B_{\rm u}^{1})_{r} & \ldots & (B_{\rm u}^{8})_{z} & (B_{\rm u}^{8})_{r}
\end{array} \right]_{4 \times 16} ,
\end{equation}
and matrix $ \mathds{B}_{\rm p} $ is
\begin{equation}
\mathds{B}_{\rm p} = \left[ [\mathds{B}_{\rm p}^{1}] \quad [\mathds{B}_{\rm p}^{2}] \quad [\mathds{B}_{\rm p}^{3}] \quad [\mathds{B}_{\rm p}^{4}] \right]_{10 \times 12} ,
\end{equation}
with each sub-matrix $ I $ being defined as follows
\begin{equation}
\vecsym{\eta} = \left[ \begin{array}{c} \eta_{rrr} \\ \eta_{zzr} \\ \eta_{\phi\phi r} \\ \eta_{rzr} \\ \eta_{rrz} \\ \eta_{zzz} \\ \eta_{\phi\phi z} \\ \eta_{rzz} \\ \eta_{r\phi\phi} \\ \eta_{z\phi\phi}
\end{array} \right] = \left[ \begin{array}{ccc}
(B_{\rm p}^{I})_{r}  & 0                    & 0                     \\
0                    & (B_{\rm p}^{I})_{r}  & 0                     \\
-(B_{\rm p}^{I})_{r} & -(B_{\rm p}^{I})_{r} & 0                     \\
0                    & 0                    & (B_{\rm p}^{I})_{r} \\
(B_{\rm p}^{I})_{z}  & 0                    & 0                     \\
0                    & (B_{\rm p}^{I})_{z}  & 0                     \\
-(B_{\rm p}^{I})_{z} & -(B_{\rm p}^{I})_{z} & 0                     \\
0                    & 0                    & (B_{\rm p}^{I})_{z} \\
4/r                  & 2/r                  & 0                     \\
0                    & 0                    & 1/r
\end{array} \right] \left[ \begin{array}{c} \varepsilon_{r}^{I} \\ \varepsilon_{z}^{I} \\ \gamma_{rz}^{I}
\end{array} \right] = \mathds{B}_{\rm p}^{I} \left[ \begin{array}{c} \varepsilon_{r}^{I} \\ \varepsilon_{z}^{I} \\ \gamma_{rz}^{I}
\end{array} \right ] ,
\end{equation}
where the spatial gradients of the shape functions are found through
\begin{equation}
(B_{\rm p}^{I})_{j} = \frac{\partial N_{\rm p}^{I}}{\partial x_{j}} = J_{ij}^{-1}\frac{\partial N_{\rm p}^{I}}{\partial \xi_{i}} .
\end{equation}
Here, the Jacobian coefficients are taken to be the same as for the quadratic interpolation in order to have a consistent spatial description (\cite{Fredriksson09}).

The stress-like quantities are collected in a vector
\begin{equation}
\vect{s} = \left[ \begin{array}{c} \vecsym{\sigma} \\ \vect{q} \\ \vect{m} \end{array} \right]
\end{equation}
where
\begin{equation}
\vecsym{\sigma}^{\rm T} = [ \sigma_{rr} \quad \sigma_{zz} \quad \sigma_{\phi\phi}  \quad \sigma_{rz}],
\end{equation}
\begin{equation}
\vect{q}^{\rm T} = [ q_{rr} \quad q_{zz} \quad q_{\phi\phi}  \quad q_{rz}],
\end{equation}
\begin{equation}
\vect{m}^{\rm T} = [ m_{rrr} \quad m_{zzr} \quad \ldots \quad m_{z\phi\phi} ],
\end{equation}
and the indices for $\vect{m}^{\rm T}$ appear in the same order as in the $ \vecsym{\eta} $-vector. The components of the standard tractions and the higher order moment tractions are $ T_{i} = \sigma_{ij}n_{j} $ and $ M_{ij} = m_{ijk}n_{k} $, respectively, where the latter may be given a matrix representation as
\begin{equation}
\left[ M_{ij} \right] = \left[ \begin{array}{ccc}
m_{rrr}n_{r} + m_{rrz}n_{z} & m_{rzr}n_{r} + m_{rzz}n_{z}  & m_{r\phi\phi}n_{\phi} \\
m_{zrr}n_{r} + m_{zrz}n_{z} & m_{zzr}n_{r} + m_{zzz}n_{z}  & m_{\phi z \phi}n_{\phi}     \\
m_{\phi r \phi}n_{\phi}     & m_{z\phi\phi}n_{\phi}        & m_{\phi\phi r}n_{r} + m_{\phi\phi z}n_{z}
\end{array} \right].
\end{equation}
Since $ M_{ij} $ is symmetric, it is convenient to collect all traction components in a vector as
\begin{equation}
\vect{t} = \left[ \begin{array}{c} \vect{T} \\ \vect{M} \end{array} \right]  = \left[ \begin{array}{c} \sigma_{rr}n_{r} + \sigma_{rz}n_{z} \\ \sigma_{rz}n_{r} + \sigma_{zz}n_{z}  \\ m_{rrr}n_{r} + m_{rrz}n_{z}  \\ m_{zzr}n_{r} + m_{zzz}n_{z} \\ m_{\phi\phi r}n_{r} + m_{\phi\phi z}n_{z} \\  m_{rzr}n_{r} + m_{rzz}n_{z}  \end{array} \right].
\end{equation}

\subsection{Finite element discretization of interface formulation}

\noindent We distinguish between two sides of an interface element: (1) is the particle side and (2) is the matrix side. Since particles are assumed to be elastic, plastic strain field is present only on side (2). The interpolation of this field using nodal values is done through
\begin{equation}
\vecsym{\varepsilon}^{\rm p(2)} = \mtrx{N}_{\rm p} \vect{d}_{\rm p}^{\rm (2)},
\end{equation}
where $ \mtrx{N}_{\rm p} $ has the same structure as the corresponding matrix for bulk elements. The interface contribution to the force balance is then equal to
\begin{equation}
\vect{f}_{\Gamma} = \int_{S^{\Gamma}} \mtrx{N}_{\rm p}^{\rm T}\vect{M}_{\Gamma}{\rm d}S,
\end{equation}
in which $\vect{M}_{\Gamma}$ is the work conjugate to plastic strain at the interface.

\subsection{Enforcing equilibrium}

\noindent The residual force vector is obtained by discretizing the force balance equation over the entire domain: 
\begin{align}\label{eqn:FEequilibrium}
\vect{r} & = \vect{f}_{V} + \vect{f}_{S^{\Gamma}} - \vect{f}_{S^{\rm ext}} = \\
  & = \int_V \mathds{B}^{\rm T} \vect{s}~{\rm d}V + \int_{S^{\Gamma}} \mtrx{N}_{\Gamma}^{\rm T}\vect{T}_{\Gamma}{\rm d}S - \int_{S^{\rm ext}} \mathds{N}^{\rm T} \vect{T}~{\rm d}S. \notag
\end{align}
The nonlinear set of equations in \eref{eqn:FEequilibrium} is solved incrementally by a consistent Euler-backward scheme. Considering a load step between $ (t) $ and $ (t+\Delta t) $ and denoting an iteration within a load step by superscript $ k $, linearization of \eref{eqn:FEequilibrium} by Taylor expansion at iteration $ k $ gives
\begin{equation}\label{eqn:FEequLinearized}
\vect{r}^{k+1} = \vect{r}^{k} + \frac{\partial \vect{r}^{k}}{\partial \mathbf{d}} \Delta\mathbf{d}^{k+1}.
\end{equation}
The external forces are independent of $ \mathbf{d} $, hence
\begin{equation}\label{eqn:partialderivforKtan}
\frac{\partial \vect{r}^{k}}{\partial \mathbf{d}} = \int_V \mathds{B}^{\rm T} \frac{\partial \mathbf{s}^{k}}{\partial \boldsymbol{\epsilon}} \mathds{B}~{\rm d}V + \int_{S^{\Gamma}} \mathds{N}_{\Gamma}^{\rm T}\frac{\partial \mathbf{T}_{\Gamma}^{k}}{\partial \boldsymbol{\upsilon}}\mathds{N}_{\Gamma} ~{\rm d}S,
\end{equation}
where the partial derivatives can be found from the constitutive relationships and use of the Euler-backward method and expressed as the consistent material point stiffness matrices $ \mathds{D} $ such that
\begin{equation}
\int_V \mathds{B}^{\rm T} \mathds{D}_V \mathds{B}~{\rm d}V + \int_{S^{\Gamma}} \mathds{N}_{\Gamma}^{\rm T} \mathds{D}_{\Gamma} \mathds{N}_{\Gamma} ~{\rm d}S = \mathds{K}_{\rm tan},
\end{equation}
which defines the tangent stiffness $ \mathds{K}_{\rm tan} $. By the choice $ \vect{r}^{k+1} = \vect{0} $, the iterative change is found as
\begin{equation}
\Delta \mathbf{d}^{k+1} = - (\mathds{K}_{\rm tan})^{-1} \mathbf{r}^{k}.
\end{equation}

\section{Gradients of an axi-symmetric plastic strain field in a cylindrical coordinate system}
\label{appendB}
\noindent The gradient of the plastic strain field in a cylindrical coordinate system is briefly outlined here, following the standard procedure for covariant differentiation of tensor fields, e.g., see \cite{ycfung}. In cylindrical coordinates $ x^{i} = \{ r,z,\phi \} $, the physical components of the strain tensor and the displacement vector, respectively, are related as
\begin{equation}
\left[ \begin{array}{c} \varepsilon_{rr}  \\ \varepsilon_{zz} \\ \varepsilon_{\phi\phi}  \\ \gamma_{rz} \\ \gamma_{r\phi} \\ \gamma_{\phi z} \end{array} \right] = \left[ \begin{array}{ccc} \partial_{r} & 0 & 0  \\ 0 & \partial_{z} & 0 \\ 1/r & 0 & \partial_{\phi}/r  \\ \partial_{z}& \partial_{r} & 0  \\ \partial_{\phi}/r & 0 & \partial_{r} - 1/r \\ 0 & \partial_{\phi}/r & \partial_{z} \end{array} \right] \left[ \begin{array}{c} u_{r} \\ u_{z} \\ u_{\phi} \end{array} \right],
\end{equation}
where $ \gamma_{ij} = 2\varepsilon_{ij}$ and $ \partial_{i} $ denotes the partial differentiation operator with respect to coordinate $ x^{i} $. In an axi-symmetric problem (2D), $ u_{\phi} = 0$ and variables are independent of $ \phi $, hence $ \partial_{\phi} = 0 $.

The gradient of the plastic strain tensor with respect to a set of curvilinear coordinates $\xi^{k}$ may in general be expressed as
\begin{equation}\label{eqn:grad2ndTensor}
\boldsymbol{\nabla}\boldsymbol{\varepsilon}^{\rm p} =  \left[ \dfrac{\partial \varepsilon^{\rm p}_{ij}}{\partial \xi^{k}} - \Gamma_{ki}^{l}\varepsilon^{\rm p}_{lj} - \Gamma_{kj}^{l}\varepsilon^{\rm p}_{il} \right] \mathbf{g}^{i}\otimes\mathbf{g}^{j}\otimes\mathbf{g}^{k} = \eta_{ijk} \mathbf{g}^{i}\otimes\mathbf{g}^{j}\otimes\mathbf{g}^{k} ,
\end{equation}
where $ \mathbf{g}^{k} $ represent the contravariant basis vectors, and where 
\begin{equation}\label{Christoffel}
\Gamma^{i}_{kj} = \frac{1}{2} g^{im}\left( \frac{\partial g_{mk}}{\partial \xi^{j}} + \frac{\partial g_{mj}}{\partial \xi^{k}} - \frac{\partial g_{kj}}{\partial \xi^{m}} \right)
\end{equation}
is the Euclidean Christoffel symbols. In a cylindrical coordinate system $ \{ r,z,\phi \} $, the only nonzero symbols are $ \Gamma^{3}_{13} = \Gamma^{3}_{31} = 1/r $ and $ \Gamma^{1}_{33} = -r $. As mentioned above, from axi-symmetric conditions it follows that $ \varepsilon_{r\phi}^{\rm p} = \varepsilon_{\phi r}^{\rm p} = \varepsilon_{z\phi}^{\rm p} = \varepsilon_{\phi z}^{\rm p} = 0 $, and that $ \partial\varepsilon_{ij}^{\rm p}/\partial\phi = 0 $. In summary, there are only 10 non-zero and independent components of $ \boldsymbol{\eta}$. These are listed in Table \ref{EtaTable}. As can be seen, two additional terms ($\eta_{r \phi \phi}$, $\eta_{z \phi \phi}$) arise in comparison to a planar 2D problem formulated in a Cartesian basis.
\begin{table}[h]
    \begin{center}		
    \begin{tabular}{lclcl}
		\hline
        $ \langle\eta_{rrr}\rangle = \partial_{r}\varepsilon_{rr}^{\rm p} $ & $ \quad $ & $ \langle\eta_{rrz}\rangle = \partial_{z}\varepsilon_{rr}^{\rm p} $ \\
				$ \langle\eta_{zzr}\rangle = \partial_{r}\varepsilon_{zz}^{\rm p} $ & $ \quad $ & $ \langle\eta_{zzz}\rangle = \partial_{z}\varepsilon_{zz}^{\rm p} $ & $ \quad $ & $ \langle\eta_{r \phi \phi}\rangle = 2(\varepsilon_{rr}^{\rm p} - \varepsilon_{\phi\phi}^{\rm p})/r $ \\
				$ \langle\eta_{\phi\phi r}\rangle = \partial_{r}\varepsilon_{\phi\phi}^{\rm p} $ & $ \quad $ & $ \langle\eta_{\phi\phi z}\rangle = \partial_{z}\varepsilon_{\phi\phi}^{\rm p} $ & $ \quad $  & $ \langle\eta_{z \phi\phi}\rangle = \gamma_{zr}^{\rm p}/r $ \\
        $ \langle\eta_{rzr}\rangle = \partial_{r}\gamma_{rz}^{\rm p} $ & $ \quad $ & $ \langle\eta_{rzz}\rangle = \partial_{z}\gamma_{rz}^{\rm p} $ \\  
			\hline
    \end{tabular}\caption{Non-zero physical components, $ \langle\eta_{ijk}\rangle$, of the gradient of plastic strain tensor in an axi-symmetric, cylindrical coordinate system.}		
\label{EtaTable}
    \end{center}
\end{table}

\section{Choice of visco-plastic parameters and mesh}
\label{appendC}
\noindent The current study aims at obtaining rate-independent solutions. Keeping this in mind, the accuracy of the overall numerical solution will depend on the choice of visco-plastic parameters and on the condition at the particle/matrix interface. The latter is important, as it governs the strength of the plastic strain gradient that develops at the interface. For instance, below the micro-hard limit ($\alpha_0 < \sqrt{2/3}$), the plastic strain gradient that evolves at the particle/matrix interface is rather weak. However, if $\alpha_0 > \sqrt{2/3}$ or if a pure micro-hard condition is applied at the particle/matrix interface, the evolving plastic strain gradient is extremely strong. The presence of strong plastic strain gradients significantly affects the degree of accuracy that can be obtained for a given set of visco-plastic parameters.

\begin{figure}[!htb]
	\begin{center}
		\includegraphics[width=0.96\textwidth]{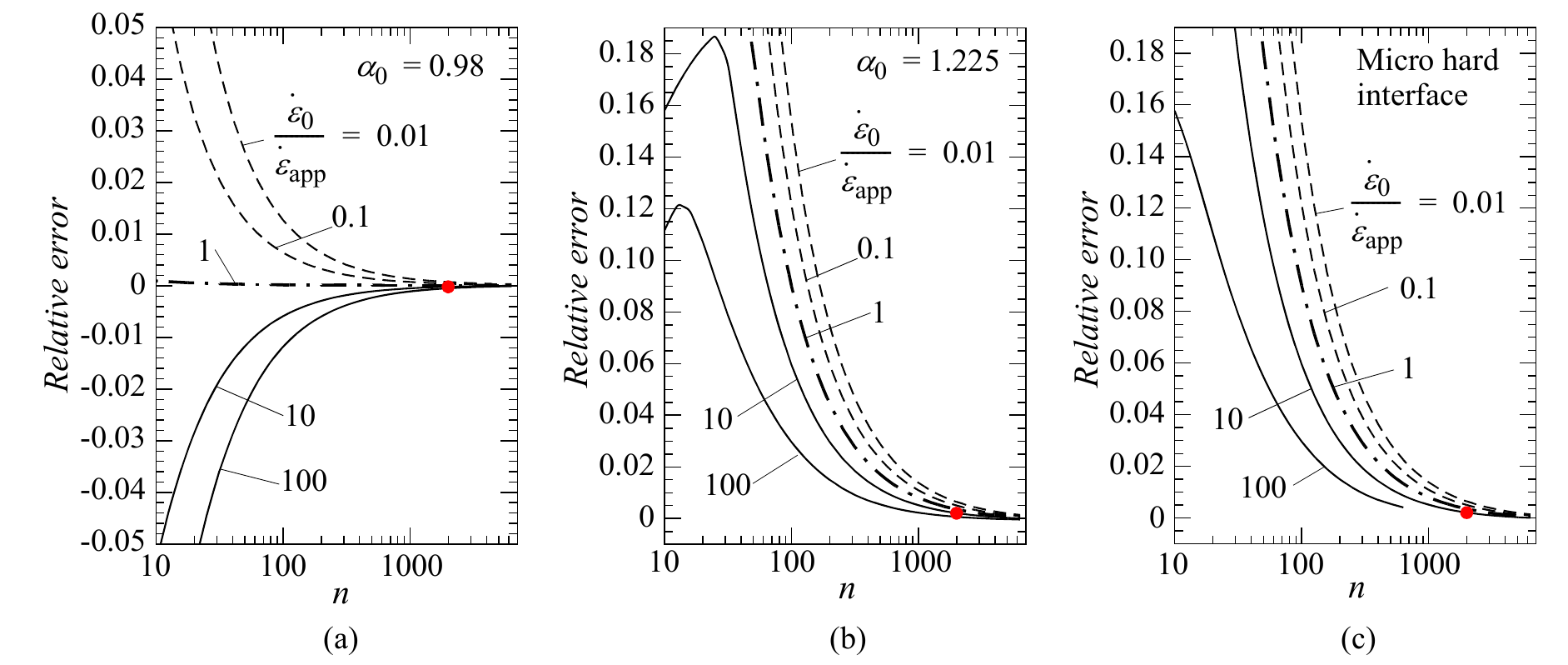}
		\caption{Relative error of the overall solution versus the visco-plastic exponent $n$ for five different ratios between the visco-plastic reference strain rate and the average effective applied strain rate ($\dot{\varepsilon}_{0}/\dot{\varepsilon}_{\rm app}$), and for three different interface conditions. (a) $\alpha_0 = 0.8$, (b) $\alpha_0 = 1.0$ and in (c) a purely micro-hard condition is applied at the particle/matrix interface.}
		\label{fig:Visco_param_conv}
	\end{center}
\end{figure}

A limited convergence study was carried out to guide the choice of the visco-plastic parameters $n$ and $\dot{\varepsilon}_{0}$. The result is shown in Figure \ref{fig:Visco_param_conv}, where the relative error in the increase in yield stress is plotted versus the exponent $n$. The relative error was calculated as $(\sigma_{\rm p}-\sigma_{\rm p,conv})/\sigma_{\rm p,conv}$, where $\sigma_{\rm p,conv}$ was evaluated with $n = 6310$ and $\dot{\varepsilon}_{0}/\dot{\varepsilon}_{\rm app} = 10$, where $ \dot{\varepsilon}_{\rm app} $ is the strain rate applied on the boundary of the unit cell. The result suggests that when $\alpha_0 < \sqrt{2/3}$, accurate solutions may be obtained  if $\dot{\varepsilon}_{0}/\dot{\varepsilon}_{\rm app} = 1$ regardless of $n$ in the range studied. By contrast, if a micro-hard condition prevails at the interface, $\dot{\varepsilon}_{0}$ and especially $n$ should be chosen sufficiently high to ensure an accurate solution, as can be seen in Figure \ref{fig:Visco_param_conv}(b,c). Based on the outcome from this convergence study, the parameter choice: $\dot{\varepsilon}_{0}/\dot{\varepsilon}_{\rm app} = 10$ and $n = 2000$, was employed in the present study, which is marked by a red dot in Figure \ref{fig:Visco_param_conv}.

In addition, a limited mesh convergence study was performed with parameters $ f = 0.02 $ and $ L_{\rm p}/\ell = 0.1 $, and where an extremely fine mesh was used as a converged reference solution. For a micro-hard interface, the relative error (defined above) was approximately given by $ 2.4 (\Delta r / r_{\rm p})^{1.3} $, where $\Delta r$ is equal to the radial extent of the element closest to the particle in the matrix. E.g., if $ \Delta r/ r_{\rm p} = 0.002$, the relative error is expected to be less than 0.001. For an interface with $\alpha_0 < \sqrt{2/3}$, a rather coarse mesh with $ \Delta r/ r_{\rm p} < 0.1$ proved to be sufficient.

\end{appendices}

\bibliography{ReferencesToA}     

\end{document}